\pdfoutput=1
\documentclass[a4paper,onecolumn]{article}
\RequirePackage[T1]{fontenc}
\usepackage[a4paper,
            bindingoffset=0.2in,
            left=0.2in,
            right=0.25in,
            top=0.5in,
            bottom=0.7in,
            footskip=.4in]{geometry}
\RequirePackage{graphicx}
\RequirePackage{mathptmx}      
\usepackage{amstext}
\usepackage{amssymb,mathtools}

\RequirePackage{flushend}
\RequirePackage[authoryear,comma,sort&compress]{natbib}
\RequirePackage[colorlinks,citecolor=blue,urlcolor=blue,linkcolor=blue]{hyperref}
\usepackage{microtype}
\usepackage{tikz}
\usepackage{lineno}
\usepackage{float}
\usetikzlibrary{decorations.markings}
\usetikzlibrary{arrows.meta}
\usepackage{caption}
\providecommand{\keywords}[1]{\textbf{\textit{Keywords:}} #1}
\begin{document}

\title{Three-Body Problem in Modified Dynamics}

\author{Hossein Shenavar}

\author{Hossein Shenavar\thanks{E-mail: h.shenavar@mail.um.ac.ir}
\\
Department of Physics, Ferdowsi University of Mashhad, P.O. Box 1436, Mashhad, Iran.
}

\maketitle

\begin{abstract}
General properties of the three-body problem in a model of  modified dynamics are investigated. It is shown that the three-body problem in this model shares some characters with the similar problem in Newtonian dynamics.  Moreover, the  planar  restricted three-body problem is solved analytically for this type of extended gravity and it is proved that under certain conditions, which generally happen at galactic and extragalactic scales, the orbits around $L_{4}$ and $L_{5}$ Lagrange points are stable. Furthermore, a code is provided to compare the behavior of orbits in the  restricted three-body problem under Newtonian and modified dynamics. Orbit integrations based on this code show contrasting orbital behavior under the two dynamics and specially exhibit in a  qualitative way that the rate of ejections is smaller in the modified dynamics  compared to Newtonian gravity. These results could help us to search for observational signatures of extended gravities.
\end{abstract}


\section{Introduction}
The three-body problem, which is written based on Newton's second law and Newton's law of universal gravitation for three particles, is arguably the oldest problem of physics without a general analytic solution ( however; see the proposal of \cite{sundman1913memoire} and also \cite{BARROWGREEN2010164} ). Although some special solutions have been found under rather specific assumptions and approximations, a general solution is still under investigation. This is while three-body systems  could be found in various scales such as  the solar system ( satellites, comets, planets etc. ), binary stars, binary black holes,  galaxies etc. \cite{2006tbp..book.....V}. Even the appearance of stellar tidal tails - which for instance has been reported  around the Large and the Small Magellanic Clouds  in the Gaia  data \citep{2017MNRAS.466.4711B} - is interpreted as a consequence of the three-body interactions \citep{2017MNRAS.472.2554B}. 

 Admittedly, the case of galactic three-body problem, hereafter GTBP, is somehow particular among the examples of the three-body problem because the usual size of galaxies is quite comparable to their distance in their group. If the size of the central galaxy and its distance to its companion are shown by $l$ and $L$ respectively, the  tidal force due to the companion on the central object is proportional to $l/L$ which is much larger in galaxy-galaxy interactions  compared to other systems, e.g. star-star and star-planet systems. In addition, the mass discrepancy problem \citep{2012LRR....15...10F,2014dmp..book.....S} shows   itself vividly at galactic scales which makes the galaxy-galaxy interactions even more interesting. Historically, GTBP has been the source of many astrophysical challenges and controversies, apart from the mathematical and numerical problem involved \citep{1972ApJ...178..623T}. Debates on whether a companion galaxy could excite the spiral wave in a disk, on the formation of galactic tails and bridges or on the question of whether the merger of galaxies is the origin of all ellipticals, still continues. See \cite{2002gaco.book.....C}, chapter 7, for an excellent review. The existence of the dark matter halo around galaxies could  complicate the matter even more.

The interaction between galaxies is usually investigated within the context of the standard model of cosmology, also named $\Lambda CDM$ model. This model aims to explain the physics of the cosmos by including  observable ingredients such as baryons, radiation etc.,  as well as two components which are yet to be explained, i.e. cold dark matter $CDM$ and the cosmological constant $\Lambda$. Besides numerous successes of this model in explaining the cosmos at large scale \citep{2022arXiv220104741T}, the search for the detection of various dark matter candidates is still ongoing \citep{2010ARA&A..48..495F,2016JPhG...43a3001M} while there are some astrophysical challenges at smaller scales. For example, one may name the missing satellite problem, the too-big-to-fail problem, a too empty local void and the core-cusp problem \citep{2017ARA&A..55..343B}. On the other hand, there are some evidence that a modification of gravity or dynamics could solve the missing mass problem; though, many issues need to be addressed yet \citep{2012PhR...513....1C,2012LRR....15...10F}.

The main aim of this work is to provide a framework by which we could compare the special modified dynamics under consideration here with Newtonian gravity, and other extended gravities, through careful examination of three-body systems. Especially, we are searching for generic differnces between different gravity models as well as scales of time and size for which one would expect to see observable effects.  To do so, we will investigate the three-body problem in a modified dynamical model.  The basic idea behind the model that we use here  is to explore the possibility that the expansion of the universe might affect the local physics. If the frame of reference is determined relative to the very distant stars, i.e. the average distribution of matter at cosmic scales, then one might expect that the cosmic evolution of matter  influence the local dynamics.  To build such model, \cite{2016Ap&SS.361...93S,2019Univ....6....1S} impose    Neumann boundary condition, instead of Dirichlet boundary condition, on cosmological perturbation equations and derive the consequences. We will review the properties of this model in the next section. There, we also show that the three-body problem in this modified dynamics (MOD) shares some of the basic characteristics of the three-body formulation in Newtonian dynamics. Then, in Sec. \ref{Restricted}, we will solve the restricted three-body problem (RTBP) in MOD and derive the modified stability criterion for $L_{4}$ and $L_{5}$  Lagrange points. Furthermore, we will present a simple MATHEMATICA code to compare restricted three-body systems in Newtonian and MOD dynamics ( Sec. \ref{Lab} ).  Some general remarks and thoughts on the problem are provided in Sec. \ref{Conclusion}. The MATHEMATICA codes related to orbit integration in this work are available on GitHub\footnote{https://github.com/Shenavar/ThreeBodyinMOD}.

\section{General Properties of the Model} 
\label{Review}
According to the current view toward the theory of gravity, i.e. Einstein's general relativity,  the interaction of two local point masses is independent of the cosmic matter at large scales.  However, Einstein's theory could also contain a type of top-down causation \citep{2013SHPMP..44..242E} through boundary conditions. Einstein field equations contain  ten nonlinear partial differential equations with four constraints; thus, in fact they consist of six degrees of freedom. This set of partial differential equations includes large classes of solutions; though, most of these solutions are unphysical. In order to find the unique physical solution which is in accordance with the nature of the problem, one must also prescribe a suitable boundary condition.  In fact, in case one observes contradictions between astrophysical phenomena and Einstein's field equations, one might suspect the field equations as well as the prescribed boundary condition. \cite{wheeler1964mach} had interpreted the  implementation of boundary conditions on Einstein field equations  as Mach's principle; though, there are other - sometimes conflicting - statements of Mach's principle too \citep{Barbour:1995iu,Bondi:1996md}.

In the present work, our model presumes a connection between global and local physics which is derived by imposing Neumann boundary condition on cosmological perturbation equations, i.e. Taylor expansion of Einstein field equations presuming a flat Friedmann-Lemaitre-Robertson-Walker (FLRW) spacetime disturbed by the perturbation of the local universe \citep{2016Ap&SS.361...93S,2019Univ....6....1S}.  Thus, this model might be considered as a Machian model. As a result of imposing Neumann boundary condition on $i,j \neq i$ components of the cosmological perturbation equations, i.e. $\partial_{i}\partial_{j}\left( \Phi -\Psi \right)=0$  while $\Phi $ is the gravitational potential and $\Psi $ is the 3-curvature perturbation,  the scalar fields would differ by a time-dependent constant: $\Phi - \Psi =c_{1}(t)$. Here, the  boundary is defined by the particle horizon and it encompasses all the objects which are already in causal contact with the central object under consideration, i.e. source of the local gravitational field. By implementing this new boundary condition,  a new term proportional to $2c_{1} cH(t)$, $c$ being the speed of light and $H(t)$ being the Hubble parameter, appears in the  equation of motion of photons \citep{2019Univ....6....1S}. Since the measurement process in general relativity is defined based on photons and freely falling particles, i.e. the method of geometrodynamic clocks introduced by \cite{marzke1964gravitation}, a modification in the equation of motion of photons would essentially change the equation of motion of massive particles. This new term is strictly a dynamical term, i.e. it appears on the dynamical side of the Newton's second law. However, since using the formalism of potential theory is more convenient,   one could transfer this new term to the gravity side of the equation of motion and derive the modified force as \citep{2016Ap&SS.361..378S}
\begin{eqnarray} \label{Force}
m\frac{d^{2}\vec{r}}{dt^{2}} &=& \vec{F}(\vec{r})  \\  \nonumber  
\vec{F}(\vec{r}) &=& Gm \sum^{N}_{i=1}m_{i}\frac{\vec{r}_{i}-\vec{r}}{ \mid \vec{r}_{i} - \vec{r} \mid^{3}}+\frac{2c_{1}a_{0}m}{M} \sum^{N}_{i=1}m_{i}\frac{ \vec{r}_{i}-\vec{r}}{\mid \vec{r}_{i} - \vec{r} \mid}
\end{eqnarray} 
for a system of particles. Here,  $\vec{F}(\vec{r})$ is the force exerted on a particle with mass $m$ which is located at $\vec{r}$, due to a system of particles each with mass $m_{i}$ and located at $\vec{r}_{i}$. Also,  $M=\sum^{N}_{i=1}m_{i} $ is the total mass of the system of particles, $N$ is their total number and  $a_{0}\equiv cH_{0} = 6.59 \times 10^{-10} m s^{-2} $ is the fundamental acceleration of the model.  In what follows, we   use  $c_{1}=0.065$ in our calculations and codes. According to   previous results, this value seems to be an upper bound of the parameter $c_{1} $  \citep{2016Ap&SS.361...93S, 2016Ap&SS.361..378S}.

In showing a dependency to a fundamental acceleration $a_{0}$, the present model is similar to MOdified Newtonian Dynamics (MOND) proposed by Milgrom which presumes a modification in Newton's second law at small accelerations \citep{Milgrom:1983ca,Milgrom:1983pn,Milgrom:1983zz}. For example, the local stability of galactic disks in both models depends on the ratio of surface density of the disk to a critical surface density defined as $\Sigma_{\dagger}=a_{0}/G =  9.9~ kg m^{-2}=4.7\times 10^{9}~M_{\odot}kpc^{-2}$ \citep{2018MNRAS.475.5603S}. In addition, both models expect baryonic Tully-Fisher relation as a galactic scaling rule; however, MOD predicts that there is another scaling relation too \citep{2021arXiv210907156S}. However, the two models are distinct in their mathematical operation because MOD is linear while MOND is nonlinear. The linearity of MOD makes analytical calculation much simpler as we will see in this work. To see this point, compare the present results with the MONDian three-body problem, for example, in predictions for LISA Pathfinder \citep{2010CQGra..27u5014B}. See also \cite{2010PhRvD..82j3001Z}.  It is interesting that there has been some attempts  from the beginning of MOND paradigm \citep{Milgrom:1983ca}, and later on \citep{1999PhLA..253..273M}, to interpret MOND as a Machian model. 

A particular difficulty in dealing with theories of extended gravities - all classes of modified gravities + modified dynamics \citep{2021MNRAS.503.2833R} - is the issue of testing these theories and then excluding them based on precise experiments. Unfortunately, due to large error bars in astrophysical data at large scales, it might be argued that these data  are unable to definitely exclude a large class of extended gravities, at least at present time. Thus, we need careful, controllable experiments ( as the best option ) or  probes of solar system ( as the next best option ) to distinguish between various extended gravities. For the former approach, see \cite{2007PhRvL..98o0801G,2010ARA&A..48..495F,2013PhRvD..87j7101D} and for the latter see \citet{2011MNRAS.415.1266I,2013MNRAS.432.3431P,2013AstL...39..141P}. Also, \cite{2020Ap&SS.365..154P} has recently proposed that the net gravitational fields in Sun-Jupiter and Sun-Neptune saddle regions are smaller than $10^{-10} ms^{-2}$, making these points suitable candidates to test  extended gravities which work below a fundamental acceleration, e.g.  MOND and MOD. See also \cite{2012PhRvD..86d4002G}.

Equation \ref{Force} is the starting point of our following formulation. In the rest of this section, we summarize some of the basic properties of the two-body and three-body problem governed by Eq. \ref{Force}. We mention that the results of the current formalism  could be simply reduced to Newtonian gravity by putting $c_{1}=0$. Also, one could easily check the significance of MOD's effects in a system   by  estimating the ratio of the MOD's fifth  force to Newtonian force: $2c_{1}a_{0}/(GM/R^{2})$, in which $M$ is the total mass of the system and $R$ is its characteristic radius. For example, one could see that in a typical globular cluster (TGC) with mass $2 \times 10^{5} M_{\odot}$ and half-mass radius $3~pc$, this ratio approaches $\sim 0.05$ which is small but not negligible. These estimations for a typical globular cluster are derived from \cite{2008gady.book.....B}, page 31.   In addition, the same ratio for TGC at the tidal radius is about $\sim 3.7$ which shows that the fifth force must be included here ( This effect is supposedly important in the physics of globular clusters and specially with regard to the question: Why don't the globular clusters collapse?). Moreover, for a typical open cluster (TOC) with mass  $300 M_{\odot}$ and half-mass radius $2~pc$ the ratio reads $\sim 16$ while at the tidal radius of TOC we have a ratio of $\sim 205$. These estimations  show that the fifth force effects needs to be carefully evaluated at these scales. In larger scales, i.e. galactic and extragalactic scales, it is straightforward to draw the same conclusion. In the case of three-body problem too, we define a parameter $\epsilon$ below which estimates the expected effects of MOD's fifth force. See Tables \ref{table:sampleS} and  \ref{table:sampleG} below.

 Since the additional term of MOD is a constant acceleration, i.e. a long range interaction term,  an important question is: "to what distance one should include the action of the masses?"   As mentioned above, the present model assumes the particle horizon $ R_{p}$ as the boundary of the systems. Thus, all the objects within the particle horizon act on the central object because they have  already been in causal contact with this mass. We split the mass distribution within the particle horizon into two categories: the homogeneous and isotropic background mass and the local mass distribution which is inhomogeneous and anisotropic. The MOD effect due to the homogeneous and isotropic background mass is mostly negligible since it is proportional to $r/R_{p}$ in which $r$ is the distance to the central mass. See Appendix in \cite{2016Ap&SS.361..378S} for a proof of this statement ( Problems dealing with large structure formation are exceptions because this factor is not negligible for them. ). Regarding the local objects, some systems   are in deep Newtonian regime; thus,  the MOD force is quite small compared to the Newtonian force. A good example of this type of physics is the solar system. In such systems, we treat the MOD force due to nearby objects, i. e. the Sun and the planets, as a  perturbative force with varying direction while the MOD force due to far away objects ( such as the  Galaxy, Andromeda etc. ) would be considered as a   force with fixed direction ( since the force due to these galaxies only varies very slowly with cosmic time.  ) . Therefore, one needs to solve the problems perturbatively starting from the Newtonian solution as the zeroth-order solution.  On the other hand, in systems for which the typical acceleration is comparable to the fundamental acceleration of   $a_{0}$ ( such as some star clusters, galaxies and galaxy clusters ), we need to estimate or calculate the MOD force ( depending on the time interval of the problem ).  

The fact that ellipses in MOD precess \citep{2016Ap&SS.361..378S}, in contrast to the Newtonian ones, is probably our best hope to test the model with the most accurate data available now ( Indeed, these tests could put an upper bound on $c_{1}$. ). However, candidate objects to test MOD needs to be carefully chosen because, until now, the model has been derived under the special assumptions that the systems are spherical, they  satisfy weak field regime and  the background metric is expanding. Although, some systems ( for example some  binary systems ) break these conditions fully or partially. We will review the above conditions below to realize suitable candidates to test MOD.

For example, consider a binary star with total mass $m = m_{1} + m_{2}$,  the semi-major axis $a$ and the eccentricity of $e$. Since  the system is losing energy, the two objects   will move toward each other, i.e. the size of their orbit decreases,  and they eventually collapse.  To compare the effect of the cosmological expansion with the effect of emitting gravitational waves in binary systems,   one could define the dimensionless rate of change in $a$ 
$$\mathcal{H}= \frac{1}{a }\frac{da}{ dt}=-\frac{64\eta c}{5a}\left( \frac{Gm}{ac^{2}} \right)^{3}\frac{1+\frac{73}{24}e^{2}+\frac{37}{96}e^{4}}{\left( 1-e^{2} \right)^{7/2}}$$
as the contraction rate of the system.    Here, $\eta =m_{1}m_{2}/ m^{2} $. See Eq. (12.84)  of \cite{2014grav.book.....P} ( page 649 ). In the case of Hulse-Taylor binary, this rate could be estimated as $\mathcal{H}  \approx -5.8 \times 10^{-17}~/s$. Data are derived from \cite{2016ApJ...829...55W} for this binary. The absolute value  of this contraction rate is larger than the Hubble constant $H_{0} \approx 70~km/s/Mpc=2.3\times 10^{-18}/s$; however, they are quite comparable in magnitude. If the contraction rate is found to be much larger than the  Hubble constant; then, one could neglect the expanding universe and impose the Neumann boundary condition within a new particle horizon defined for the geometry of the binary system. This could happen in the last stages  of the inspiral of the two objects. It should be mentioned that in this case, a new Neumann constant should be devised and the fifth forth of the system could be repulsive ( depending on the sign of the new Neumann constant ).

On the other hand, in the case of wide binaries, i.e. binary stars which are hundreds or thousands of AU apart, the magnitude of the contraction  rate would be much smaller than the Hubble constant; thus, MOD presumptions would be in the safe zone and  we would be able to test the model. However, then the isolation of the system ( considering the long range behavior of the MOD's fifth force ) needs to be carefully examined. Although, the effect of the other objects could be modeled by an external force for problems dealing with short range of time. See \cite{2018MNRAS.480.2660B,2019MNRAS.482.3453B,2019MNRAS.487.5291B} for tests of gravitational law using wide binaries.

  Thus, there are two regimes which need to be treated separately.  Curiously, binary systems such as Hulse-Taylor binary sit between these  two regimes.    Thus, they provide another interesting challenge in testing models of extended gravity based on the fundamental acceleration $a_{0}$.

When the system is contracting and no longer spherical, we could still ( in principle ) impose a Neumann boundary condition on the geometry of a binary system. However, then the second theoretical challenge arises as the problem of timing ( measuring time based on geometrodynamic clocks ). This is also a huge challenge which is beyond the present work. But the essence of the problem is that we should have estimations of each term in the light's geodesic equation to read the time and finally find a new equation of motion for massive particles.

Although, the fact is that even in its present form,  the free parameter of the model $c_{1}$ could be   confined - with enough accuracy --  using the post Newtonian approximation and solar system data. Indeed, to the second order of approximation ( 2PN ), post-Newtonian formalism suggests the following formula
$$\Delta \phi = 2\pi + 6\pi ( \frac{GM}{c^{2}a(1-e^{2})})   + \frac{3 \pi}{2} (18 + e^{2}) ( \frac{GM}{c^{2}a(1-e^{2})})^{2} + ... + \frac{8c_{1}a_{0}Ca^{2}}{GM}(1-e^{2})$$
for the the azimuthal angle increase  in the course of a complete radial cycle. Here, $a$ is the  semi-major axis of the ellipse, $e$ is the eccentricity of the orbit, $M$ is the mass of sun and  $C$ is the integral in Eq. (12) of \cite{2016Ap&SS.361..378S} which is dependent to $e$. The first three terms could be found, for example, in Eq. (5.198)  of \cite{2014grav.book.....P} ( page 275 ) and the last term is reported by \cite{2016Ap&SS.361..378S}.  By  an order-of-magnitude analysis of terms in the above equation, the reader could easily see that the third and the last  term on the right hand side would be quite close in the case of inner planets; thus, they provide the possibility to confine $c_{1}$.   However, it should be mentioned that including the gravitational  effects due to other objects of the solar system is also necessary for a complete analysis. This matter is beyond the scope of the present work.

\subsection{Two-body Orbits in MOD}
Imagine a test particle $m$  which rotates around a very heavier object with mass $M$ at distance $r$.  Since the modified term in the equation of motion, i.e. the term  proportional to $a_{0}$, is also a central force,  the angular momentum $\vec{L}$ of the test particle about any axis through the mass $M$ is  constant. Knowing this, one could readily derive the modified equation of motion for this two-body problem as 
\begin{eqnarray}
m\ddot{r}-mr\dot{\theta}^{2} &=& -\frac{GMm}{r^{2}}-2c_{1}a_{0}\frac{m}{M+m}  \\    \nonumber
mr\ddot{\theta}+2m\dot{r}\dot{\theta} &=& 0
\end{eqnarray}
 in polar coordinates $r$ and $\theta$. By multiplying the second equation by $r$, it could be shown that this is equivalent to the equation of conservation of angular momentum 
\begin{eqnarray}
\frac{dL}{dt}=\frac{d}{dt}(mr^{2}\dot{\theta})=0.
\end{eqnarray}
Now we can find the angular velocity as 
$\dot{\theta}=L/mr^{2}$ and then put it into the first equation of motion
\begin{eqnarray}
m\ddot{r} &=& \frac{L^{2}}{mr^{3}} -\frac{GMm}{r^{2}}-2c_{1}a_{0}\frac{Mm}{M+m}
\end{eqnarray}
which is equivalent to solving the equation of motion with the next effective potential energy 
\begin{eqnarray}
    \psi &=& \frac{L^{2}}{2mr^{2}} -\frac{GMm}{r} +2c_{1}a_{0}\frac{Mm}{M+m}r
\end{eqnarray}
See the plot of $\psi $ in Fig. 1 of \cite{2016Ap&SS.361..378S}. One of the most important consequences of this modified term is that, under this new equation of motion the  Laplace-Runge-Lens (LRL)
vector is not a constant anymore. This means that the line of apsides changes direction now which  is a crucial point in our analysis of the three body problem. See Fig. \ref{fig:Equilateral}  below. The   averaged precession rate due to the new term is derived in \cite{2016Ap&SS.361..378S}. There, it has also been  reasoned that due to Bertrand's theorem, there is no other constant of the motion in MOD beside energy and the three elements of angular momentum vector, i.e. one constant less than the Kepler problem.
\begin{figure*}[htbp]
\centering
\includegraphics[ width=\textwidth , height=0.5\textwidth]{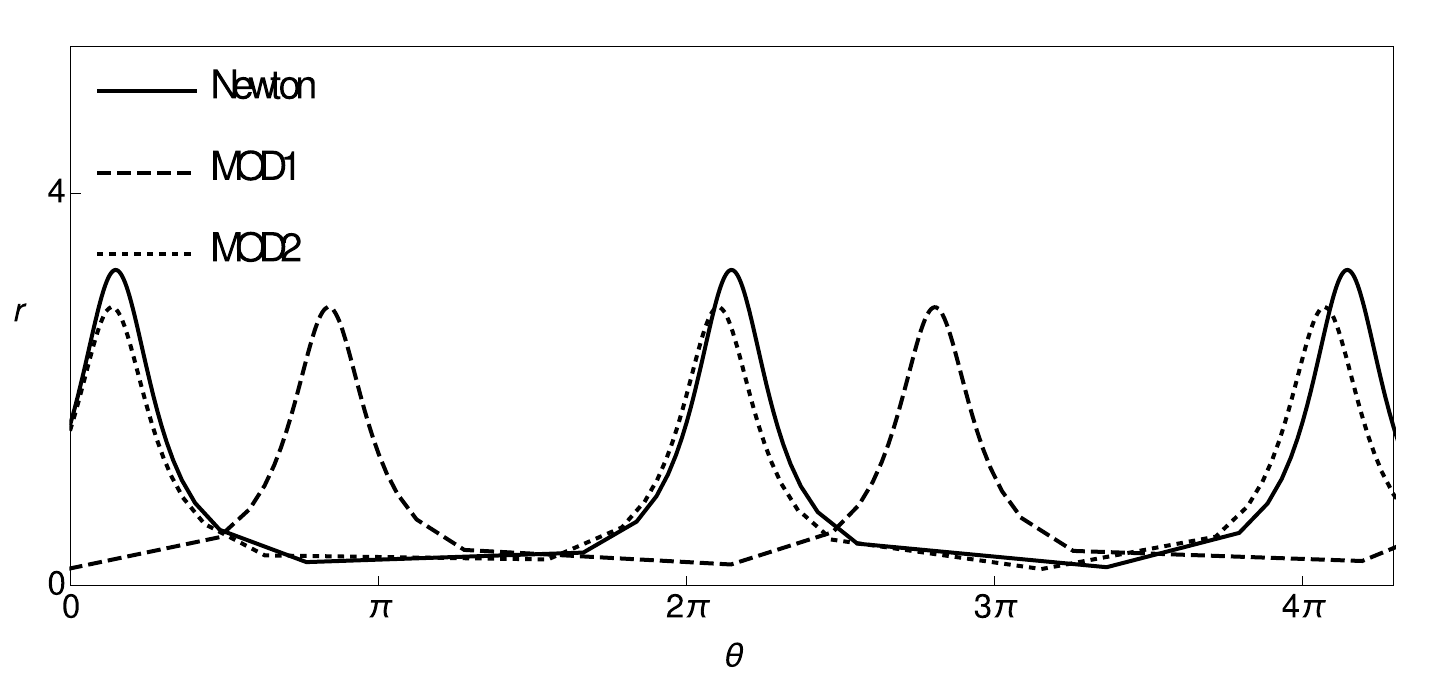}    \\ 
 \includegraphics[width=0.31 \textwidth]{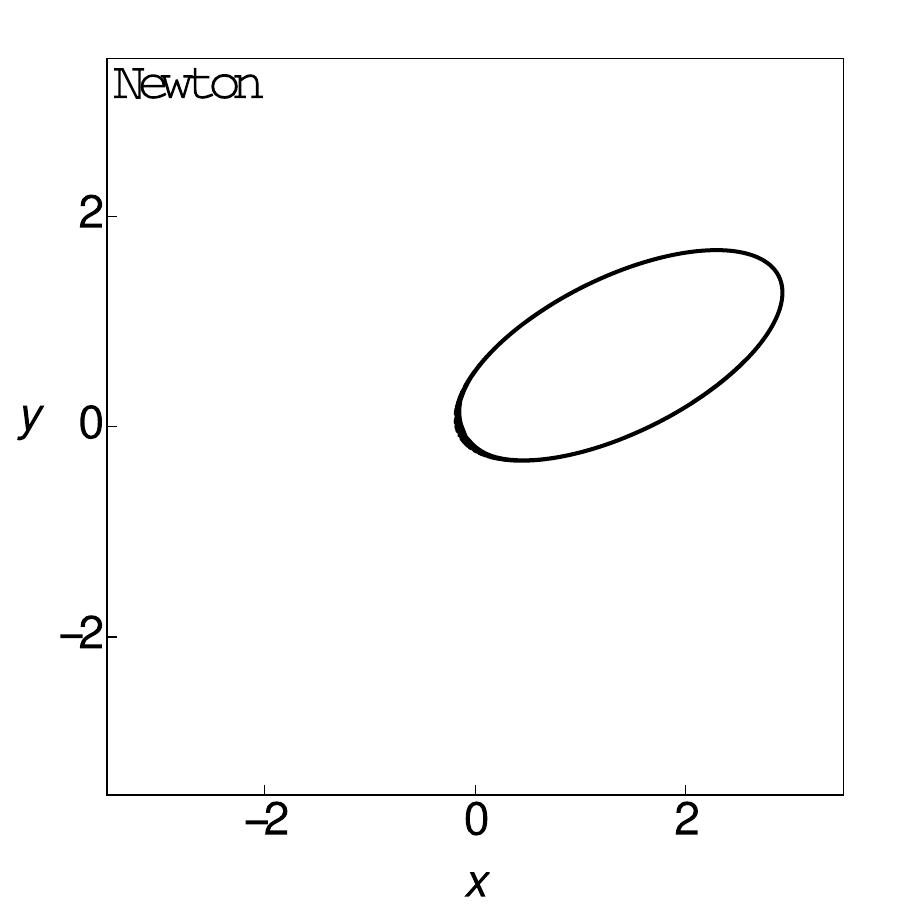}   \includegraphics[width=0.31 \textwidth]{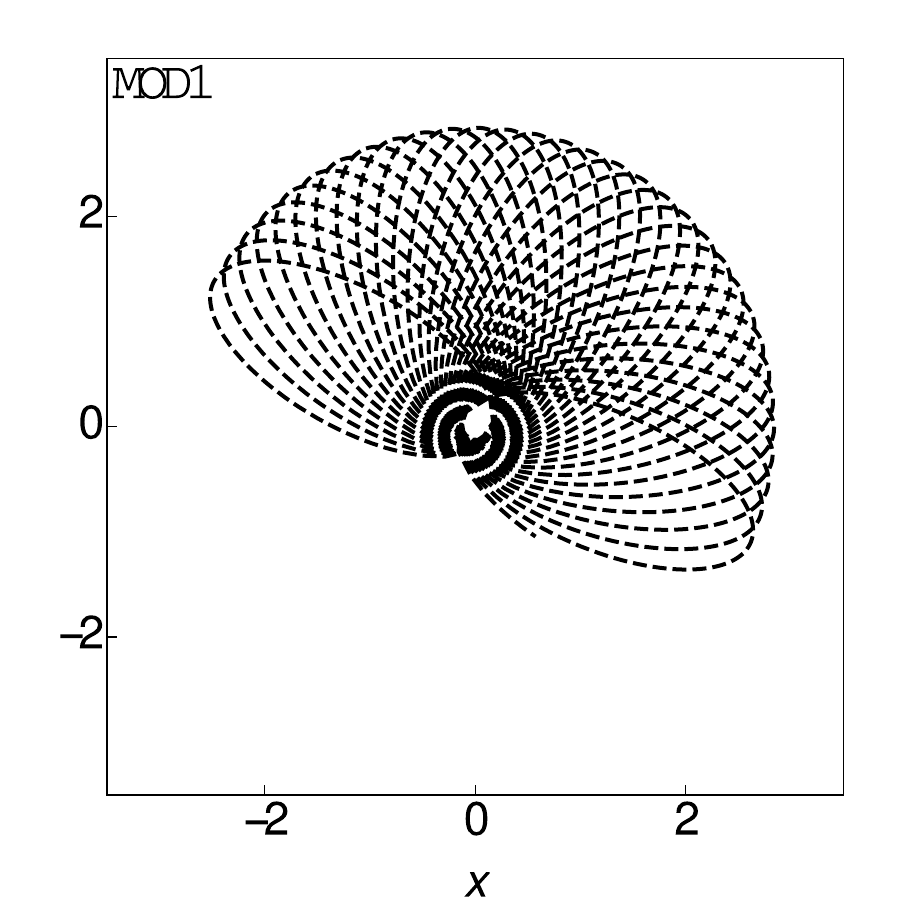}  \includegraphics[width=0.31\textwidth]{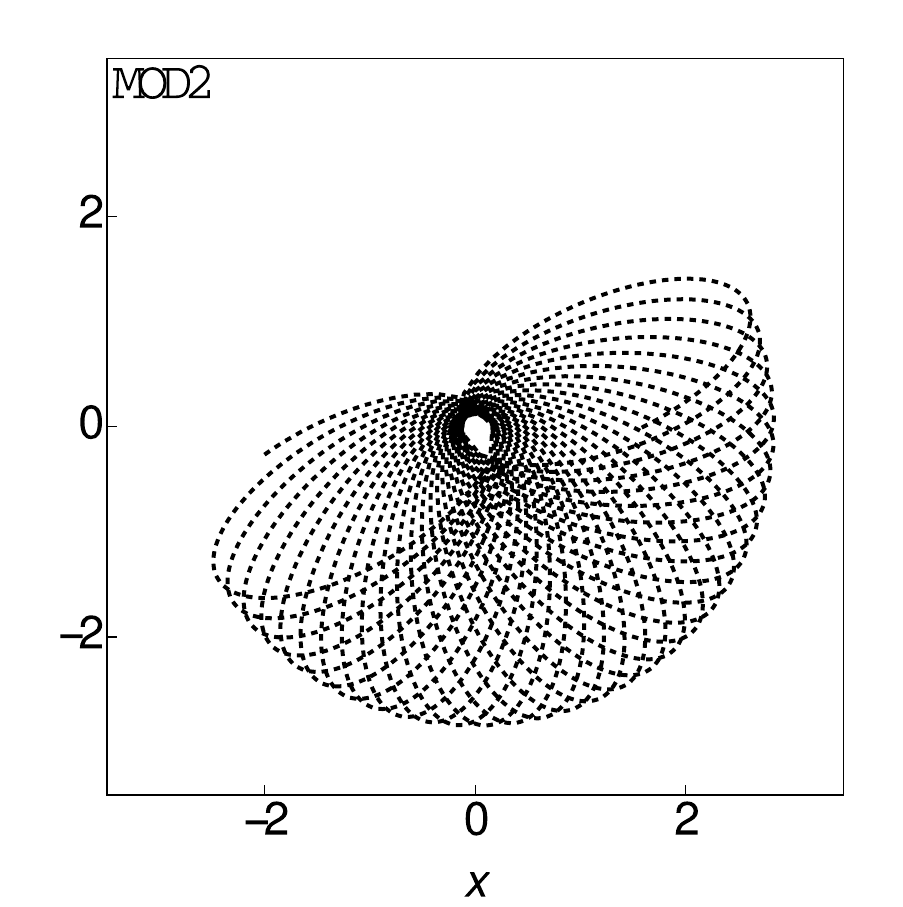}  
\\  \caption{Aperiodic bounded orbits according to Newtonian and MOD models. Upper panels: The distance of the test particle ( in units of $\mathcal{D}$ ) from the central mass is plotted as a function of time ( in units of $\mathcal{T}$ ) within Newtonian model as well as MOD1 and MOD2. Bottom: The corresponding orbits are shown here (  The coordinates  $x$ and $y$ are in units of $\mathcal{D}$. ).  }
  \label{fig:orbits}
\end{figure*}

In different regimes, this equation of motion could be solved numerically. To see this, we suppose a test particle with negligible mass and following units ( which are specially suitable for galactic systems )
\begin{eqnarray} \label{units}
\mathcal{T} &=& 1.2 \times 10^{6} ~~years ~~~\Rightarrow ~~~~\text{Time unit} \\  \nonumber
\mathcal{D} &=& 0.5~~~kpc ~~~~~~~~~~~~~~\Rightarrow ~~~~\text{Distance unit} \\  \nonumber
\mathcal{M}     &=& 2 \times 10^{10}  ~~M_{\odot}~~~~ ~~~\Rightarrow~~~~ \text{Mass unit} \\  \nonumber
\mathcal{V} &=& 406 ~~km/s~~~~~~~~~~~~\Rightarrow~~~~\text{Velocity unit} \\  \nonumber
G &=& 1. ~~~~~\mathcal{D}^{3}/\mathcal{M}\mathcal{T}^{2} \\  \nonumber
a_{0} &=& 0.061~~\mathcal{D}/\mathcal{T}^{2} \\  \nonumber
\end{eqnarray}
   We solve the equations of motion numerically, presuming $\theta(t=0)=0$, $\dot{r}(t=0)=0.5~\mathcal{V}$ and the motion starts at a point where the total energy is half the total energy at $r=1.0 \mathcal{D}$.  In this way, both Newtonian and MOD  models provide two solutions with  positive $r$. We   show only one of the Newtonian orbits while the  plots related to  MOD are displayed by MOD1 and MOD2 in Fig. \ref{fig:orbits}.  In addition, we presume that the central object has a mass of $0.5~\mathcal{M}$ and that the angular momentum per unit mass is equal to $L/m = 0.4~\mathcal{D}\mathcal{V}$. Putting these together, the derived solutions for both models are presented in Fig. \ref{fig:orbits} which shows a closed ellipse for the Newtonian model and  precessing elliptical orbits for MOD.  Also, MOD orbits sweep smaller regions compared to Newtonian orbits.  See the top panel in Fig. \ref{fig:orbits}. It is also worth noting that MOD1 and MOD2 orbits are similar paths with opposite phases.

 It should be mentioned that if a two-body system is accelerated by other sources of gravity, then one would have the accelerated Kepler problem which shows an interesting phenomenology \cite{2007CeMDA..99...31N}. This is especially relevant in the case of modified Kepler problem in MOD because the fifth force of MOD is a long range force. However, this issue is beyond the scope of the present work. 

\subsection{General Form of the Equation of Motion in three-body problem and their symmetries}
One of the routes to solve the three body problem, and probably the most fruitful path yet found, is to try to  reduce the problem to the two-body problem. Here, we will show how this is also possible in MOD model. In fact, as we see below, the method could be generalized to any potential of the form $\propto m_{1}m_{2}r^{n}$, which is proportional to the two masses, but not to potentials like $\Lambda r^{2}$ (Although, see \cite{2016ARep...60..397E} who solve the three-body problem in $\Lambda CDM $ model by presuming equal masses for the three galaxies. ).  

For a system of three objects, one can find the equations of motion as follows:
\begin{eqnarray}  \label{general3}
\ddot{\vec{r}}_{1} &=& -G \left( m_{2}\frac{\vec{r}_{12}}{r^{3}_{12}}+m_{3}\frac{\vec{r}_{13}}{r^{3}_{13}} \right)-\frac{2c_{1}a_{0}}{M}\left( m_{2}\frac{\vec{r}_{12}}{r_{12}}+m_{3}\frac{\vec{r}_{13}}{r_{13}} \right)  \\  \nonumber
\ddot{\vec{r}}_{2} &=& -G \left( m_{3}\frac{\vec{r}_{23}}{r^{3}_{23}}+m_{1}\frac{\vec{r}_{21}}{r^{3}_{21}} \right)-\frac{2c_{1}a_{0}}{M}\left( m_{3}\frac{\vec{r}_{23}}{r_{23}}+m_{1}\frac{\vec{r}_{21}}{r_{21}} \right)   \\  \nonumber
\ddot{\vec{r}}_{3} &=& -G \left( m_{1}\frac{\vec{r}_{31}}{r^{3}_{31}}+m_{2}\frac{\vec{r}_{32}}{r^{3}_{32}} \right)-\frac{2c_{1}a_{0}}{M}\left( m_{1}\frac{\vec{r}_{31}}{r_{31}}+m_{2}\frac{\vec{r}_{32}}{r_{32}} \right)
\end{eqnarray}
in which $\vec{r}_{ij}\equiv \vec{r}_{i}-\vec{r}_{j}$ is the relative distance between particles $i$ and $j$. The set of equations \ref{general3} consists of 18 degrees of freedom ( dof ); however, since the coordinates are related through the next equation
$$m_{1}\ddot{\vec{r}}_{1}+m_{2}\ddot{\vec{r}}_{2}+m_{3}\ddot{\vec{r}}_{3}=0$$
the dof reduces to 15. This constraint also shows that the center of mass of the system moves with constant velocity.

In addition, one could simply show that the energy and angular momentum are conserved; thus, the dof still reduces to 11. Furthermore, if we presume that the center of mass defines the  coordinate's origin, dof would be 8. Moreover, when the orbits lie  within a fixed plane, dof becomes 6. 

\begin{figure*}[htbp]
\centering
\includegraphics[width=0.45\textwidth]{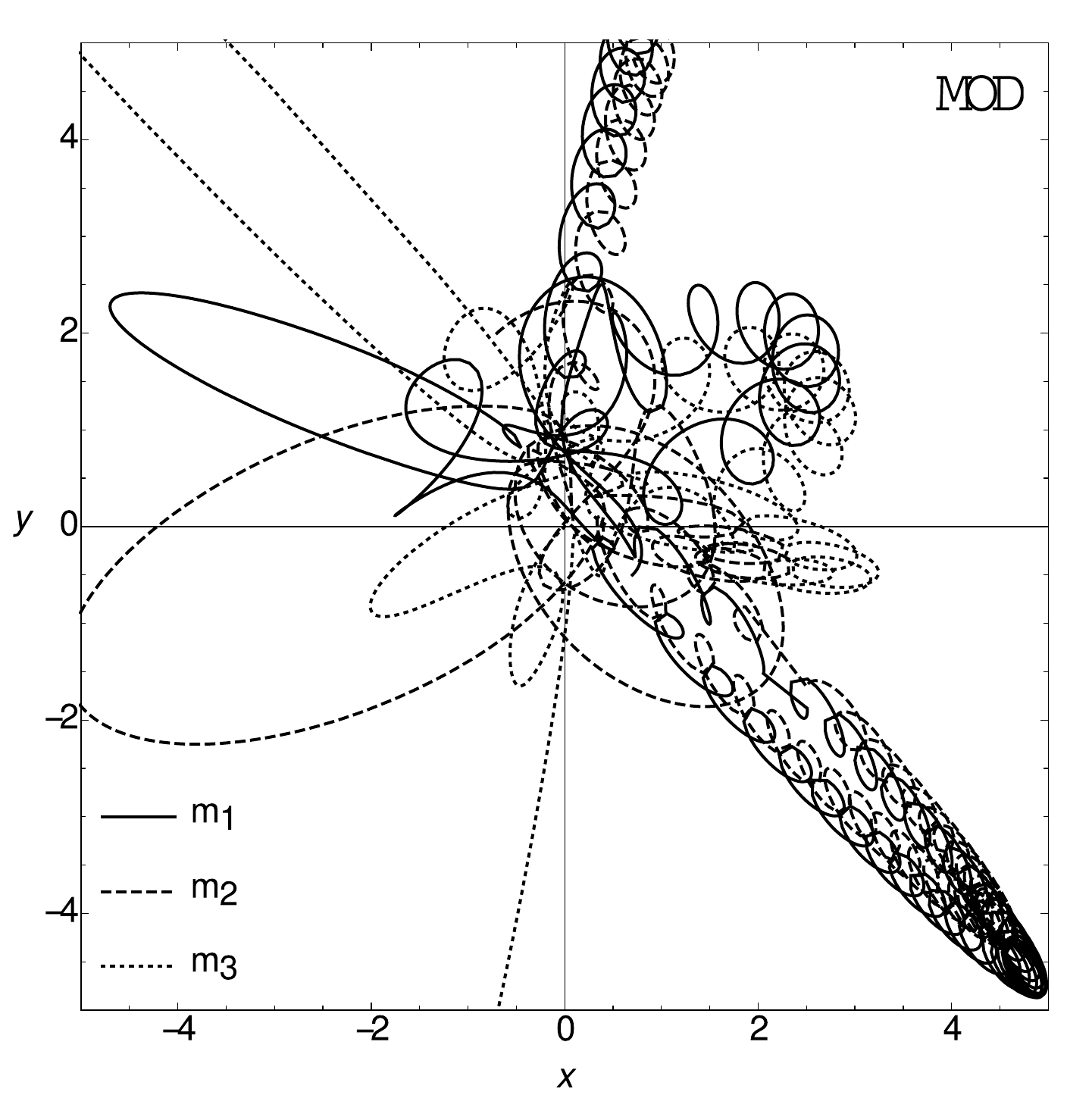} \includegraphics[width=0.45\textwidth]{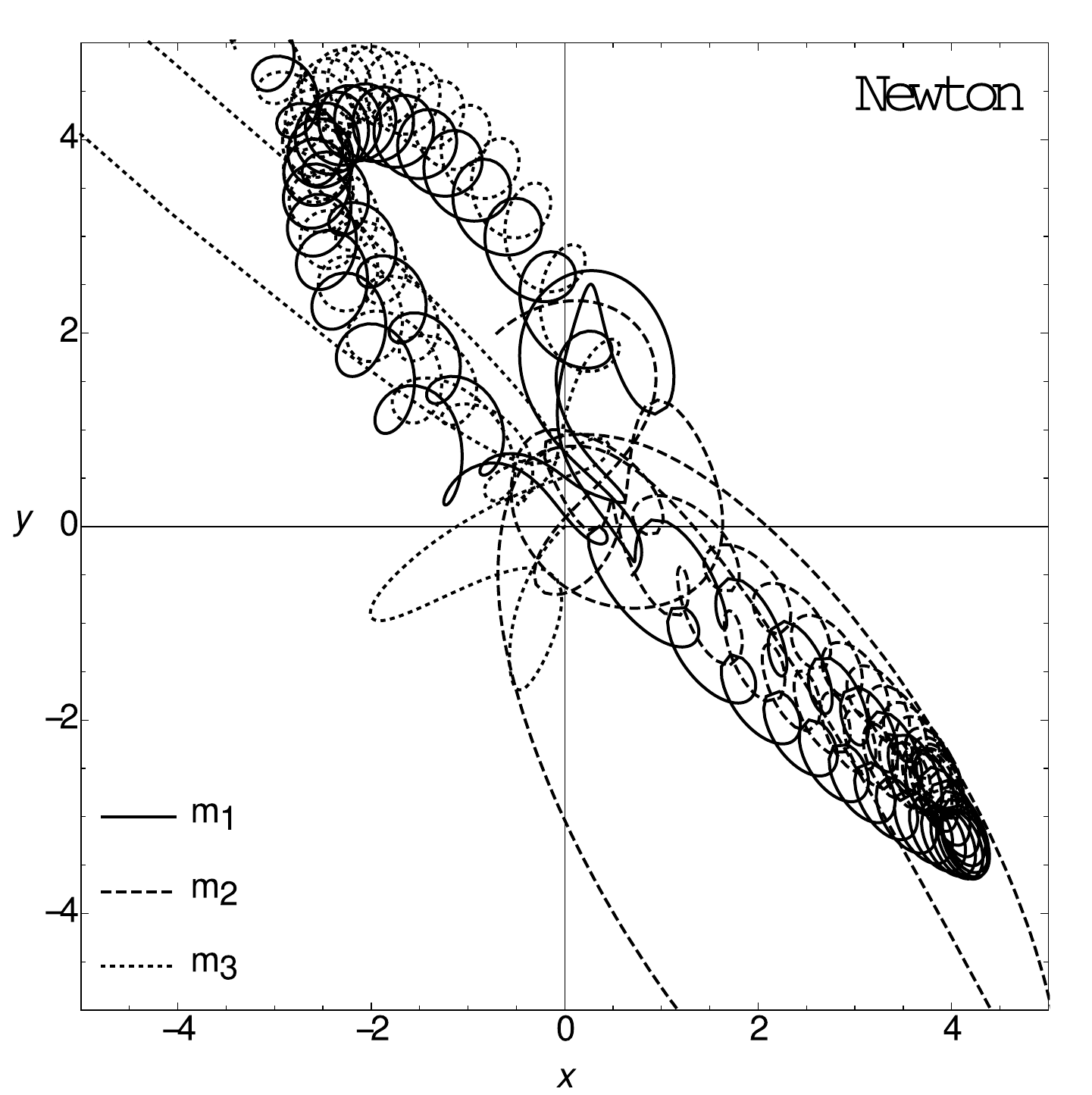}
\\
  \caption{The time evolution ($200~\mathcal{T}$) of three equal mass  galaxies ( all three with mass  $\mathcal{M}$ ) in two dimensions and according to Newtonian and MOD models with exactly the same initial conditions derived from \citet{10.2307/2661357}. The plots show that the fate of galaxies in a group could entirely change under modified dynamics. The coordinates $x$ and $y$ are in units of $\mathcal{D}$. }
  \label{fig:3bequalmass}
\end{figure*}

The problem would still be very difficult to solve; though, it could be tackled numerically. For example, using the units of \ref{units}, we have solved the set of equations \ref{general3} for three equal masses ( = unit mass $\mathcal{M}$ ) and an evolution time of $200~ \mathcal{T}$. See Fig. \ref{fig:3bequalmass}.  The initial conditions are derived from \cite{10.2307/2661357} and they are the same for Newtonian and MOD models.  However, the particles under MOD and Newtonian dynamics show different trajectories.    In addition, we mention that Newtonian solution obeys scaling, i.e. three-body solutions show the same behaviors at different scales which is due to a symmetry in their equation of motion ( see below ). However, MOD three-body solutions are manifestly scale dependent.

\begin{figure*}[htbp]
\centering
\includegraphics[width=0.45\textwidth]{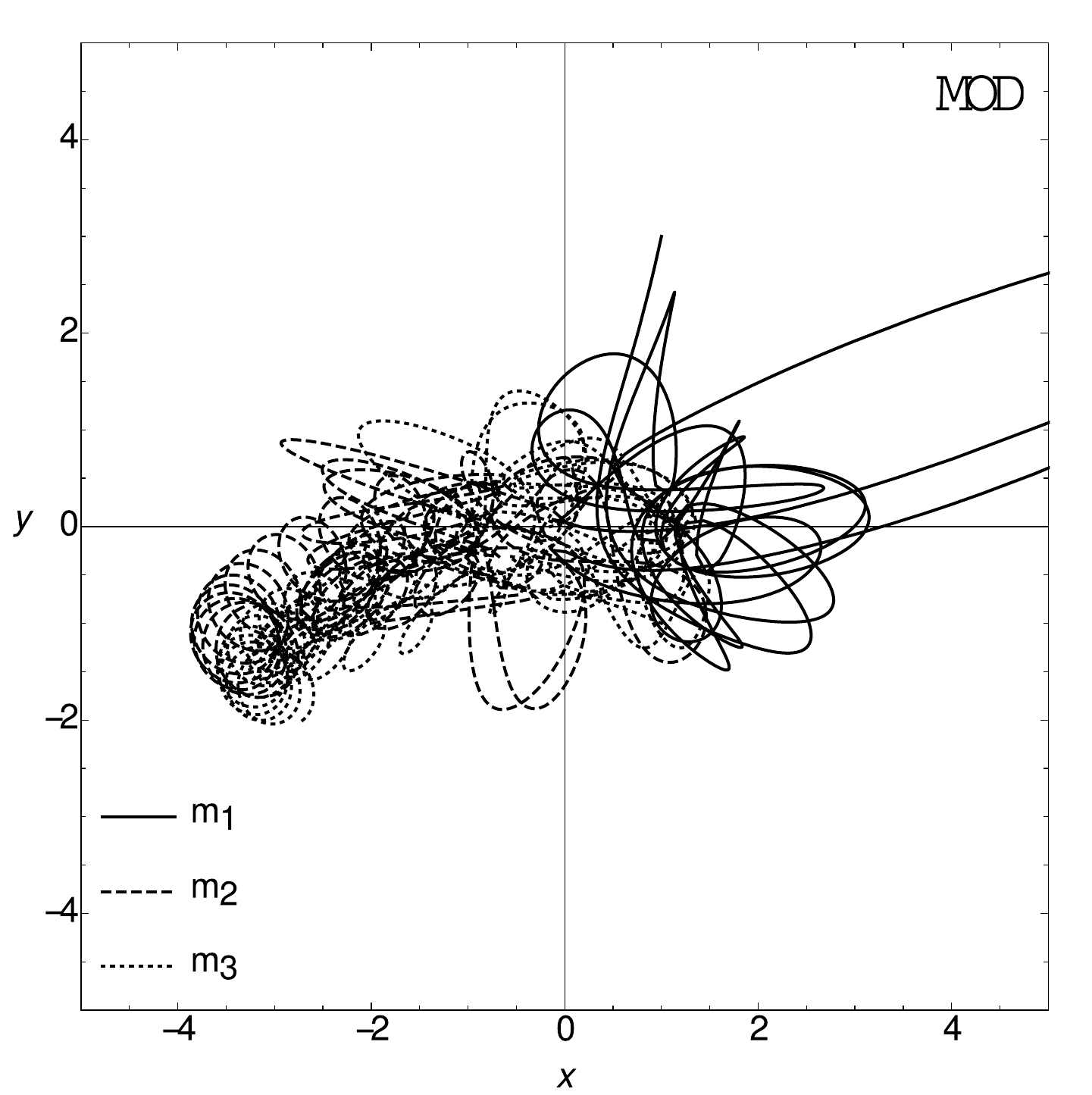} \includegraphics[width=0.45\textwidth]{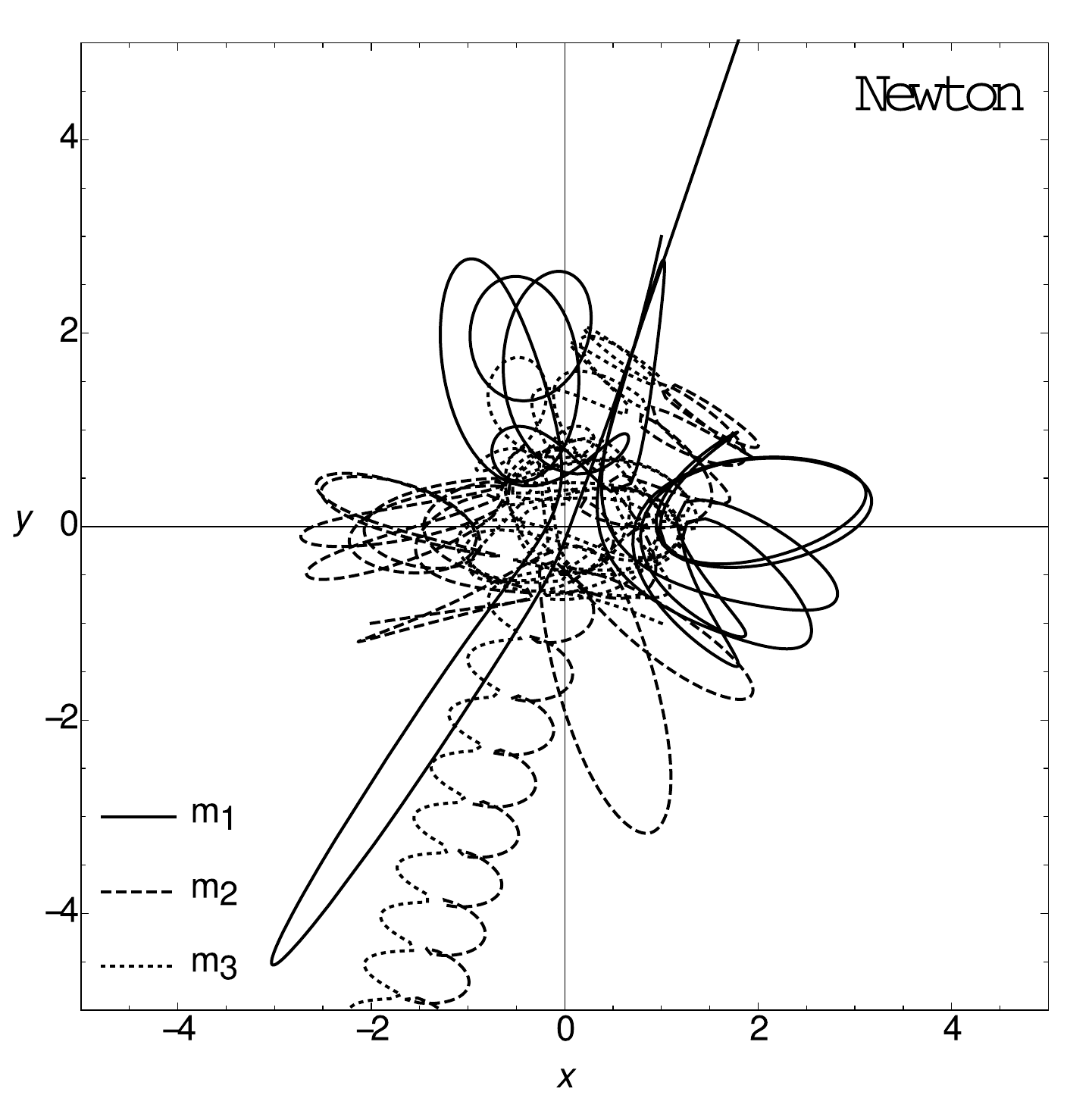}
\\
  \caption{The time evolution ($100~\mathcal{T}$) of Pythagorean three-body problem \citep{1967AJ.....72..876S} according to Newtonian and MOD models.  The coordinates $x$ and $y$ are in units of $\mathcal{D}$.}
  \label{fig:Pythagorean}
\end{figure*}

Another problem that we have solved numerically is the Pythagorean three-body problem for which  three particles with masses $m_{1}=3,~m_{2}=4,~m_{3}=5$ are initially  placed, with zero velocity, at positions $(1,~3)$, $(-2,~-1)$ and $(1,~-1)$ in $xy$ plane respectively, i.e. a right angle triangle \citep{1967AJ.....72..876S}. You may see the time evolution ( $100~\mathcal{T}$ ) of this problem in Fig. \ref{fig:Pythagorean}. As these two plots show, the particle $m_{1}$ in the Newtonian model ejects from the system while the other two particles form a couple.  On the other hand, the three-body system in MOD model lasts longer and there is no ejection within the time evolution of 100 $\mathcal{T}$; though, the ejection occurs at about   $100 \mathcal{T}$.   This is most probably due to the strong attractive force of MOD ( see the discussions below ).

\subsection{Equilateral  Triangle Solution}
This solution has been first introduced by \cite{1973CeMec...8....5B}. The symmetries of the problem would be most apparent in relative coordinates, 
\begin{eqnarray}
\vec{s}_{1}=\vec{r}_{3}-\vec{r}_{2}  \\  \nonumber
\vec{s}_{2}=\vec{r}_{1}-\vec{r}_{3}  \\  \nonumber
\vec{s}_{3}=\vec{r}_{2}-\vec{r}_{1} \\  \nonumber
\end{eqnarray}
in which the three $\vec{s}_{i}$ coordinates are related through $\vec{s}_{1}+\vec{s}_{2}+\vec{s}_{3}=0$. This is also known as the Lagrangian formulation of the equations of motion. Using these coordinates, the  equations of motion could readily be derived    as follows:
\begin{eqnarray}
\ddot{\vec{s}}_{1} &=& -G\left( m_{1}\vec{W}+M\frac{\vec{s}_{1}}{s^{3}_{1}} \right)-\frac{2c_{1}a_{0}}{M}\left( m_{1}\vec{V}+M\frac{\vec{s}_{1}}{s_{1}} \right)   \\   \nonumber
\ddot{\vec{s}}_{2} &=& -G\left( m_{2}\vec{W}+M\frac{\vec{s}_{2}}{s^{3}_{2}} \right)-\frac{2c_{1}a_{0}}{M}\left( m_{2}\vec{V}+M\frac{\vec{s}_{2}}{s_{2}} \right)   \\   \nonumber
\ddot{\vec{s}}_{3} &=& -G\left( m_{3}\vec{W}+M\frac{\vec{s}_{3}}{s^{3}_{3}} \right)-\frac{2c_{1}a_{0}}{M}\left( m_{3}\vec{V}+M\frac{\vec{s}_{3}}{s_{3}} \right)   \\   \nonumber
\end{eqnarray}
in which
\begin{eqnarray}
\vec{W} &=& \frac{\vec{r}_{12}}{r^{3}_{12}} +
\frac{\vec{r}_{23}}{r^{3}_{23}}
+\frac{\vec{r}_{31}}{r^{3}_{31}}=-\left( \frac{\vec{s}_{1}}{s^{3}_{1}}  +\frac{\vec{s}_{2}}{s^{3}_{2}}  +\frac{\vec{s}_{3}}{s^{3}_{3}}   \right)      \\   \nonumber
\vec{V} &=& \frac{\vec{r}_{12}}{r_{12}} +
\frac{\vec{r}_{23}}{r_{23}}
+\frac{\vec{r}_{31}}{r_{31}} =-\left( \frac{\vec{s}_{1}}{s_{1}}  +\frac{\vec{s}_{2}}{s_{2}}  +\frac{\vec{s}_{3}}{s_{3}}   \right)
\end{eqnarray}
Now we could see that when $\vec{W}=0$ and $\vec{V}=0$, then the system of equations for $\vec{s}_{i}$ decouples  into a set of three similar two-body equations, i.e. three modified Kepler equations of the form introduced above.  This condition happens when $s_{1}=s_{2}=s_{3}$ since $\vec{s}_{1}+\vec{s}_{2}+\vec{s}_{3}=0$. In other words, in this solution the three particles remain at the vertices of an equilateral triangle, however, the size and orientation of the  triangle in the plane will, in general, change in time  as the particles move.

\begin{figure*}[htbp]
\centering
\includegraphics[width=0.45\textwidth]{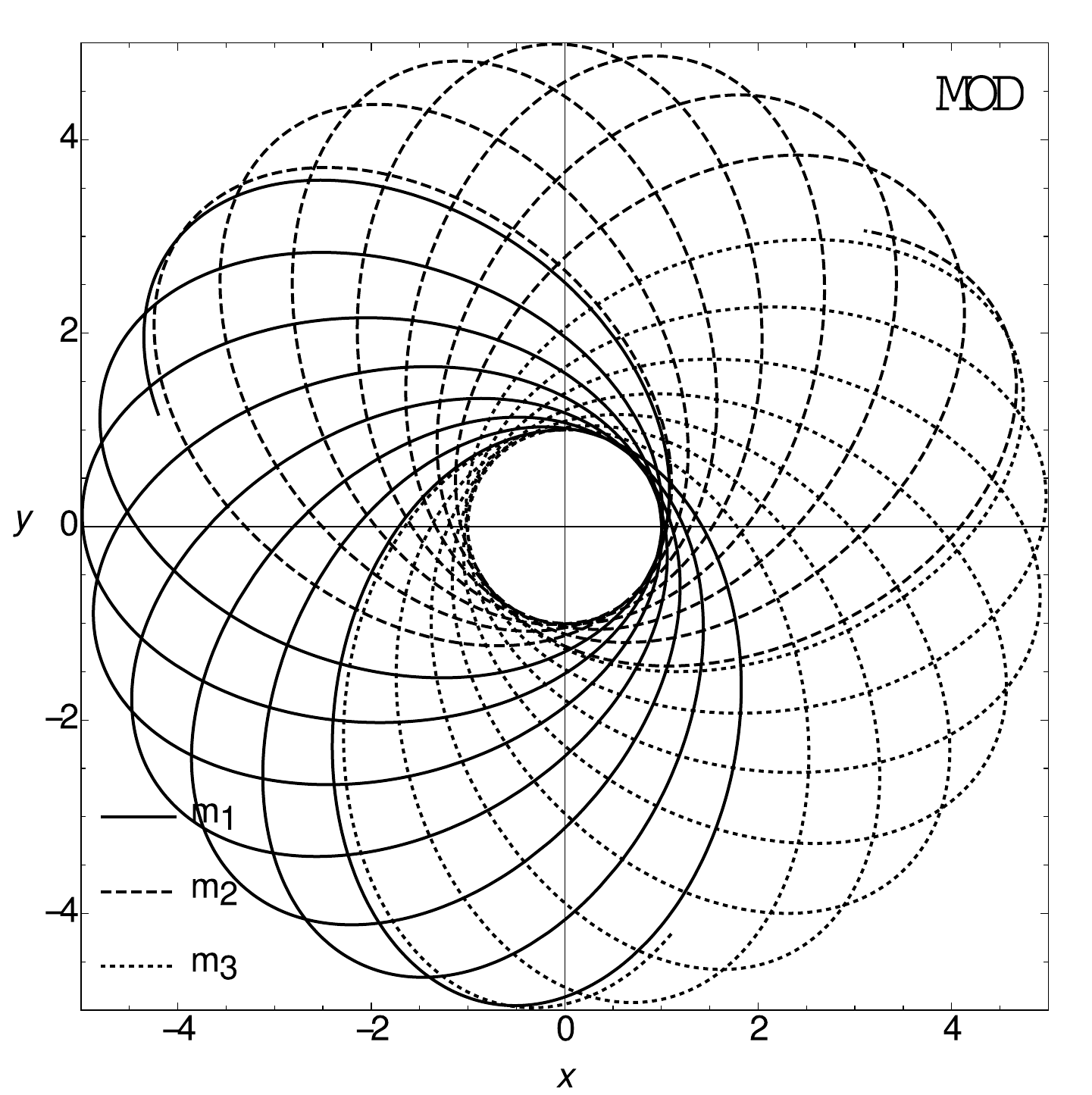} \includegraphics[width=0.45\textwidth]{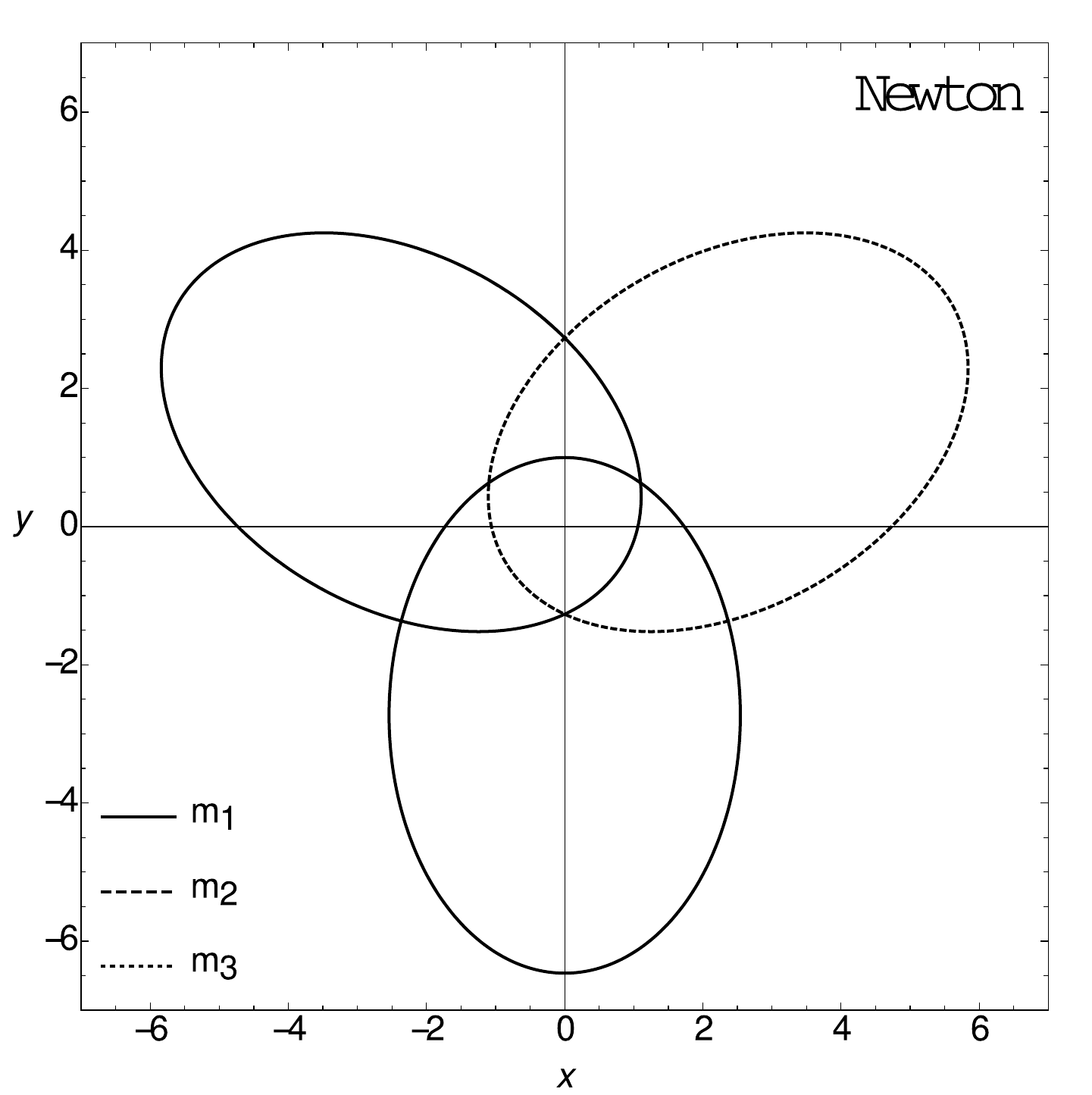}
\\
  \caption{The time evolution ( $300~\mathcal{T}$ ) of the equilateral triangle solution for three equal mass galaxies ( all three galaxies with mass  $\mathcal{M}$ ) in two dimensions and according to Newtonian and MOD models with exactly the same initial conditions, i.e. initial positions and velocity. The coordinates $x$ and $y$ are in units of $\mathcal{D}$. Please note that due to the overlap of the Newtonian orbits on the right hand side, the trajectories of $m_{1}$ and $m_{2}$ are almost indistinguishable.}
  \label{fig:Equilateral}
\end{figure*}

To explain the difference between Newtonian and MOD models in the three-body problem, we have built a closed, equilateral triangle solution in Newtonian dynamics and its counterpart in Fig. \ref{fig:Equilateral}. Again, we have the same unit masses $\mathcal{M}$ and similar initial conditions ( time evolution $=300~\mathcal{T}$ ).  In this case, the Newtonian orbits are fixed while the precession of the MOD orbits ( due to LRL nonzero rate ) makes the distinction between the two dynamics very clear.  Although, as it is clear from Fig. \ref{fig:Equilateral}, all three masses still stay at the  vertices of an equilateral triangle.

The equilateral  triangle solution is a rather general solution of the three body problem because, as it might be easily seen now, any new  potential energy of the general form of $\propto m_{1}m_{2}r^{n}$ will produce a term proportional to $   \vec{s}_{1}/s^{n-2}_{1}  + \vec{s}_{2}/ s^{n-2}_{2}  + \vec{s}_{3}/s^{n-2}_{3} $ in the equations of motion which still vanishes providing $s_{1}=s_{2}=s_{3}$. In fact, here the proportionality of the potential energy to $m_{1}m_{2}$ plays the crucial role in decoupling of the equations.  See \cite{1993PhRvL..70.3675M} for a general discussion on these potentials. For example, we could now realize that  for a force due to the cosmological constant $\Lambda$ the situation is different and the equilateral  triangle solution could not appear anymore.

\subsection{Third body orbiting a binary}
In some cases, the three-body problem results in the formation of a brief or permanent binary. See Fig. \ref{fig:3bequalmass} or the Newtonian case in Fig. \ref{fig:Pythagorean}. In this subsection we want to investigate a simplified version of a binary in three-body systems.  Imagine that particles $m_{1}$ and $m_{2}$ form a binary while the particle $m_{3}$ is orbiting them at a relatively large distance, e.g. a planet orbiting a binary star. In  this situation, we use the binary's  centre of mass $\vec{r}_{b}=\left( m_{1}\vec{r}_{1}+  m_{2}\vec{r}_{2}\right)/\left( m_{1}+m_{2} \right)$ as a reference point, and then measure the position of $m_{3}$ relative to $\vec{r}_{b}$, i.e. $\vec{R}=\vec{r}_{3}-\vec{r}_{b}$. This vector, in addition to the relative position vector of the binary, i.e. $\vec{\varrho}=\vec{r}_{21}$, form the Jacobi system of coordinates for which we can derive
\begin{eqnarray}
\ddot{\vec{\varrho}} &=& -G\left( (m_{1}+m_{2})\frac{\vec{\varrho}}{\varrho^{3}}+m_{3}\left[\frac{\vec{r}_{31}}{r^{3}_{31}} -\frac{\vec{r}_{32}}{r^{3}_{32}}   \right] \right)    -\frac{2c_{1}a_{0}}{M}\left( (m_{1}+m_{2})\frac{\vec{\varrho}}{\varrho}+m_{3}\left[\frac{\vec{r}_{31}}{r_{31}} -\frac{\vec{r}_{32}}{r_{32}}   \right] \right)   \\ \nonumber
\ddot{\vec{R}} &=& -GM \left( \frac{m_{1}}{m_{1}+m_{2}}\frac{\vec{r}_{31}}{r^{3}_{31}} + \frac{m_{2}}{m_{1}+m_{2}}\frac{\vec{r}_{32}}{r^{3}_{32}}  \right)    -2c_{1}a_{0}  \left( \frac{m_{1}}{m_{1}+m_{2}}\frac{\vec{r}_{31}}{r_{31}} + \frac{m_{2}}{m_{1}+m_{2}}\frac{\vec{r}_{32}}{r_{32}}  \right) 
\end{eqnarray}
Assuming that the third body is indeed at a large distance compared to the relative distance of the binary, i.e. if $|\vec{R}|  \approx |\vec{r}_{13} | \approx |\vec{r}_{23}  | $ and  $|\vec{\rho}| \ll |\vec{R}|$,  we arrive at the next separated two-body equations
\begin{eqnarray}
\ddot{\vec{\varrho}} &=& -G  (m_{1}+m_{2})\frac{\vec{\varrho}}{\varrho^{3}}
 -2c_{1}a_{0}\frac{(m_{1}+m_{2})}{M}  \frac{\vec{\varrho}}{\varrho}  \\   \nonumber   
\ddot{\vec{R}} &=& -GM \frac{\vec{R}_{3}}{R^{3}_{3}} - 2c_{1}a_{0} \frac{\vec{R}_{3}}{R_{3}}
\end{eqnarray}
for which we observe that due to MOD term, the line of apsides in the orbit of the binary changes direction. Also, the same happens for the motion of the third body with respect to binary's center of mass.

\begin{figure*}
\centering
\includegraphics[width=0.45\textwidth]{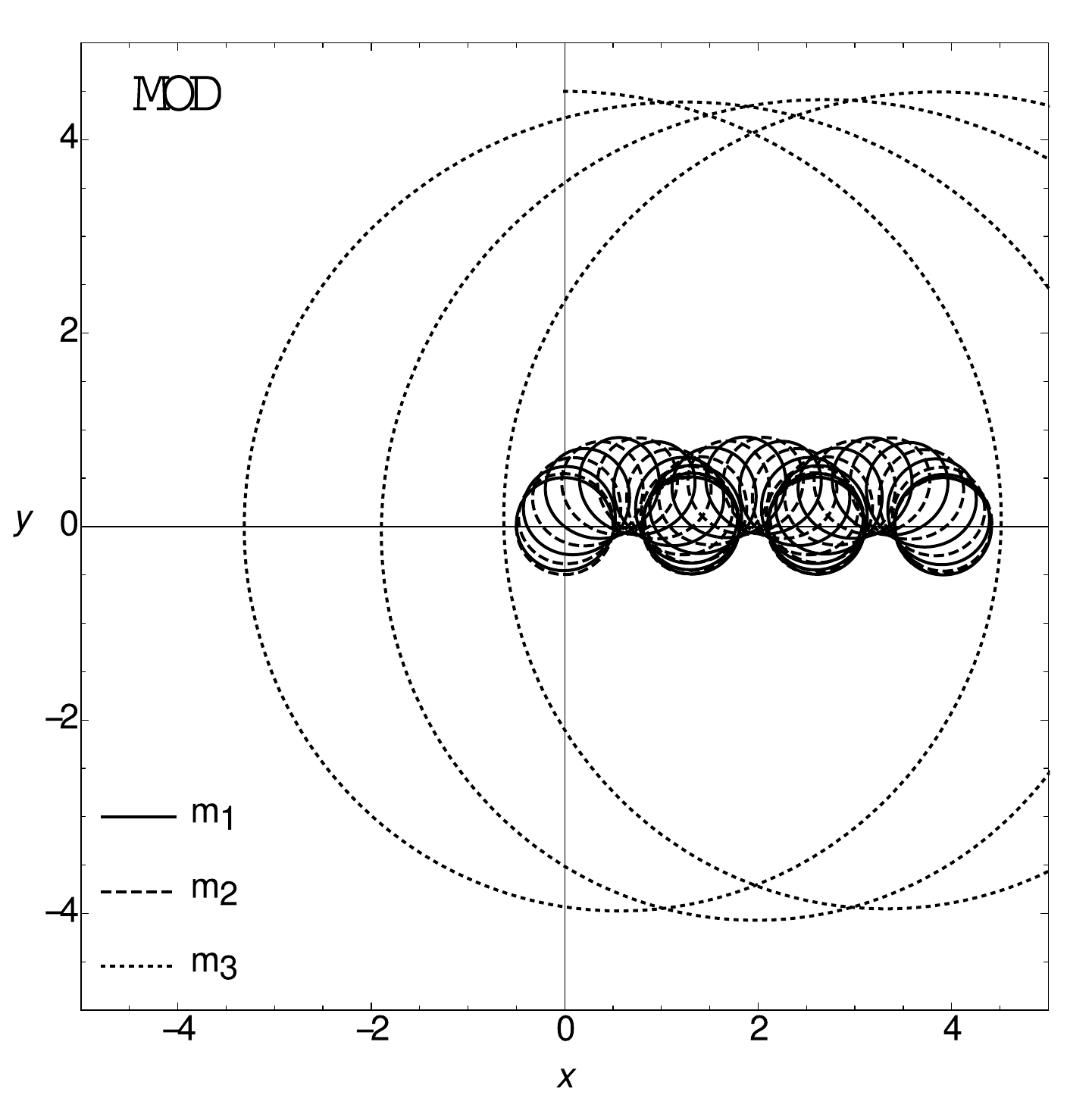} \includegraphics[width=0.45\textwidth]{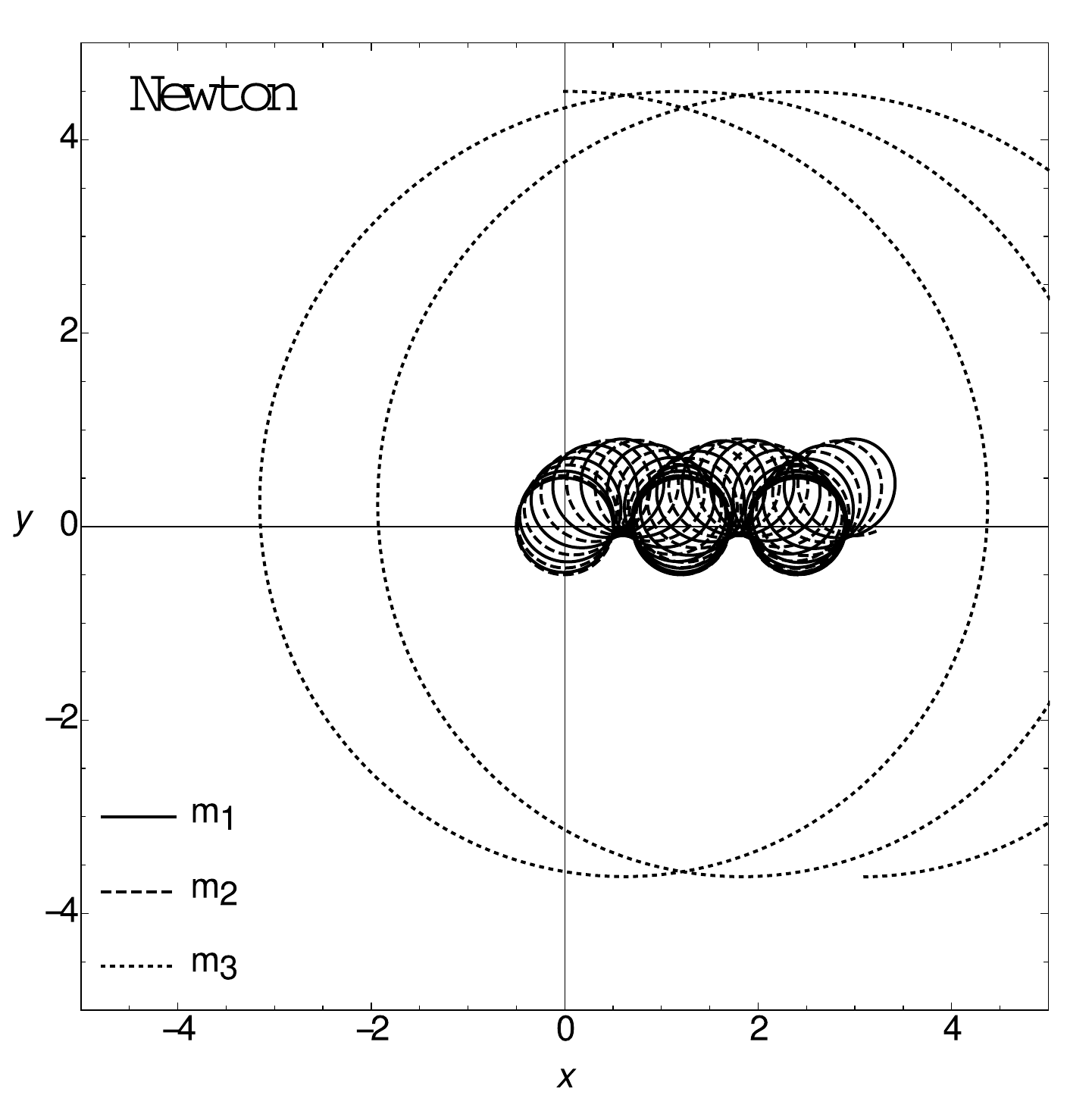}
\\
\includegraphics[width=0.45\textwidth]{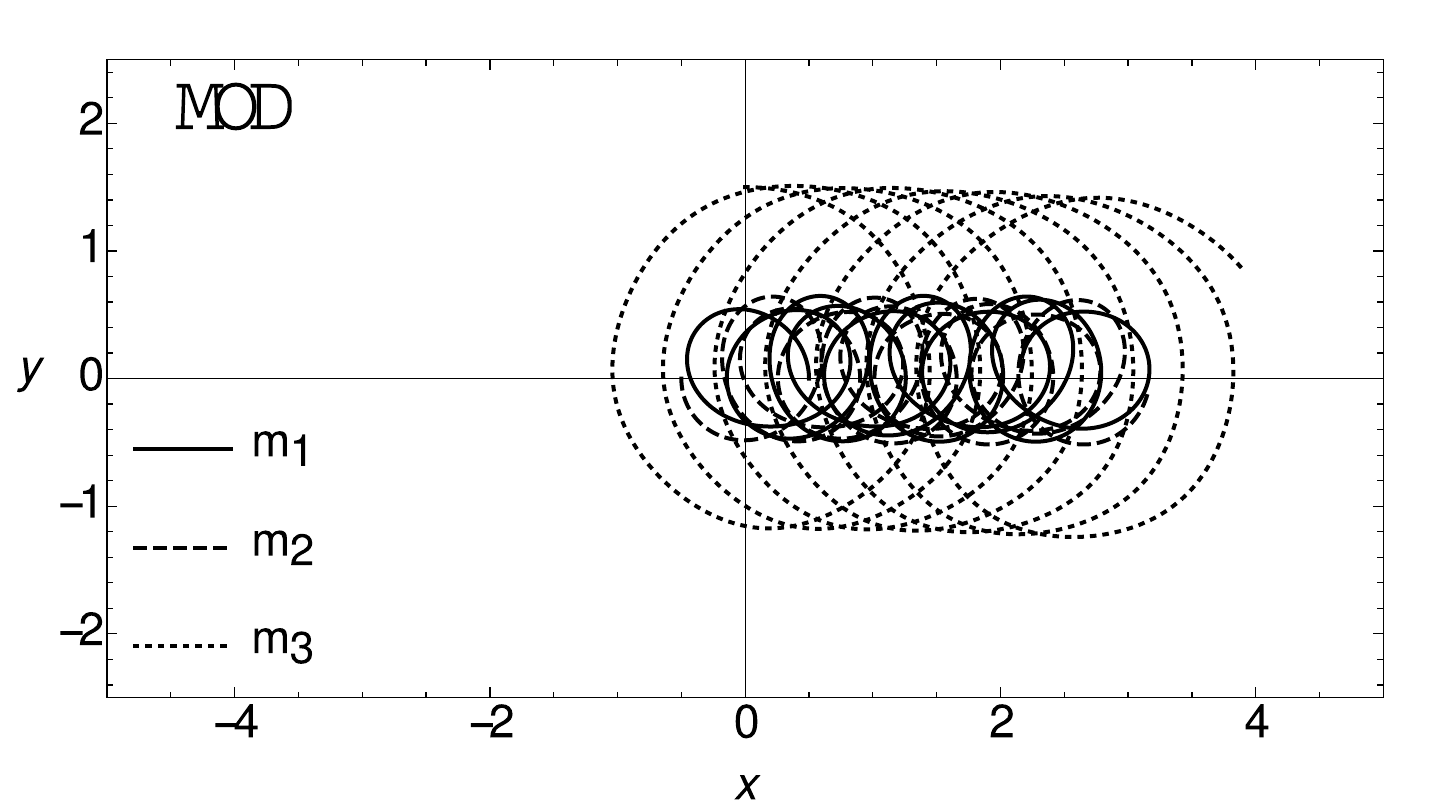} \includegraphics[width=0.45\textwidth]{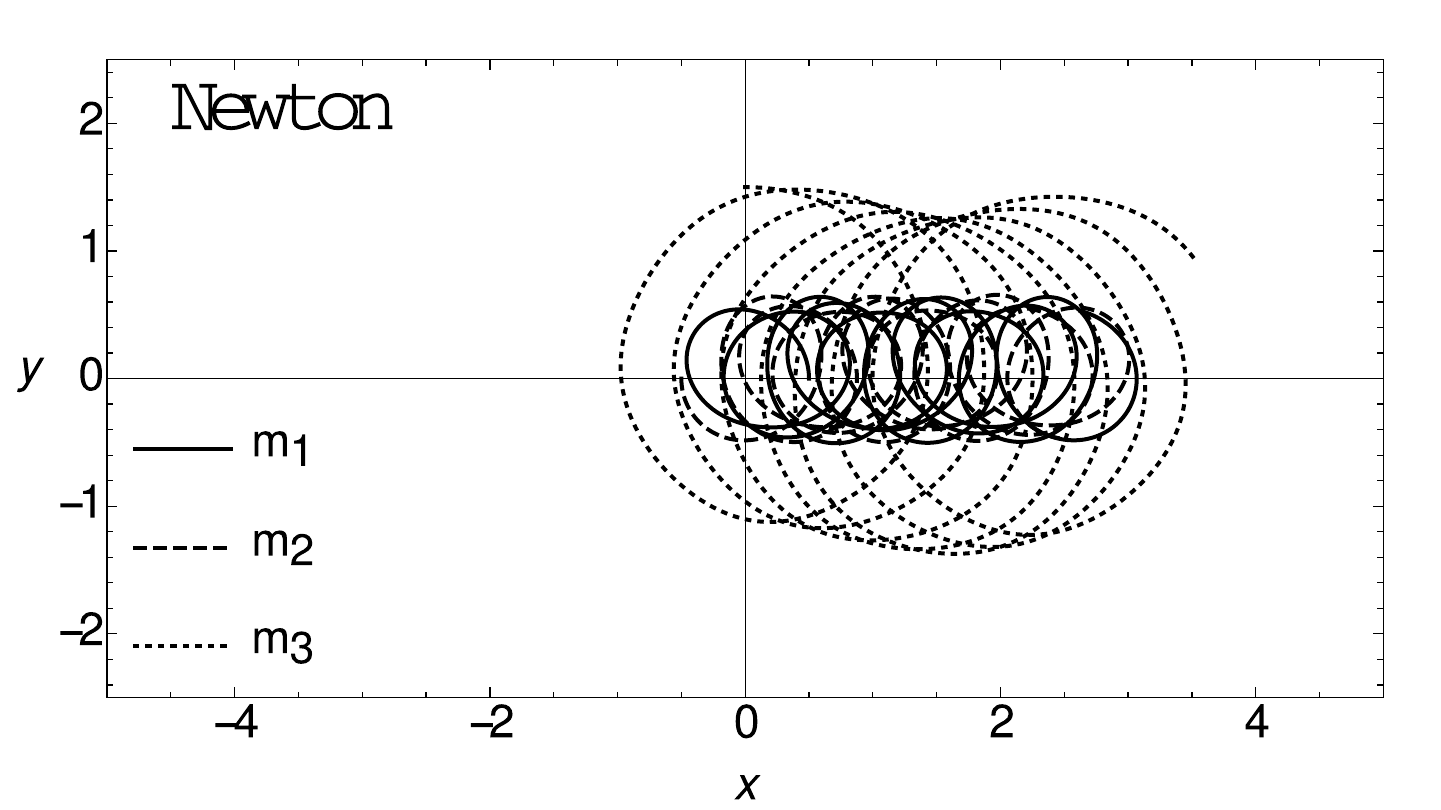}
\\
\includegraphics[width=0.45\textwidth]{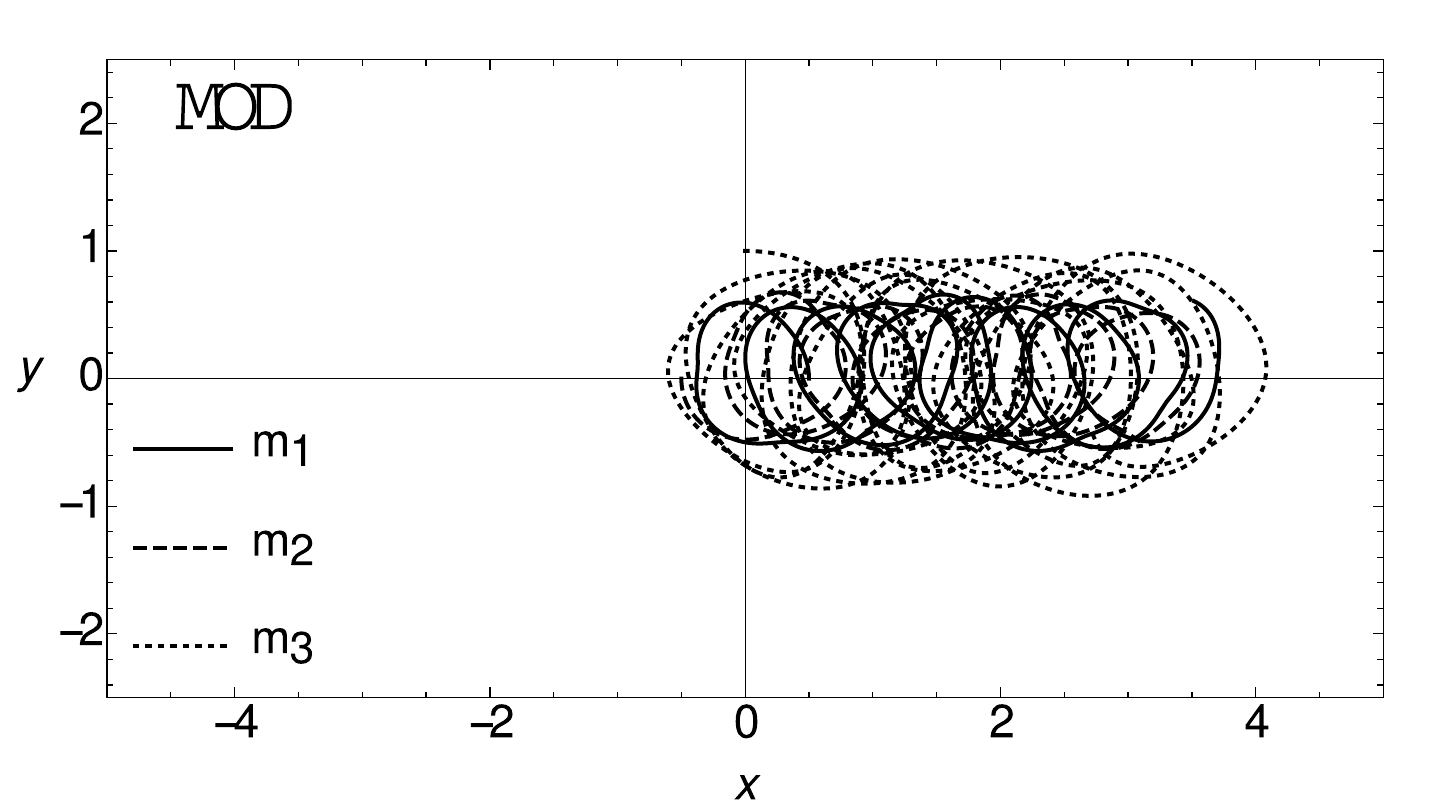} \includegraphics[width=0.45\textwidth]{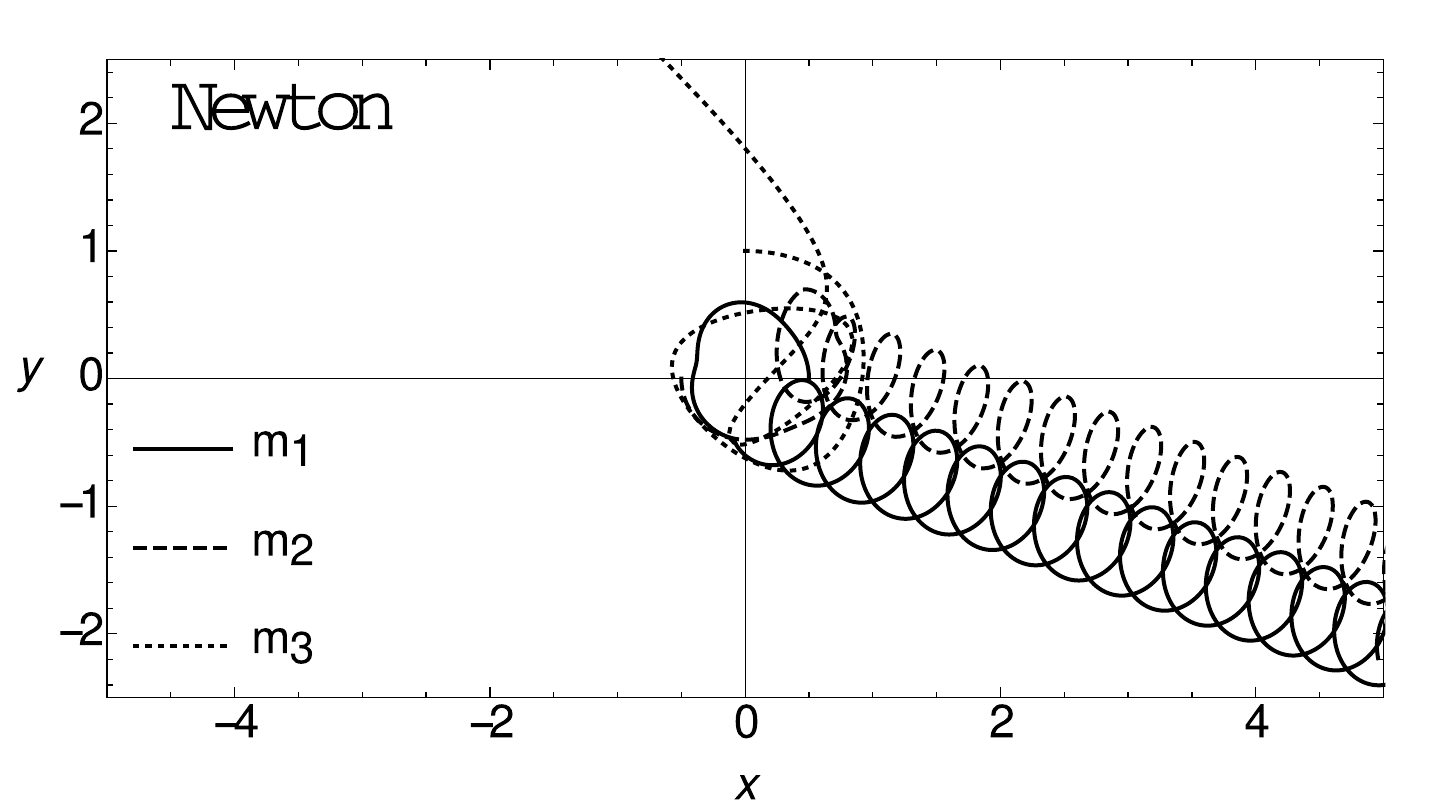}
\\
\caption{Upper panels: The time evolution ($300~\mathcal{T}$) of a binary system, each with mass  $0.1~\mathcal{M}$, and a third object with mass $0.01~\mathcal{M}$. The initial distance of the third particle is $4.5~\mathcal{D}$ on the $y$ axis.  The  Middle panels: The initial distance of the third particle is decreased to $1.5~\mathcal{D}$ on the $y$ axis. Lower panels:  The initial distance of the third particle is reduced to $1.0~\mathcal{D}$. The irregularity of the orbits increases from top to bottom; however, the lower MOD panel is still bounded while Newtonian evolution shows an ejection. The coordinates $x$ and $y$ are in units of $\mathcal{D}$.}
  \label{fig:Binary}
\end{figure*}

We have built three models consisting of a binary, each particle in the binary having a mass of  $0.1~\mathcal{M}$, and a third object with mass $0.01~\mathcal{M}$ in Fig. \ref{fig:Binary}. The masses are chosen smaller compared to the previous plots because we wanted the effect of MOD term to be more obvious, i.e. less Newtonian attraction. In plots related to MOD, the particles $m_{1}$ and $m_{2}$ in the binary start their motion from the initial  positions of  $\left( d,~ 0 \right)$ and $\left( -d,~ 0 \right)$ and initial velocities of $\left(0,~  d\omega \right)$ and $\left(0,~ -d\omega \right)$ respectively where the angular frequency $\omega$ is equal to 
$$\omega=\left( \frac{G(m_{1}+m_{2})}{8d^{3}}+\frac{2c_{1}a_{0}}{2d} \right)^{1/2}.$$
The third particle, however, is located initially at $\left(0,~d_{3} \right)$ and has an initial velocity of $\left( d_{3}\omega_{3},~0 \right)$ in which 
$$\omega_{3}=\left( \frac{G(m_{1}+m_{2})}{d^{3}_{3}}+\frac{2c_{1}a_{0}}{d_{3}} \right)^{1/2}.$$
The angular frequencies in Newtonian model are derived in the same way and by putting $c_{1}=0$. Using this set of initial conditions, the three-body system is "almost" in  dynamical equilibrium.

To see the effect of the modified dynamics, we have derived the time evolution of the three-body systems for different values of $d_{3}$ while keeping $d=0.5~\mathcal{D}$ fixed. For the upper panels in Fig. \ref{fig:Binary}, we have $d_{3}=4.5~\mathcal{D}$, i.e. $m_{3}$ is relatively far from the binary. As a result, the motions are  ordered in this plot, at least within the time evolution  of $300~\mathcal{T}$. Then, we reduce the distance of the third particle to $d_{3}=1.5~\mathcal{D}$ and the motions within the binary in both Newtonian and MOD becomes less regular while the motion of $m_{3}$ is still relatively ordered ( $150~\mathcal{T}$ ). See the middle  panels in Fig. \ref{fig:Binary}. Surely, in this case the orbit of $m_{3}$ is more regular in MOD than Newtonian model. Finally, when we decrease the initial distance of $m_{3}$ to $d_{3}=1.0~\mathcal{D}$, the irregularity of the orbits in MOD increases even more while we observe an ejection within Newtonian dynamics. See the lower  panels in Fig. \ref{fig:Binary}. We will argue in Sec. \ref{Lab} that the increased regularity in MOD orbits, compared to Newtonian ones,  is  due to the long range nature of MOD's fifth force.

\subsection{Different Regimes of Scaling}
Within the Newtonian gravity, the solutions of the three-body problem ( and also the N-body problem ) could be scaled freely. In fact, if $r_{1}$, $r_{2}$ and $r_{3}$ are the solutions of the three-body problem, then it could be proved that $q^{2}r_{1}$, $q^{2}r_{2}$ and $q^{2}r_{3}$ are also the solutions of the same problem if the time scale is redefined as $t~\Rightarrow~q^{3}t$. Here $q$ is a real number.

However, as Eqs. \ref{general3} show,   an important feature of the MOD model is that it breaks this scaling of the Newtonian model. This happens mainly because the new terms in the equations of motion transform differently than the Newtonian force. Also, the term due to modified dynamics is  proportional to the Hubble parameter which is itself time-dependent. Although, in the case that the modified term dominates the equation of motion, and where the motion is short enough that the Hubble parameter could be considered a constant, then we would have a new regime of scaling ( This case probably rarely happens because the modified term dominates the extragalactic scales in which the change in Hubble parameter is not negligible. ). In this case, if $r_{1}$, $r_{2}$ and $r_{3}$ are the solutions of the three-body problem dominated by a linear potential, then it could be proved that $qr_{1}$, $qr_{2}$ and $qr_{3}$ are also the solutions of the same equations if the time scale is redefined as $t~\Rightarrow~q^{1/2}t$. 

The main point here is the breaking of the three-body  scaling at galactic scales due to MOD term. In other words, although all the plots on the right hand side of Figs. \ref{fig:3bequalmass} to \ref{fig:Binary} could be re-scaled freely, i.e. they could happen at any scale, the same is not true for the plots on the left (MOD plots). The MOD plots are solely true for the galactic mass, time and size units specified above.  Although, when the MOD terms becomes negligible compared to the Newtonian terms, the scaling is recovered again. 

For systems with more particles, the scaling is still valid in Newtonian dynamics ( unless some no-gravitational forces are introduced into the problem ). This might be considered as a puzzle for Newtonian dynamics. The reason is that in Newtonian dynamics, any solution to an N-body problem could be scaled up ( or down ) to build another solution at arbitrary larger ( or smaller ) scales. However, we know that gravitational systems, e.g. planetary systems, star clusters, different types of galaxies and galaxy clusters, form and behave distinctly, i.e. they are physically different. Thus, in Newtonian dynamics, one is left with different initial geometries, mass distribution,  various populations  and non-gravitational forces within the systems to derive the desired physics at different scales. In modified dynamics, on the other hand, the scaling of  the  N-body problem is inherently broken. Therefore, systems at various acceleration  scales ( or similarly, at different surface densities ) are expected to act distinctly. The exact form of the behavior, however, depends on the equation of motion of the extended gravity under consideration and  needs to be studied through careful numerical experiments.

In the following, we will solve the most practical case of the three-body problem, i.e. the planar restricted three-body problem, in MOD. 

\section{Planar  Restricted Three-Body Problem}
\label{Restricted}
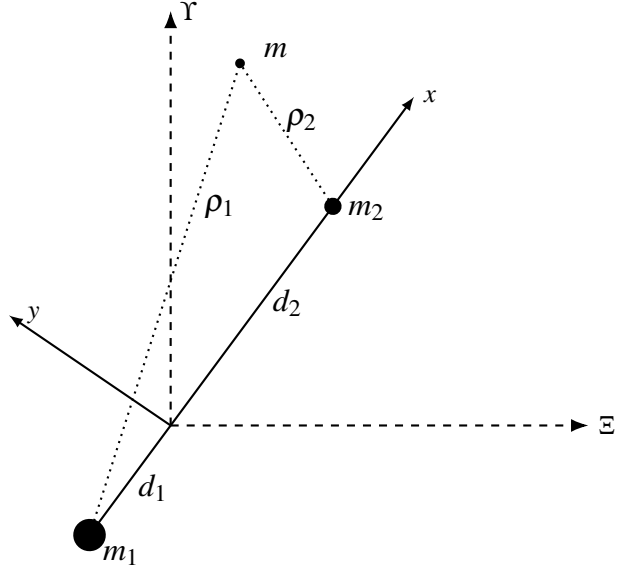
\begin{figure}[htbp]
\begin{center}
    \begin{tikzpicture}[thick] 
    \draw[-Latex, dashed] (0,0) -- (5.5,0)  node[right, text width=5em] {$\Xi$};
    \draw[-Latex, dashed] (0,0) -- (0,5.5)  node[right, text width=5em] {$\Upsilon$};

    \begin{scope}[rotate=20,draw=black]
        \draw[-latex] (0,0) -- (4.5,3)  node[right, text width=5em] {$x$};
        \draw[thick] (0,0) -- (-1.5,-1);
        \draw [thick,fill=black] (-1.5,-1) circle (0.2cm);
        \node [text=black] at (-1.2,-1.4) {\large $m_{1}$};
        \node [text=black] at (-0.5,-0.7) {\large $d_{1}$};
        \draw [thick,fill=black] (3,2) circle (0.1cm);
        \node [text=black] at (3.4,1.8) {\large $m_{2}$};
        \node [text=black] at (2,1) {\large $d_{2}$};
        \draw [thick,fill=black] (2.5,4.2) circle (0.05cm);
        \node [text=black] at (3,4.2) {\large $m$};
        \draw[-latex] (0,0) -- (-1.5,2.1)  node[right, text width=5em] {$~y$};
        \draw [thick,dotted] (-1.5,-1) -- (2.5,4.2);
        \node [text=black] at (1.6,2.5) {\large $\rho_{1}$};
        \node [text=black] at (3,3.2) {\large $\rho_{2}$};
        \draw [thick,dotted] (3,2) -- (2.5,4.2);
    \end{scope}
\end{tikzpicture}
\end{center}
\caption{A test particle $m$ is in the field of two massive objects $m_{1}$ and $m_{2}$. In our analysis, we employ the rotating coordinates $\left( x, y \right)$ instead of the fixed coordinates $\left( \Xi, \Upsilon \right)$. }
\label{Fig3Body}
\end{figure}

Here, we consider planar RTBP and we will use the method of \cite{desloge1982classical}, Chapter 62, because Desloge's method is general enough to be extended to the modified dynamics under investigation here. We consider the two more massive  objects in planar RTBP as $m_{1}$ and $m_{2}$  and the test particle as $m$. See Fig.  \ref{Fig3Body}. The two massive systems might be extended objects such as galaxies; however, for the sake of simplicity, we consider them as point masses in this research. In addition, we assume that the force between the objects is given by Eq. \ref{Force}. Also, $m_{1}$ and $m_{2}$ are in circular motion around their ( at rest ) center of mass. Moreover, the particle $m$ is moving in the same plane as the other two particles. If the effect of the test particle on $m_{1}$ and $m_{2}$ is neglected, as we assume here, then one could simply derive that:
\begin{eqnarray}
\frac{Gm_{1}m_{2}}{(d_{1}+d_{2})^{2}}+2c_{1}a_{0}\frac{m_{1}m_{2}}{M}=m_{1}d_{1}\omega^{2}=m_{2}d_{2}\omega^{2}
\end{eqnarray}
in which $d_{1}$ and $d_{2}$ are the distance of the masses $m_{1}$ and $m_{2}$ to the center respectively, while $\omega$ is the angular speed of the system.  Furthermore, in this section we have $M=m_{1}+m_{2}$  since we neglect the mass of $m=m_{3}$. From the above equation, one could derive the following results 
\begin{eqnarray} \label{eqs}
m_{1}d_{1} &=& m_{2}d_{2}\\ \nonumber
\frac{m_{2}}{M} &=& \frac{d_{1}}{d_{1}+d_{2}}\\ \nonumber
\frac{m_{1}}{M} &=&  \frac{d_{2}}{d_{1}+d_{2}}\\ \nonumber
\omega^{2} &=&  \frac{G (m_{1}+m_{2})}{(d_{1}+d_{2})^{3}}+\frac{2c_{1}a_{0}}{d_{1}+d_{2}}
\end{eqnarray}
which are the same as the Newtonian counterparts, except for the last equation.

As we will see in following, if we use the rotating coordinates $\left( x, y \right)$ instead of the fixed ones $\left( \Xi, \Upsilon \right)$, then, the resulting equations of motions would be more convenient to solve. In the rotating frame, the distances between the test particle $m$ and the masses $m_{1}$ and $m_{2}$ are given by 
\begin{eqnarray} \label{coords3}
\rho^{2}_{1} = (x + d_{1})^{2} + y^{2}\\  \nonumber 
\rho^{2}_{2} = (x - d_{2})^{2} + y^{2}
\end{eqnarray}
as Fig. \ref{Fig3Body} shows. Using these new coordinates, it is possible to rewrite the kinetic energy as
\begin{equation}
T=\frac{1}{2}m\left( \dot{\Xi}^{2} + \dot{\Upsilon}^{2}  \right) = \frac{1}{2}m\left( \dot{x}^{2} + \dot{y}^{2} -2\omega\dot{x}y +2\omega x \dot{y} +\omega^{2}\left[ x^{2}+y^{2}  \right]  \right)
\end{equation}
while the potential energy could be found as \citep{2016Ap&SS.361..378S}
\begin{equation}
\phi=-\frac{Gmm_{1}}{\rho_{1}} -\frac{Gmm_{2}}{\rho_{1}}+2c_{1}a_{0} \frac{m}{M}\left(m_{1}\rho_{1}  + m_{2}\rho_{2} \right).
\end{equation}
Now it is straightforward to derive the equations of motion from Lagrange equation. However, it is even more convenient to define an effective potential $U$ instead of the potential energy $\phi$  as follows:
\begin{eqnarray} \label{Upotential}
U &\equiv & \frac{\phi}{m}-\frac{1}{2}\omega^{2}\left( x^{2}+y^{2} +d_{1}d_{2} \right)\\ \nonumber
~&=& -\frac{Gm_{1}}{\rho_{1}} -\frac{Gm_{2}}{\rho_{2}}+ \frac{2c_{1}a_{0}}{M}\left(m_{1}\rho_{1}  + m_{2}\rho_{2} \right) - \frac{1}{2(d_{1}+d_{2})}\left( \frac{GM}{(d_{1}+d_{2})^{2}} + 2c_{1}a_{0} \right) \left( \frac{d_{1} \rho^{2}_{2}+d_{2} \rho^{2}_{1} }{d_{1}+d_{2}} \right)
\end{eqnarray}
The benefit of defining $U$ is that this potential includes all terms dependent to the coordinates $\left( x, y \right)$,  or $\left( \rho_{1}, \rho_{2} \right)$, 
 while the terms dependent to $\dot{x}$ and $\dot{y}$ are separated now. Then, one could derive the equation of motion of the test particle as follows:
\begin{eqnarray} \label{eom}
\ddot{x}-2\omega \dot{y} = -\frac{\partial U}{\partial x} \\ \nonumber
\ddot{y}+2\omega \dot{x} = -\frac{\partial U}{\partial y}
\end{eqnarray}

Eqs. \ref{eom} define a dynamical system and to understand the bahavior of this set of equations, we need to study the potential $U$ ( The situation here is somewhat more complicated than the problem of a single particle in a potential $U$ since here we have a coupled system of equations with the additional  terms $2\omega \dot{x}$ and $2\omega \dot{y}$ also playing roles. ). The derivatives of $U$ with respect to the coordinates $ \rho_{1}$ and $ \rho_{2}$ could be derived as
\begin{eqnarray} \label{dUrho}
\frac{\partial U}{\partial \rho_{1}}= \frac{Gm_{1}}{\rho^{2}_{1}} \left( 1-\frac{\rho^{3}_{1}}{(d_{1}+d_{2})^{3}} \right) +2c_{1}a_{0}\frac{m_{1}}{M} \left( 1-\frac{\rho_{1}}{d_{1}+d_{2}} \right)\\ \nonumber 
\frac{\partial U}{\partial \rho_{2}}= \frac{Gm_{2}}{\rho^{2}_{2}} \left( 1-\frac{\rho^{3}_{2}}{(d_{1}+d_{2})^{3}} \right) +2c_{1}a_{0}\frac{m_{2}}{M} \left( 1-\frac{\rho_{2}}{d_{1}+d_{2}} \right)
\end{eqnarray}
from which one may obtain the first and second derivatives of $U$ with respect to the coordinates $\left( x, y \right)$:
\begin{eqnarray} \label{derivatives}
U_{x} &=& \frac{Gm_{1}\left( x+ d_{1} \right)}{\rho^{3}_{1}}+\frac{Gm_{2}\left( x- d_{2} \right)}{\rho^{3}_{2}}  +\frac{2c_{1}a_{0}}{M} \left( m_{1}\frac{ x+ d_{1}  }{\rho_{1}}+  m_{2} \frac{ x- d_{2}  }{\rho_{2}} \right) -\omega^{2}x \\ \nonumber
U_{y} &=& \frac{Gm_{1}y}{\rho^{3}_{1}}+\frac{Gm_{2}y}{\rho^{3}_{2}}  +\frac{2c_{1}a_{0}}{M} \left( \frac{ m_{1}  }{\rho_{1}}+  \frac{ m_{2}  }{\rho_{2}} \right)y -\omega^{2}y\\   \nonumber
U_{xx} &=& Gm_{1} \left( \frac{ 1 }{\rho^{3}_{1}}-3 \frac{ ( x+ d_{1} )^{2}}{\rho^{5}_{1}} \right) +Gm_{2} \left( \frac{ 1 }{\rho^{3}_{2}}-3 \frac{ ( x- d_{2} )^{2} }{\rho^{5}_{2}} \right)  \\    \nonumber
 &+& \underbrace{ \frac{2c_{1}a_{0}m_{1}}{M\rho_{1}} \left( 1- \frac{ ( x+ d_{1} )^{2} }{\rho^{2}_{1}} \right)
+\frac{2c_{1}a_{0}m_{2}}{M\rho_{2}} \left( 1- \frac{ ( x- d_{2} )^{2} }{\rho^{2}_{2}} \right)} - \omega^{2}   \\    \nonumber
 U_{yy} &=& Gm_{1} \left( \frac{ 1 }{\rho^{3}_{1}}-3 \frac{ y^{2}}{\rho^{5}_{1}} \right) +Gm_{2} \left( \frac{ 1 }{\rho^{3}_{2}}-3 \frac{ y^{2} }{\rho^{5}_{2}} \right)  + \frac{2c_{1}a_{0}m_{1}}{M\rho_{1}} \left( 1- \frac{ y^{2} }{\rho^{2}_{1}} \right)
+\frac{2c_{1}a_{0}m_{2}}{M\rho_{2}} \left( 1- \frac{ y^{2} }{\rho^{2}_{2}} \right)
 -\omega^{2}   \\    \nonumber
U_{xy} &=& -3\frac{Gm_{1}\left( x+ d_{1} \right)y}{\rho^{5}_{1}}-3\frac{Gm_{2}\left( x- d_{2} \right)y}{\rho^{5}_{2}}  -\frac{2c_{1}a_{0}}{M} \left( m_{1}\frac{ x+ d_{1}  }{\rho^{3}_{1}}+  m_{2} \frac{ x- d_{2}  }{\rho^{3}_{2}} \right)y\\    \nonumber
\end{eqnarray}
From now on, we will show derivatives with respect to $\rho_{1}$ and $\rho_{2}$ using $\partial / \partial \rho_{i}$ while we will use indices to show derivation with respect to $x$ and $y$; thus, $U_{xy}=\partial^{2}U/\partial x \partial y$ etc, .

\subsection{Jacobian Integral}
By multiplying the equations of motions \ref{eom} by $\dot{x}$ and $\dot{y}$ respectively, and then adding the results, one can derive:
\begin{equation}
    \frac{1}{2} \frac{d}{dt}\left( \dot{x}^{2}+\dot{y}^{2}\right)+\frac{d}{dt}U=0 
\end{equation}
which clearly indicates a constant of the motion.  This constant could be considered as the energy in the rotating frame, or
\begin{equation}
    \frac{1}{2} \left( \dot{x}^{2}+\dot{y}^{2} \right) +U = E
\end{equation}
though, in the literature related to dynamical astronomy it is more common to use the Jacobi integral $C=-2E$ instead of $E$. The value of this constant could be determined based on initial conditions, i.e. initial position and velocity, of the motion of the test particle.  Although, it should be mentioned that for the sake of simplicity, here we assume that the term $c_1 a_0$ is constant. This assumption is useful for our simple toy models; however, in reality it only works when we are dealing with short periods of time ( compared to $1/H_{0}$).

\begin{figure*}[ht]
\centering
\includegraphics[width=0.95\textwidth]{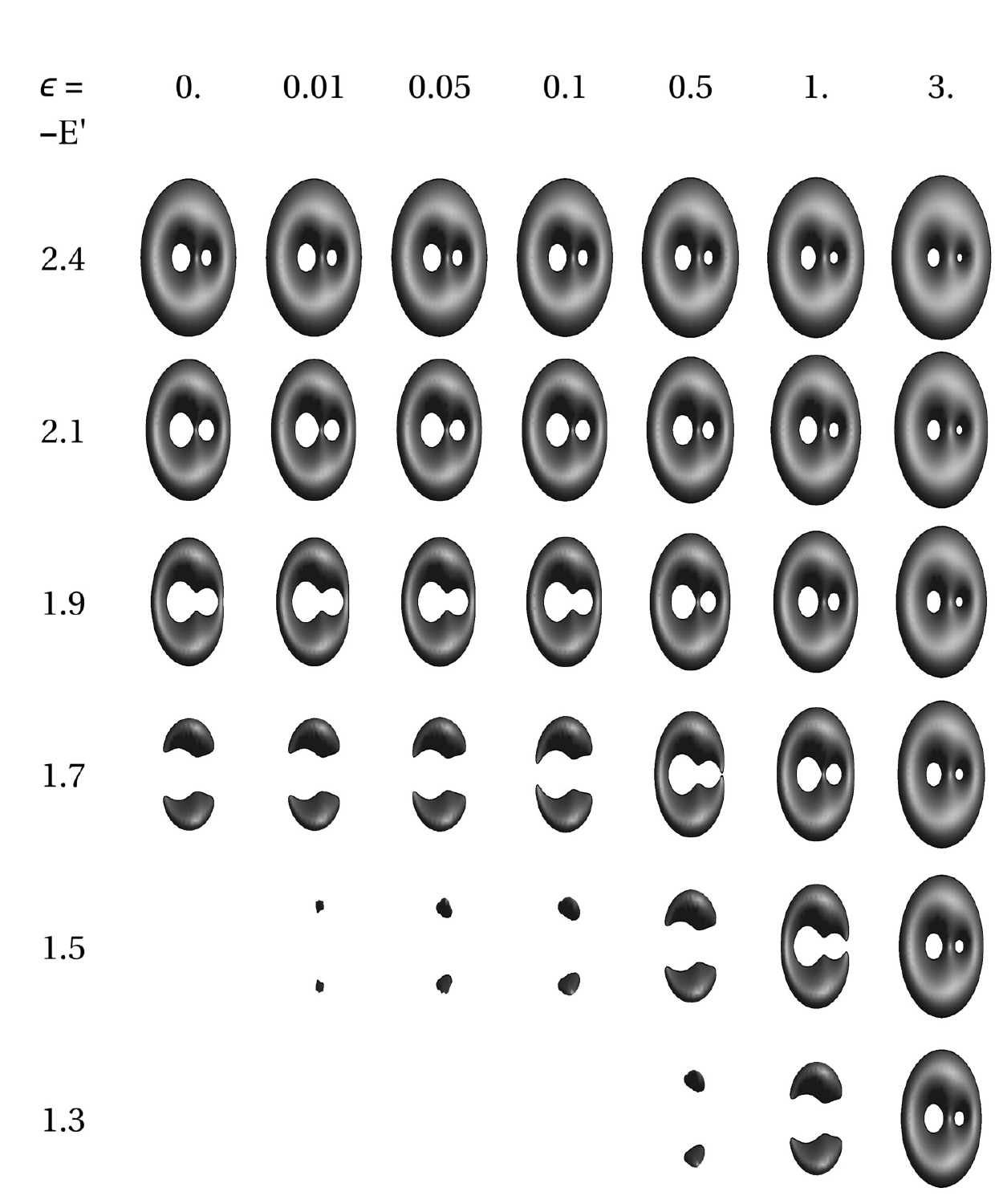}
\vspace{-10pt}
\caption{ The effective potential, in units of $GM/(d_{1}+d_{2})$, for different values of the dimensionless energy  in the rotating frame $E^{\prime} \equiv E/(GM/(d_{1}+d_{2}))$ and the ratio $\epsilon$, in the case $\mu = 0.3$.
\label{fig:effectiveU}}
\end{figure*}

Since the kinetic energy $\frac{1}{2} \left( \dot{x}^{2}+\dot{y}^{2} \right)$ cannot be negative, the motion of the test particle is  restricted to the region of space where the condition $\dot{x}^{2}+\dot{y}^{2}= 2 \left(E-U \right)  \geq 0$ is strictly true. This condition could be interpreted as follows: When a test particle with an energy in the rotating frame $E$   moves in a bounded subspace which is defined by condition $E > U$, then the particle will always remain in that region. This situation could happen when the test particle is very close to either of the masses $m_{1}$ and $m_{2}$, i.e. when it is deep in the potential well of one of the massive objects. See Fig. \ref{fig:effectiveU}. However, when the subspace is not closed, the test particle could eventually escape from the initial  region and move to another region of the subspace.

To see clearly what happens here, we  plot the  effective potential $U$, in units of $GM/(d_{1}+d_{2})$:
\begin{eqnarray} \label{dimensionlessU}
\frac{U}{GM/(d_{1}+d_{2})}  ~&=& -\frac{1-\mu}{\rho^{\prime}_{1}} -\frac{\mu}{\rho^{\prime}_{2}}  \\  \nonumber 
~&+& \epsilon \left[(1-\mu)\rho^{\prime}_{1}  + \mu \rho^{\prime}_{2} \right] - \frac{1}{2}\left(1+\epsilon \right) \left[ (1-\mu) \rho^{\prime~ 2}_{1} + \mu \rho^{\prime~ 2}_{2} \right]
\end{eqnarray}
in which $\mu=m_{2}/M$, $\rho^{\prime}_{i}=\rho_{i}/(d_{1}+d_{2})$ is a dimensionless scale for $i=1,~2$, and  
\begin{equation} \label{epsilon}
    \epsilon \equiv 2c_{1}\frac{a_{0}/G}{M/(d_{1}+d_{2})^{2}}
\end{equation}
is a parameter which determines the ratio of the MOD acceleration compared to the Newtonian one. We have plotted the dimensionless effective potential \ref{dimensionlessU} in Fig. \ref{fig:effectiveU}
for $\mu=0.3$, various values of the dimensionless  energy  in the rotating frame $E^{\prime} \equiv E/(GM/(d_{1}+d_{2}))$ and different values of $\epsilon$.

It is quite interesting that the addition of the specific modified potential of MOD does not change the shape of $U$ and  the reason for this is relatively easy to understand. Imagine a test particle near the mass $m_{2}$ in a way that we have $\rho^{\prime}_{2} \ll 1$. Therefore,  the term $\mu/\rho^{\prime}_{2}$ dominates other terms in Eq. \ref{dimensionlessU} ( It even dominates the modified potential terms proportional to $\epsilon$ ). In other words, in a very close distance to  $m_{2}$, the Newtonian potential well dominates the dynamics. On the other hand, far from the objects, i.e. when $  \rho^{\prime}_{1} \gg 1 $ and $  \rho^{\prime}_{2} \gg 1 $, the second-order terms related to the centrifugal force dominate both Newtonian and MOD potentials. These are the terms of the form $\propto \rho^{\prime~2}_{i} $ in Eq. \ref{Upotential} and when they dominate, the test particle would be in another potential well.  Of course, the presence of the MOD term causes the former potential well to be deeper. Thus, we see that in Fig. \ref{fig:effectiveU} by increasing the parameter $\epsilon$, i.e. by increasing the MOD term compared to the Newtonian term, the forbidden regions extend. For example, for $E^{\prime}=-1.5 $ and $\epsilon = 1.0$ we still would have a large realm of forbidden subspace while there is no forbidden subspace in the Newtonian gravity ($\epsilon = 0$) for this energy.  However, as stated before, the figure also shows that the form of the potential stays the same. In fact we will prove in the next part that  the number of the critical points does not change either. We would have three saddle points ($L_{1}$, $L_{2}$ and $L_{3}$) along the axis of $x$ while there are two symmetric maximum points ( $L_{4}$ and $L_{5}$) off the $x$ axis. 

Now, let's check the realms of possible motion \citep{koon2000dynamical}. Let's designate the energy at every Lagrange point $L_{i}$ as $E^{\prime}_{i}$ and the energy of the test particle by $E^{\prime}_{p}$.
Then, if   $E^{\prime}_{p}<E^{\prime}_{1}$, the  test particle could not transfer between the two interior potential wells ( around $m_{1}$ and $m_{2}$). The existence of the MOD term, i.e. $\epsilon >0$, only makes this condition stronger because the attractive force becomes stronger with increasing $\epsilon$. See Fig. \ref{fig:effectiveU} with $E^{\prime}=-2.1$, for example, and various values of $\epsilon$. In addition, when $E^{\prime}_{1} < E^{\prime}_{p} < E^{\prime}_{2}$, there appears a path between the two masses on the $x$ axis. Thus, the test particle could now transfer between regions around $m_{1}$ and $m_{2}$; however, it could not go to the exterior region which extends to infinity. Also, we could see that with increasing $\epsilon$, the neck might close up again, i.e. the net attractive force becomes stronger and now the test particle is confined again. Thus, it is possible for a test particle in Newtonian gravity with a specific energy to cross the neck between $m_{1}$ and $m_{2}$; however, the  particle with the same energy in MOD might be incapable of crossing  between $m_{1}$ and $m_{2}$ regions. To see this point, in Fig. \ref{fig:effectiveU} compare the row with $E^{\prime}=-1.9$ and various values of $\epsilon$. There is also the possibility of $E^{\prime}_{2} < E^{\prime}_{p} < E^{\prime}_{3}$ in which the particle could transfer between the interior and the exterior realms. Again, by increasing $\epsilon$, this possibility might close up.

Moreover, as we will prove in the following, the $L_{4}$ and $L_{5}$ Lagrange points happen at $\rho^{\prime}_{1}=\rho^{\prime}_{2}=1$. Thus, in these points we would have  $E^{\prime}_{4}=E^{\prime}_{5}=-3/2+\epsilon /2$. Now, when $E^{\prime}_{3} < E^{\prime}_{p} < E^{\prime}_{4}$, the test particle could move through all $x$ axis and has access to almost entire space, unless  forbidden areas around $L_{4}$ and $L_{5}$. It is easy to see that by increasing $\epsilon$, the forbidden subspace around these two points grows. Finally, when $E^{\prime}_{p} > E^{\prime}_{4}$, the particle has access to the entire space. Again, a particle with a specific energy could be free to move in Newtonian dynamics but still banned from some areas in MOD ( with the same energy ).

\subsection{Lagrange Points}
The equilibrium points of the motion are the points in which the test particle remains at rest in the rotating frame of reference. According to Eqs. \ref{eom}, this situation happens when $\frac{\partial U}{\partial x}=0 $ and $\frac{\partial U}{\partial y}=0$. Writing in terms of $\rho_{1}$ and $\rho_{2}$, one could see that these two equations are equal to
\begin{eqnarray} \label{sys1}
U_{x} &=& \frac{\partial U}{\partial \rho_{1}}\frac{\partial \rho_{1}}{\partial x}+\frac{\partial U}{\partial \rho_{2}}\frac{\partial \rho_{2}}{\partial x}=0\\ \nonumber
U_{y} &=& \frac{\partial U}{\partial \rho_{1}}\frac{\partial \rho_{1}}{\partial y}+\frac{\partial U}{\partial \rho_{2}}\frac{\partial \rho_{2}}{\partial y}=0
\end{eqnarray}
This homogeneous system of equations provides two sets of solutions ( in total, five distinctive points of $L_{i}$ ). The first set of solutions are obtained when 
\begin{eqnarray}
\begin{large}
\det\begin{pmatrix} \frac{\partial \rho_{1}}{\partial x} & \frac{\partial \rho_{2}}{\partial x} \\ \frac{\partial \rho_{1}}{\partial y} & \frac{\partial \rho_{1}}{\partial y} \end{pmatrix}
\end{large}
\neq 0,
\end{eqnarray}
thus; these solutions must satisfy the following equations
\begin{equation}
\frac{\partial U}{\partial \rho_{1}}=\frac{\partial U}{\partial \rho_{2}}=0.
\end{equation}
Please note that the solutions to this set of equations depends on the type of the extended gravity potential which is under consideration since it depends on $U$. Using Eqs. \ref{dUrho} one could see that there are in general nine distinct solutions for this system of equations
\begin{eqnarray} \label{L4point}
\frac{\rho_{1}}{d_{1}+d_{2}}&=&\frac{\pm \sqrt{-4 \epsilon -3}-1}{2 (\epsilon +1)},~~~~\frac{\rho_{2}}{d_{1}+d_{2}}=1  \\  \nonumber
\frac{\rho_{1}}{d_{1}+d_{2}}&=&\frac{\pm \sqrt{-4 \epsilon -3}-1}{2 (\epsilon +1)},~~~~\frac{\rho_{2}}{d_{1}+d_{2}}=\frac{\pm \sqrt{-4 \epsilon -3}-1}{2 (\epsilon +1)}  \\  \nonumber
\frac{\rho_{2}}{d_{1}+d_{2}}&=&\frac{\pm \sqrt{-4 \epsilon -3}-1}{2 (\epsilon +1)}, ~~~~ \frac{\rho_{1}}{d_{1}+d_{2}}=1  \\  \nonumber
\frac{\rho_{1}}{d_{1}+d_{2}} &=&\frac{\rho_{2}}{d_{1}+d_{2}}=1 ~~~~~~\Leftarrow~~~~~\text{the only real solution.}\\  \nonumber
\end{eqnarray}
 which only one of them, i.e. the last one, is real since the quantity $\sqrt{-4 \epsilon -3}$ is, by definition, imaginary. This real solution arises within the context of Newtonian theory too and it could be described as follows:  if an equilateral triangle is constructed with masses $m_{1}$ and $m_{2}$ placed at the two vertices of the triangle, then the equilibrium point would occur at the third vertex. The equilibrium point in the direction of positive ( negative ) y is named the fourth ( fifth ) Lagrange point $L_{4}$ ($L_{5}$). 
 
 The second set of solutions arises when the determinant of coefficients in the system of equations \ref{sys1} is null:
\begin{eqnarray}
\begin{large}
\det\begin{pmatrix} \frac{\partial \rho_{1}}{\partial x} & \frac{\partial \rho_{2}}{\partial x} \\ \frac{\partial \rho_{1}}{\partial y} & \frac{\partial \rho_{1}}{\partial y} \end{pmatrix}
\end{large}
= \begin{large}
\det\begin{pmatrix} \frac{x+d_{1}}{\rho_{1}} & \frac{x-d_{2}}{\rho_{2}}  \\ \frac{y}{\rho_{1}} & \frac{y}{\rho_{2}}
\end{pmatrix}
\end{large}
= 0
\end{eqnarray}
which is true only in the case of $y=0$, i.e. this set of solutions occurs on the $x$ axis. Please note that this  constraint is independent of the type of the extended gravity under study, i.e. no relation to $U$.  According to Eqs. \ref{derivatives}, on the $x$ axis we have $\frac{\partial U}{\partial y}|_{y=0} =0$; thus, we should investigate the zeros of $\frac{\partial U}{\partial x} $ to find the extremum points: 
\begin{eqnarray} \label{Ux}
U_{x} &=& \frac{Gm_{1}\left( x+ d_{1} \right)}{|x+ d_{1}|^{3}}+\frac{Gm_{2}\left( x- d_{2} \right)}{|x- d_{2}|^{3}}   \\ \nonumber  
&+&  \underbrace{ \frac{2c_{1}a_{0}}{M} \left( m_{1}\frac{ x+ d_{1}  }{|x+ d_{1}|}+  m_{2} \frac{ x- d_{2}  }{|x- d_{2}|} \right)} -\omega^{2}x  =0
\end{eqnarray}
This equation, being of third order, must have three solutions ( This might change in other extended gravities. ). To find these solutions, one could break down the $x$ axis into three parts, i.e. $\left( -\infty, -d_{1} \right)$, $\left( -d_{1}, d_{2} \right)$ and $\left( d_{2}, +\infty \right)$, and then analyze the behavior of $U_{x}$ separately in these sub-intervals. Before doing so, we note that the term due to modified dynamics, i.e. the underlined term in Eq. \ref{Ux}, only adds a constant to $U_{x}$ in each sub-intervals. This constant would be of the general form of $\frac{2c_{1}a_{0}}{M}\left( \pm m_{1}\pm m_{2}\right)$. Also, we note that $U_{xx}$ is everywhere negative 
\begin{equation}
U_{xx}=-\frac{2Gm_{1}}{|x+ d_{1}|^{3}}-\frac{2Gm_{2}}{|x- d_{2}|^{3}}-\omega^{2}    
\end{equation}
and goes to $-\infty$ near $x=-d_{1}$
and $x=d_{2}$.   An important observation here is the disappearance of the MOD term in $U_{xx}$. In fact, it could be easily seen that any additional potential of the form $\propto r^{n} $ produces a term in $U_{xx}$  of the form of ( for every gravitating particle $i$ )
$$U\propto r^{n}~~~~~~\Rightarrow~~~~~~U_{xx}\propto \rho^{n-2}_{i}\left( 1+ (n-2)\frac{(x \pm d_{i})^2}{\rho^{2}_{i}} \right) $$
 which  only in the case of $n=1$ is identically zero on the $x$ axis. Thus, MOD does not change the behavior of the $U_{xx}$ on the $x$ axis, and similar to Newtonian gravity, $U_{xx}$ in MOD is strictly negative on $y=0$ axis. 

Since $U_{xx}$ is negative  everywhere on $y=0$, it turns out that $U_{x}$ is decreasing in all three sub-intervals. Let's first consider the sub-interval $\left( d_{2}, +\infty \right)$. When $x \rightarrow d^{+}_{2}$, the function   $U_{x}$ diverges to $+\infty$ while for $x \rightarrow  +\infty$, $U_{x}$ goes to $-\infty$. Thus, there must be a zero for $U_{x}$ in $\left( d_{2}, +\infty \right)$. We will call this point the second Lagrange point $L_{2}$.   A similar situation  happens when we consider the other two sub-intervals. For example, within the sub-interval $\left( -\infty, -d_{1} \right)$, we have  $U_{x} \rightarrow +\infty$ when $x \rightarrow  -\infty$ while  the function   $U_{x}$ diverges to $-\infty$ when $x \rightarrow -d^{-}_{1}$ Thus, there must be another zero for $U_{x}$ in $\left(-\infty, -d_{1} \right)$. This point is called the third Lagrange point $L_{3}$ while the one within $\left( -d_{1}, d_{2} \right)$ is named $L_{1}$. We will now investigate the stability of orbits around Lagrange points.

\subsection{Stability of the Lagrange Points}
We investigate the stability of orbits around equilibrium points by employing a perturbative method. If the Lagrange point under consideration is located at $(x_{0},~y_{0})$, then the equation of motion at a point $(x,~y)$  very near to the Lagrange point
\begin{eqnarray} \label{perturb}
x=x_{0}+\xi \\ \nonumber
y=y_{0}+\eta
\end{eqnarray}
would be governed by the equation of motion \ref{eom} as follows:
\begin{eqnarray}  
\ddot{\xi}-2\omega \dot{\eta} = -U_{x}\left( x_{0}+\xi,  y_{0}+\eta \right) \\ \nonumber
\ddot{\eta}+2\omega \dot{\xi} = -U_{y}\left( x_{0}+\xi,  y_{0}+\eta \right)
\end{eqnarray}
Now, we expand the right hand side of the above equation to first order in $\xi $ and $\eta$ as
\begin{eqnarray}  \label{pert}
U_{x}\left( x_{0}+\xi,  y_{0}+\eta \right) &=& U_{xx}\left( x_{0},  y_{0} \right)\xi + U_{xy}\left( x_{0},  y_{0} \right)\eta \\ \nonumber
U_{y}\left( x_{0}+\xi,  y_{0}+\eta \right) &=& U_{yy}\left( x_{0},  y_{0} \right)\eta + U_{xy}\left( x_{0},  y_{0} \right)\xi.   
\end{eqnarray}
 These expansions would be employed in following to study the stability of motion around the Lagrange points.
\vspace{10pt}

\textbf{Stability of orbits around $L_{1}$, $L_{2}$ and $L_{3}$:} For these points we would have $\rho_{1}=|x_{0}+d_{1}|$, $\rho_{2}=|x_{0}-d_{2}|$ and $y=0$; thus, from Eqs. \ref{pert} one can derive the next set of equations 
\begin{eqnarray}  
\ddot{\xi}-2\omega \dot{\eta} &=& \left( 2\Omega^{2}_{N} +\omega^{2} \right)\xi  \\ \nonumber
\ddot{\eta}+2\omega \dot{\xi} &=&- \left( \Omega^{2}_{N} + \Omega^{2}_{MOD} -\omega^{2}  \right)\eta 
\end{eqnarray}
in which
\begin{eqnarray}  
\Omega^{2}_{N} &=& \frac{Gm_{1}}{|x_{0}+d_{1}|^{3}}+\frac{Gm_{2}}{|x_{0}-d_{2}|^{3}}  \\ \nonumber
\Omega^{2}_{MOD} &=& \frac{2c_{1}a_{0}}{M} \left( \frac{m_{1}}{|x_{0}+d_{1}|} 
+\frac{m_{2}}{|x_{0}-d_{2}|} \right).
\end{eqnarray}
A critical observation 
here is that the underlined terms in the derivatives of \ref{derivatives} are equal to zero. This happens only for a linear potential form as could be checked simply. Therefore, the current model only modifies the form of the equation of motion in the direction of $\eta$.

To check possible divergence of the solutions, we will put $\xi = \xi_{0}e^{\lambda t}$ and $\eta = \eta_{0}e^{\lambda t}$ into the last equations to derive 
\begin{eqnarray}
\begin{pmatrix} 
\lambda^{2}- 2\Omega^{2}_{N} -\omega^{2} & -2\omega \lambda \\ 2\omega \lambda & \lambda^{2} + \Omega^{2}_{N} + \Omega^{2}_{MOD} -\omega^{2}
\end{pmatrix}
\begin{pmatrix} 
\xi_{0}  \\ \eta_{0}
\end{pmatrix} =
\begin{pmatrix} 
0 \\ 0
\end{pmatrix}
\end{eqnarray}
which would imply that the determinant of the coefficient matrix must be zero for any nontrivial solution to exist:
\begin{eqnarray} \label{quartic}
\lambda^{4} &+& \left( 2\omega^{2} - \Omega^{2}_{N} + \Omega^{2}_{MOD}  \right)\lambda^{2}  - \left( 2\Omega^{2}_{N}+ \omega^{2} \right)\left(  \Omega^{2}_{N}+ \Omega^{2}_{MOD} - \omega^{2} \right)=0
\end{eqnarray}
This, in turn, results in 
\begin{eqnarray}
\lambda^{2} &=&    \frac{1}{2} \left(\Omega _N^2-\Omega
   _{\text{MOD}}^2-2 \omega ^2\right)  \pm \sqrt{9 \Omega _N^4 -8 \omega ^2 \Omega _N^2  + \Omega _{\text{MOD}}^2 \left(\Omega _{\text{MOD}}^2+6 \Omega _N^2+8 \omega ^2\right)}
\end{eqnarray}
Here, we could see that one could simply derive the  Newtonian counterpart of this result by putting $\Omega_{MOD}=0$, as we expected. Compare with page 641 of \cite{desloge1982classical}, for example.

Let us investigate the nature of the solutions in this case. If we demand that the solutions have to be stable, then  $\lambda$ must be purely imaginary, i.e. $\lambda^{2}<0$. For this to be true, we must have ( from the biquadratic equation \ref{quartic} ):
$$- \left( 2\Omega^{2}_{N}+ \omega^{2} \right)\left(  \Omega^{2}_{N}+ \Omega^{2}_{MOD} - \omega^{2} \right) >0$$
or, since the term $\left( 2\Omega^{2}_{N}+ \omega^{2} \right)$ is always positive, we would have
\begin{equation} \label{condition1}
    \Omega^{2}_{N} + \Omega^{2}_{MOD} - \omega^{2} < 0
\end{equation}
On the other hand, we might find another bound on these values by considering Eq. \ref{Ux} that must be true at the Lagrange points on the $x$ axis:
\begin{eqnarray} 
U_{x}~(x_{0}) &=& \frac{Gm_{1}\left( x_{0} + d_{1} \right)}{|x_{0} + d_{1}|^{3}}+\frac{Gm_{2}\left( x_{0}- d_{2} \right)}{|x_{0}- d_{2}|^{3}}   \\ \nonumber  
&+&  \underbrace{ \frac{2c_{1}a_{0}}{M} \left( m_{1}\frac{ x_{0} + d_{1}  }{|x_{0} + d_{1}|}+  m_{2} \frac{ x_{0} - d_{2}  }{|x_{0} - d_{2}|} \right)} -\omega^{2}x_{0}  =0
\end{eqnarray}
From here we can derive
\begin{eqnarray}  \label{condition2}
\Omega^{2}_{N} &+& \Omega^{2}_{MOD} - \omega^{2}  \\  \nonumber  
&=& \frac{Gm_{1}d_{1}}{x_{0}}\left( \frac{1}{|x_{0}-d_{2}|}-  \frac{1}{|x_{0}+d_{1}|} 
 \right) \\  \nonumber 
 &\times& \left( \frac{1}{|x_{0}-d_{2}|^{2}}+  \frac{1}{|x_{0}+d_{1}||x_{0}-d_{2}|}+\frac{1}{|x_{0}+d_{1}|^{2}}+\frac{2c_{1}a_{0}/G}{M}
 \right).
\end{eqnarray}
The terms in the parenthesis in the last row are positive definite.  Also, the term on the second row  is always positive since
 \begin{eqnarray} \nonumber
 \frac{1}{x_{0}}\left( \frac{1}{|x_{0}-d_{2}|}-  \frac{1}{|x_{0}+d_{1}|} 
 \right)  ~~~~~~~~~~~~~~~~~~~~~~~~~~~~~~~~~~~~~~~\\ \nonumber 
~~~~~~~~~~ =   \begin{cases}
    \frac{-(d_{1}+d_{2})}{x_{0}(x_{0}-d_{2})(x_{0}+d_{1})},&   L_{3}:~~ x_{0} <  -d_{1}\\
    -\frac{2x_{0}+d_{1}-d_{2}}{x_{0}(x_{0}-d_{2})(x_{0}+d_{1})},&   L_{1}:~~  -d_{1} < x_{0} <  d_{2}\\
    \frac{(d_{1}+d_{2})}{x_{0}(x_{0}-d_{2})(x_{0}+d_{1})},&   L_{2}:~~ x_{0} > d_{2}.\\
\end{cases} 
 \end{eqnarray}
 The positiveness of the first and the third quantities is trivial; however, to see that the second one is also positive we note that within the interval of $-d_{1} < x_{0} <  d_{2}$, the inequalities of $2x_{0}+d_{1}-d_{2}+d_{2}>x_{0}$ and $d_{2}>x_{0}$ always hold; thus, $2x_{0}+d_{1}-d_{2}>0$ must be correct in this interval.  Therefore, we see that the conditions \ref{condition1} and \ref{condition2} are in contradiction. For this reason we realize that $\lambda$ could not be purely imaginary; hence, $\lambda^{2}$ must be positive. From here, we conclude that $\lambda$ has at least a positive root. So, one of the solutions will grow exponentially and the orbit becomes unstable near $L_{1}$, $L_{2}$ and $L_{3}$.
\vspace{10pt}

\textbf{Stability of orbits near $L_{4}$  and $L_{5}$:}
Regarding Eqs. \ref{coords3} and \ref{L4point}, for these two Lagrange points we would have
\begin{eqnarray} \label{L4position}
\rho_{1}&=&\rho_{2} ~~=~~ d_{1}+d_{2}  \\    \nonumber
x_{0}&=& \frac{ -d_{1}+d_{2}}{2}  \\    \nonumber
y_{0} &=& \pm  \frac{\sqrt{3}}{2}\left(  d_{1}+d_{2}\right)
\end{eqnarray}
from which we could derive the second-order derivatives of the potential at the point $(x_0, y_0)$  as follows:
\begin{eqnarray}
U_{xy} &=&- \sqrt{3}\alpha^{2}\left( 3+\epsilon \right) \left( \frac{m_{1}-m_{2}}{m_{1}+m_{2}} \right)   \\  \nonumber
U_{yy} &=&-3\alpha^{2} \left( 3+\epsilon \right)  \\  \nonumber
U_{xx} &=& -\alpha^{2} \left( 3+\epsilon \right) \\  \nonumber
\alpha^{2}&=&\frac{GM}{4(d_{1}+d_{2})^{3}}
\end{eqnarray}
Putting these values into the perturbation equations \ref{pert} and writing them in the matrix form
\begin{eqnarray}
\begin{large}
\begin{pmatrix} 
\lambda^{2}+ U_{xx} & -2\omega \lambda +U_{xy} \\ 2\omega \lambda +U_{xy} & \lambda^{2} + U_{yy}
\end{pmatrix}
\end{large}
 \begin{large}
\begin{pmatrix} 
\xi_{0}  \\ \eta_{0}
\end{pmatrix}
\end{large} =
 \begin{large}
\begin{pmatrix} 
0 \\ 0
\end{pmatrix}
\end{large}
\end{eqnarray}
one could find the next eigenvalue equation
\begin{eqnarray}\nonumber
0 &=& \left( \lambda^{2}+ U_{xx} \right) \left( \lambda^{2} + U_{yy} \right) - \left(  -2\omega \lambda +U_{xy}\right)\left( 2\omega \lambda +U_{xy}  \right) \\  \nonumber
~ &=& \lambda^{4}+ \left(  U_{xx}+U_{yy}+4\omega^{2} \right)\lambda^{2} +U_{xx}U_{yy}-U^{2}_{xy}  \\  \nonumber
&=& \lambda^{4}+ 4\alpha^{2}\left(1+3\epsilon \right)\lambda^{2} +4\alpha^{4}\left(1+3\epsilon \right)^{2}R^{2}  \\  \nonumber
\end{eqnarray}
in which 
$$R^{2} \equiv \frac{3(3+\epsilon)^{2}}{\left(1+3\epsilon \right)^{2}} \times \frac{m_{1}m_{2}}{(m_{1}+m_{2})^{2}}$$
is a dimensionless quantity. Be aware of a mistype in the corresponding  eigenvalue equation of \citet{desloge1982classical}, i.e. equation (75) on page 642 of the book. The maximum value of the parameter $3(3+\epsilon)^{2}/(1+3\epsilon )^{2}$ in the above equation is equal to $27$ and it  happens when $\epsilon =0$, i.e. Newtonian model, while the minimum value is equal to $1/3$ and it occurs for large values of $\epsilon$, i.e. when the fifth force of modified dynamics completely dominates the system. 

\begin{figure}[htbp]
\centering
\includegraphics[width=9cm]{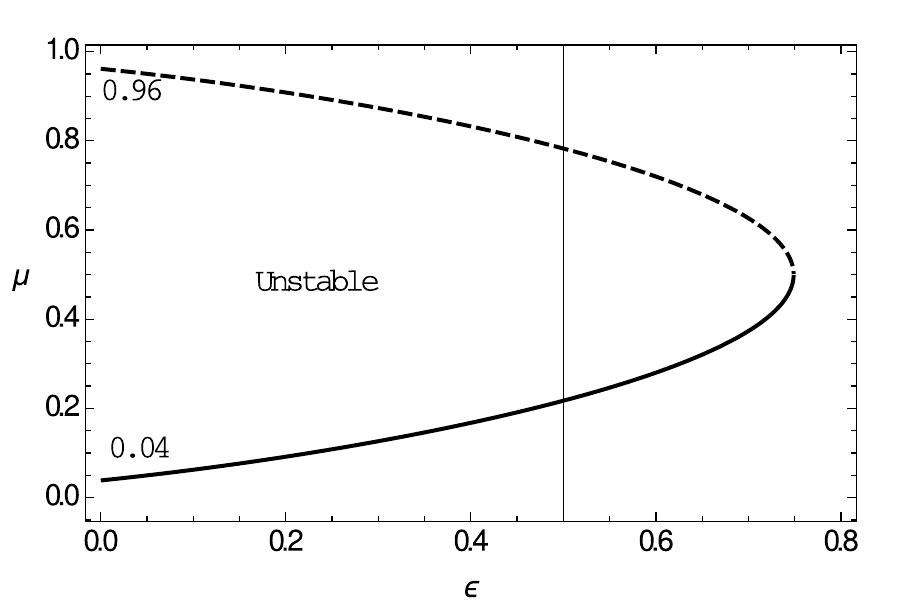}    \\
  \caption{The boundaries of stable and unstable systems in $L_{4} $ and $L_{5}$ Lagrange points, defined by $\mu$, for different values of $\epsilon$.  Systems within the bullet shape curve are unstable.  When $\epsilon$ increases, i.e. when the term proportional to $a_{0}$ begins to dominate the equation of motion, systems with larger mass ratios could  be stable in their $L_{4} $ and $L_{5}$ Lagrange points even when  $\mu >  0.04$. For example, in the case of $\epsilon \approx 0.5$, i.e. the vertical line in the above plot, systems with  a mass ratio of $\mu \lesssim 0.2 $ could still be stable.  }
  \label{fig:Stable}
\end{figure}

The solutions to the above eigenvalue equation could be simply found as:   
\begin{equation} \label{lambda2}
    \lambda^{2}=2\alpha^{2}\left(1+3\epsilon \right) \left(  -1\pm \sqrt{1-R^{2}}\right).
\end{equation}
According to this equation, the system is stable if $\lambda$ is pure imaginary, i.e. if and only if $R^{2} < 1$:
\begin{eqnarray} \label{stablecond}
R^{2} < 1~~~\Rightarrow~~~\mu < \frac{1}{2} - \frac{1}{2} \sqrt{1-\frac{4\left(1+3\epsilon \right)^{2}}{3(3+\epsilon)^{2}}}  ~~~~~~\text{or,} \\ \nonumber 
 \mu > \frac{1}{2} + \frac{1}{2} \sqrt{1-\frac{4\left(1+3\epsilon \right)^{2}}{3(3+\epsilon)^{2}}}~~~~~~~~~  
\end{eqnarray}
These two  inequalities are similar from physical point of view because it does not matter which mass defines $\mu$. A system in its $L_{4}$ and  $L_{5}$ points would be stable if the  inequality \ref{stablecond} is satisfied. For the case of Newtonian gravity, i.e. $\epsilon =0$, we say that if $\mu \lesssim 0.04 $ then the system would be stable. We have plotted the bounds on $\mu$ in Fig. \ref{fig:Stable} as a function of $\epsilon$. It is clear from this plot that the region of stable systems grows with increasing  $\epsilon$ ( The area within the bullet shape contains all systems with unstable $L_{4}$ and  $L_{5}$ points. Also, the Newtonian case is represented by the vertical axis, i.e. $\epsilon =0$ . ). For example, systems with  a mass ratio of $ \mu  \lesssim  0.2 $ could still be stable if $\epsilon \approx 0.5$. See the vertical line corresponding to $\epsilon = 0.5$ in Fig. \ref{fig:Stable}. In fact, according to Fig. \ref{fig:Stable}, systems with $\epsilon > 0.75$ are completely  stable in their $L_{4}$ and  $L_{5}$ points, i.e, $R^{2}$ is always less than 1 in these cases. Large values of $\epsilon$ happen only at galactic and extragalactic scales as we will see in following. To summarize,   $L_{4}$ and  $L_{5}$ points with $(\mu, \epsilon)$ outside of the bullet shape of Fig. \ref{fig:Stable} are stable according to MOD.

We have reported the values of $\epsilon$ and $\mu$ for various two-body configurations within the solar system in Tab. \ref{table:sampleS}. The highest values of $\epsilon$ in the solar system could be found for Sun-Neptune  and Sun-Plutoids and it is about $\epsilon \approx 10^{-5}$ ( Plutoids listed here: Pluto, Eris and Haumea ).  The case of Sun-Eris is the highest in the solar system. These values show that the dynamics of the three-body systems within solar system should be mainly governed by Newtonian model, i.e. little room for a vivid sign of this type of extended gravity in the solar system.  In addition, with the exception of Pluto-Charon, all values of $\mu$ are well below the critical value of $0.04$ for the stability of $L_{4}$ and $L_{5}$ point. This makes the stability of the orbit of any third object within  $L_{4}$ point of Pluto-Charon an interesting question; however, the parameter $\epsilon$ in this system is of the order of $10^{-8}$ which again points to the fact that the system is governed by Newtonian  dynamics ( of course, in short periods of evolution of the system ).

\begin{table*}
\centering
\begin{small}   
\caption{Two-body configurations in the solar system and two Kepler systems. In the case of solar system objects, the masses ( $m_{1} > m_{2}$ ), average orbital distance ( $d_{1} +d_{2}$ ), orbital period $T$, average velocity $V$, $\mu = m_{2}/M$ and $\epsilon = 2c_{1}(a_{0}/G)/(M/(d_{1}+d_{2})^{2})$ are derived from MATHEMATICA's PlanetData, MinorPlanetData and  PlanetaryMoonData. Also, Kepler 444 and Kepler 47 data are derived from \citet{2016ApJ...817...80D} and \citet{2012Sci...337.1511O} respectively. It appears that $\mu$ and $\epsilon$ are very low for almost all configurations in the solar systems. The parameter $\epsilon$ is the largest ( not too large though, compared to galactic systems ) in the case of Sun - Plutoids, i.e. Pluto, Eris and Haumea, and also Sun-Neptune and Kepler 444. Thus, the effect of modified dynamics is primarily expected in these systems. In addition, $\mu$  is the lowest for Pluto-Charon and Kepler 444. Due to relatively low $\epsilon$ in these configurations, none of the systems are expected to show drastic difference in the stability of their $L_{4}$ and $L_{5}$ compared to Newtonian gravity. However, the case of Kepler 444 is the most promising  due to relatively large $\mu$ and $\epsilon$ parameter. } 
\centering  
  \begin{tabular}{ l c l c c c c c }   
    \hline 
System & $m_{1}$ &  $m_{2}$ & $d_{1}+d_{2}$ & $T$ & $V$ & $\mu$ & $\epsilon$    \\
     & $(kg)$ & $(kg)$ & $(AU)$ & $(year)$ & $(km/s)$ &      \\
\hline 
\text{Sun} - \text{Mercury}  & $1.989\times 10^{30}$ & $3.301\times 10^{23}$ & 0.395 & 0.241 & 47.4 & $1.66\times 10^{-7}$ & $2.26\times
   10^{-9}$ \\
 \text{Sun} - \text{Venus} & $1.989\times 10^{30}$ & $4.867\times 10^{24} $& 0.723 & 0.615 & 35.0 & $ 2.45\times 10^{-6}$ & $7.56\times
   10^{-9}$ \\
  \text{Sun} - \text{Earth} & $1.989\times 10^{30}$ & $5.970\times 10^{24} $ & 1.000 & 1.000 & 29.8 & $3.00\times 10^{-6}$ & $1.45\times 10^{-8}$ \\
 \text{Sun} - \text{Mars} & $1.989\times 10^{30}$ & $6.417\times 10^{23} $ & 1.530 & 1.881 & 24.1 & $ 3.23\times 10^{-7} $ & $ 3.38\times
   10^{-8} $ \\
  \text{Sun} - \text{Jupiter} & $1.989\times 10^{30}$ & $ 1.898\times 10^{27}$ & 5.209 & 11.863 & 13.0 & $ 9.53\times 10^{-4}$ & $3.92\times
   10^{-7}$ \\
 \text{Sun} - \text{Saturn} & $1.989\times 10^{30}$ & $5.683\times 10^{26}$ & 9.551 & 29.447 & 9.64 & $2.86\times 10^{-4}$ & $1.32\times
   10^{-6}$ \\
 \text{Sun} - \text{Uranus} & $1.989\times 10^{30} $ & $8.681\times 10^{25} $ & 19.213 & 84.017 & 6.80 & $ 4.36\times 10^{-5} $ & $5.33\times
   10^{-6}$ \\
 \text{Sun} - \text{Neptune} & $ 1.989\times 10^{30}$ & $1.024\times 10^{26} $ & 30.070 & 164.791 & 5.43 & $ 5.15\times 10^{-5}$ &
   $ 1.31\times 10^{-5}$ 
   \\
   \\
 \text{Sun} -   \text{Pluto} & $1.989\times 10^{30}$ & $1.309\times 10^{22}$ & 40.704 & 247.921 & 4.67 & $6.58\times 10^{-9}$ & $2.39\times
   10^{-5}$ \\
 \text{Sun} - \text{Eris} & $1.989\times 10^{30}$ & $1.670\times 10^{22}$ & 74.30 & 556.9 & 3.44 & $8.4\times 10^{-9}$ & $7.98\times 10^{-5}$ \\
 \text{Sun} - \text{Haumea} & $1.989\times 10^{30} $& $4.006\times 10^{21}$ & 44.085 & 285.00 & 4.49 & $2.01\times 10^{-9}$ &$ 2.81\times 10^{-5}$
\\
\\
\text{Neptune} - \text{Proteus} & $1.0243\times 10^{26}$ & $5.03\times 10^{19}$ & $7.864\times 10^{-4}$  & $3.07\times 10^{-3}$ & 7.62 & $4.91\times 10^{-7}$
   & $1.74\times 10^{-10}$ \\
 \text{Neptune} - \text{Triton} & $1.0243\times 10^{26}$ & $2.14\times 10^{22}$ & $2.372\times 10^{-3}$ & $1.61\times 10^{-2}$ & 4.39 & $2.09\times 10^{-4}$
   & $1.58\times 10^{-9}$ \\
 \text{Neptune} - \text{Nereid} & $1.0243\times 10^{26}$ & $3.09\times 10^{19}$ & $4.725\times 10^{-2}$  & $9.86\times 10^{-1}$ & 0.934 & $3.01\times 10^{-7}$ &
   $6.27\times 10^{-7}$ \\
\\
 \text{Pluto} - \text{Nix}  & $1.309\times 10^{22}$ & $4.7\times 10^{16}$ & $3.25\times 10^{-4}$ & $6.81\times 10^{-2}$ & 0.134 & $3.6\times 10^{-6}$ &
   $2.32\times 10^{-7}$ \\
 \text{Pluto} - \text{Hydra}   & $1.309\times 10^{22}$ & $8.0\times 10^{16}$ & $4.33\times 10^{-4}$ & $1.05\times 10^{-1}$ & 0.116 & $6.4\times 10^{-6}$
   & $ 4.12\times 10^{-7} $ \\
    \text{Pluto} - \text{Charon} & $ 1.309\times 10^{22} $ & $ 1.6\times 10^{21}$ & $ 1.31\times 10^{-4}$ & $ 1.75\times 10^{-2} $ & 0.211 &$ 1.10\times
   10^{-1} $ & $ 3.35\times 10^{-8}$ \\
   \\
   \text{Kepler 444 (A- BC)} & $1.498 \times 10^{30}$ & $1.07 \times 10^{30}$ & 36.7  &  198  & ? & $4.17 \times 10^{-1}$ & $1.50 \times 10^{-5}$
   \\
   \text{Kepler 47 (A-B)} & $2.074 \times 10^{30}$ & $7.20 \times 10^{29}$ & 0.084  &  $2.04 \times 10^{-2}$  & ? & $2.58 \times 10^{-1}$ & $7.25 \times 10^{-11}$
\\
\hline 
\end{tabular}
\label{table:sampleS} 
\end{small}
\end{table*}

Although, \cite{2020Ap&SS.365..154P} has argued that  MOND could be tested in Sun-Jupiter and Sun-Neptune saddle regions, for which the net gravitational field is smaller than $10^{-10} ms^{-2}$. This could be a crucial test for MOND because, tidal stresses are expected to diverge in this model  at saddle points while they are finite in Newtonian model \citep{2006PhRvD..73j3513B}. Tidal stresses also remain finite in MOD, as one might check in a straightforward manner. Thus, there is the possibility of distinguishing between various extended gravities at these saddle points; though,  careful experimentation must be prepared to achieve this goal.

In addition, the long-range nature of the fifth force of MOD most probably affects the long-term stability of the solar system ( in spite of being a small term ).  The reason for this expectation is that the fifth force reduces the accessible states at large distances as we will explain below. The question of the stability of the solar system in extended gravities is an interesting question and needs to be addressed carefully in due time.

Fortunately, we have now growing data on other planetary systems. These systems could also be considered as a laboratory in testing extended gravities. For example,  Kepler 444, which is  a triple star system, would also be a good candidate in investigating RTBP in MOD. As Tab. \ref{table:sampleS} show, with $\mu = 0.417 $ and $\epsilon = 1.50 \times 10^{-5}$ ( Kepler 444 A as the most massive body and the combination of Kepler 444 BC as the second body ), this is the best stellar system that author could find to test RTBP in MOD. In fact, although the value of $\epsilon$ is still not quite substantial, the large value of $\mu$ and non-negligible value of $\epsilon$ would provide a chance to compare the stability of the system under MOD and Newtonian laws.  In the case of Kepler 47, however, the value of $\epsilon$ is quite small and one has to run the system for much longer times, compared to Kepler 444 system, to see a clear difference between Newton and MOD models.

Since the Newtonian force and the fifth force are quite comparable at galactic and extragalactic scales, the machinery of RTBP is supposed to produce vivid effects  at these scales. To see this point, we have reported the values of $\epsilon$ and $\mu$ for some nearby galactic systems in Tab. \ref{table:sampleG}. It could be seen from this table that in the case of Milky Way and a typical globular cluster (MW-TGC), or Milky Way and a typical open cluster (MW-TOC), the value of $\epsilon$ is quite large. This shows that the fifth force is expected to have a major effect on the evolution of the globular clusters around the Milky Way. However, in these cases the values of $\mu$ are  small and even negligible; thus one expects that RTBP would be dominated by the Newtonian force here. On the other hand, in the case of Milky Way and Large Magellanic Cloud ( MW-LMC ), or Milky Way and Small Magellanic Cloud (MW-SMC), the estimated values of $\mu$ and $\epsilon$ are quite substantial. Because the value of $\epsilon$ is quite large in these cases, it is expected that the orbits of LMC and SMC are both dominated by the fifth force, instead of the Newtonian force. Also, since the value of $\mu$ is considerable too in the cases of MW-LMC and MW-SMC, the orbits of any star around these objects would be a perfect case for investigating RTBP in MOD (  see Fig. \ref{fig:Stable} ). Although, it should be noted that a similar situation would happen for LMC-SMC system, making the problem a four-body problem indeed.

The M51a galaxy ( also known as  NGC5194 or Whirlpool galaxy ) and its companion M51b ( or NGC5194 ) are usually considered as a classic example of interacting galaxies \citep{1972ApJ...178..623T}. Here we see that $\mu$ is about 0.3 and if the two galaxies become really close, e.g. a distance of about $20~kpc$, then the Newtonian and the fifth force are of the same order. However, when the distance between M51a and M51b becomes larger, say around $100~kpc$, the share of the Newtonian force decrease drastically in comparison to the fifth force. Thus, within MOD model the companion galaxy could still affect Whirlpool even from large distances while the Newtonian force is now negligible at $ \approx 100~kpc$. The differnce between the two models could be checked by modifying the code of \cite{1972ApJ...178..623T} to include the MOD's fifth force. This is a relatively straightforward task as we report the code, and its results, in future.

\begin{table*}
\centering
\begin{small}   
\caption{Two-body configurations in galactic scales. Data related to the Milky Way (MW), typical globular cluster (TGC) and typical open cluster (TOC) are derived from \citet{2008gady.book.....B}, chapter one. In addition, properties of Large Magellanic Cloud ( LMC ) and Small Magellanic Cloud (SMC) are derived from \citet{2000glg..book.....V}. Moreover, the masses of M51a ( Whirlpool or NGC5194 ) and  M51b (NGC5195) are reported from \citet{2012ApJ...755..165M}. } 
\centering  
  \begin{tabular}{ l c l c c c }   
    \hline 
System & $m_{1}$ &  $m_{2}$ & $d_{1}+d_{2}$ &  $\mu$ & $\epsilon$    \\
     & $(M_{\odot})$ & $(M_{\odot})$ & $(kpc)$ &       \\
\hline 
   \text{MW-TGC} & $\sim  5 \times 10^{10}$ & $\sim  2 \times 10^{5}$ & $\sim 5$  &    $4.0 \times 10^{-6}$ & $\sim 0.3$
   \\
   \text{MW-TOC} & $\sim  5 \times 10^{10}$ & $\sim 3 \times 10^{2}$ & $\sim 5$ &   $6.0 \times 10^{-9}$ & $\sim 0.3$
   \\
   \\
   \text{MW-LMC} & $\sim  5 \times 10^{10}$ & $\sim 1 \times 10^{10}$ & $\sim 50$ &   $1.6 \times 10^{-1}$ & $\sim 25$
   \\
   \text{MW-SMC} & $\sim  5 \times 10^{10}$ & $\sim 7 \times 10^{9}$ & $\sim 61$   & $1.2 \times 10^{-1}$ & $\sim 45$
   \\
   \text{LMC-SMC} &$\sim 1 \times 10^{10}$  & $\sim 7 \times 10^{9}$ & $\sim 23$   & $4.1 \times 10^{-1}$ & $\sim 32$
   \\
   \\
   \text{M 51 (a-b)} & $5.8 \times 10^{10}$ & $2.5 \times 10^{10}$ & 20   & $3.0 \times 10^{-1}$ & $2.9$
   \\
      \text{M 51 (a-b)} & $5.8 \times 10^{10}$ & $2.5 \times 10^{10}$ & 100 &    $3.0 \times 10^{-1}$ & $73.4$
\\
\hline 
\end{tabular}
\label{table:sampleG} 
\end{small}
\end{table*}

 As  \cite{2000ApJ...532..922W} has explained, in the case of the effects of the Milky Way on Large Magellanic cloud structure, the   interaction between these two systems could have significant observational consequences. For example, orbits of the LMC disk  are reported to be "torqued out of the disk plane, thickening the disk and populating a spheroid". Furthermore, the problem of  the structural evolution of the companions could be studied using analytic models as well as n-body simulations. The  three-body analysis of such systems is of course a first step in understanding the observations. This type of analysis clarifies the general behavior of the test particle, i.e. stars; thus, paving the way for simple observational comparisons and understanding the results of  future n-body simulations.  Comparing  accurate n-body simulations and observational data could then be used to test models of extended gravities since these models show explicit dissimilarities in their n-body behavior. However, considering the current level of accuracy in observational data, the   methods  ( analytical vs. n-body ) and the physical assumptions ( dark matter vs. extended gravities ) that we choose  to study the problem, could make the interpretation of the final results quite debatable. 

\section{Simple PRTBP Laboratory}
\label{Lab}
This section provides a qualitative view toward the nature of orbits in RTBP within MOD model. A rigorous analysis based on a geometric framework \citep{2022arXiv220316019F}, is also possible; though, it is beyond the scopes of the current study. Here, we provide a MATHEMATICA code to test the behavior of a test particle in the case of restricted three-body problem.   This code compares Newtonian and MOD models and works as follows. We will use the units in Eqs. \ref{units}. In addition,  the particles $m_{1}$ and $m_{2}$ are moving on circles with angular frequency $\omega$ ( see Eq. \ref{eqs} ) and the motion of the third particle does not affect $m_{1}$ and $m_{2}$. You may put $c_{1}=0$ in Eq. \ref{eqs} to derive the Newtonian angular frequency of the motion.   Moreover, both $m_{1}$ and $m_{2}$ start their motion on the $x$ axis while the initial position of particle $m_{3}$ is chosen around the Lagrange point $L_{4}$, given by Eq. \ref{L4position}, and it is perturbed by a random function times some scale length. This choice of the initial position for $m_{3}$ is based on Eqs. \ref{perturb}, i.e. $(x_{0},~y_{0})$ being the coordinate of $L_{4}$ while $(\xi,~\eta)$ is the displacement perturbation. Also, the particles $m_{1}$ and $m_{2}$ rotate counterclockwise while the test particle could rotate in both directions ( This is done by changing a phase factor in the code. ). Finally, the initial angular  speed of $m_{3}$ is chosen  randomly between zero and twice the real part of $\lambda$ in Eq. \ref{lambda2} ( For the Newtonian case, one should put $\epsilon =0$ in Eq. \ref{lambda2}. ).

The set of equations of motion with the above mentioned initial conditions, is solved numerically using  NDSolve in MATHEMATICA 10.0. Because the scaling of the three-body problem  is broken in MOD, the parameter space of the problem is much larger compared to the Newtonian case. However, according to the above discussions, and specially Tables \ref{table:sampleS} and \ref{table:sampleG}, we would expect to see vivid difference between the two models in galactic scales. Thus, we presume $m_{1}=10~ \mathcal{M}$, $m_{2}=7~ \mathcal{M}$, $d=20~ \mathcal{D}$ and a simulation  time of $3000~\mathcal{T}$.

\begin{figure*}
\centering
\includegraphics[width=17.5cm]{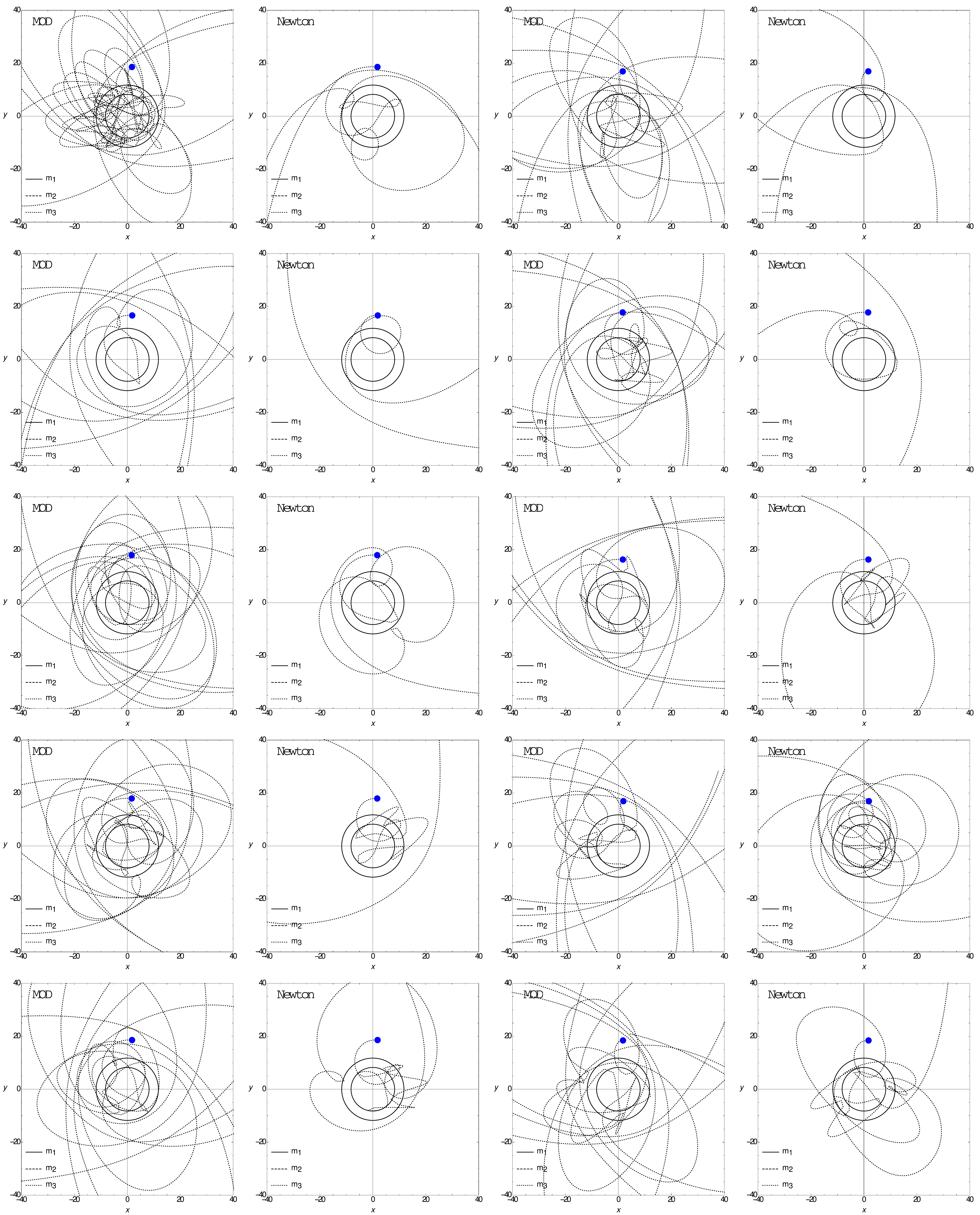}
\\
  \caption{The evolution of a test particle around two massive objects of $m_{1}=10~ \mathcal{M}$ and $m_{2}=7~ \mathcal{M}$ with the distance of $d=20~ \mathcal{D}$. The evolution time is $3000~\mathcal{T}$, the initial position of $m_{3}$ is chosen randomly to be within $(x_{0}\pm 0.1x_{0},~y_{0} \pm 0.1y_{0})$ and this particle rotates initially in the same direction of $m_{1} $ and $m_{2}$.  The ejection in Newtonian models is more common than the MOD models.}
  \label{fig:mu04pl01}
\end{figure*}

\begin{figure*}
\centering
\includegraphics[width=17.5cm]{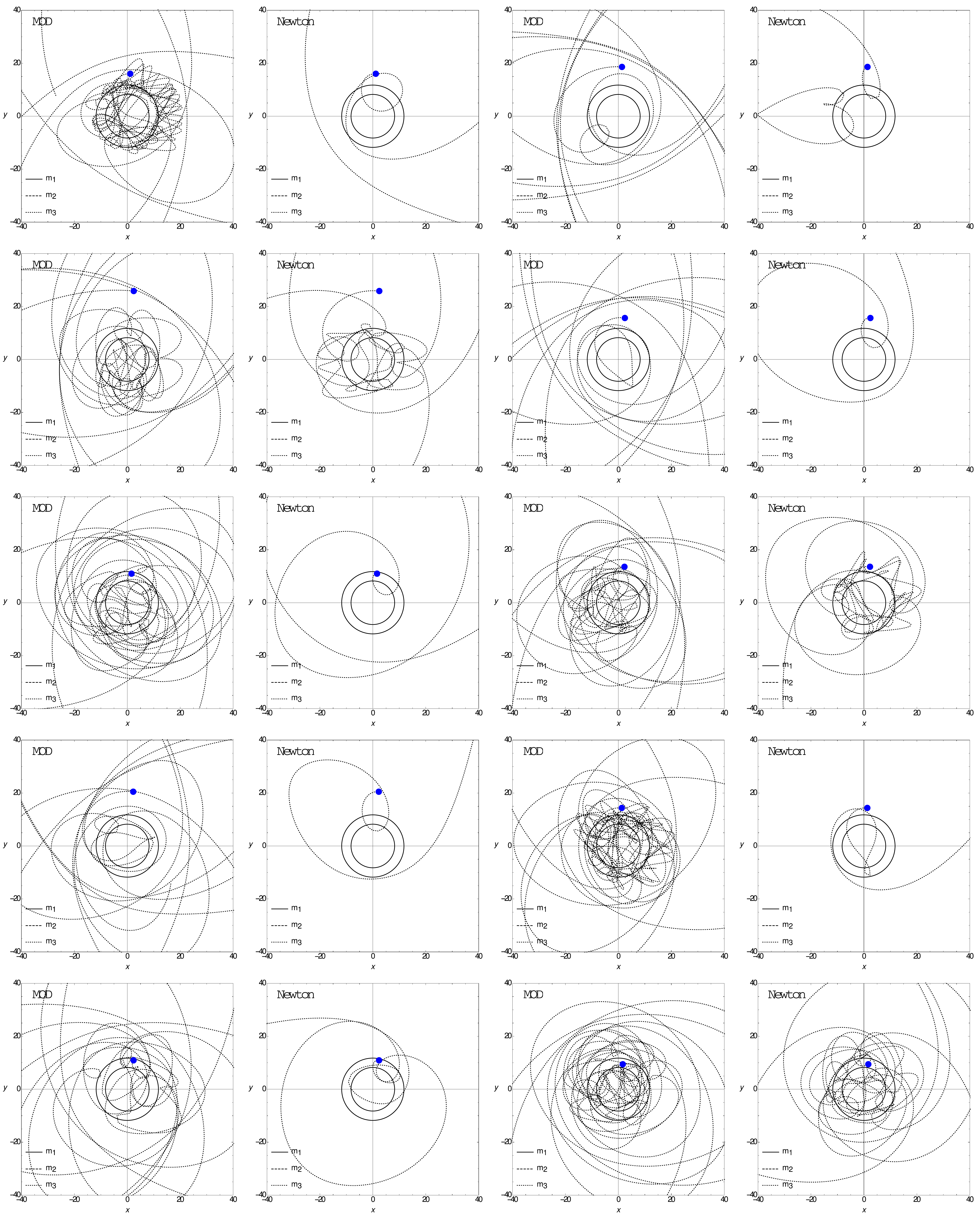}
\\
  \caption{The evolution of a test particle around two massive objects of $m_{1}=10~ \mathcal{M}$ and $m_{2}=7~ \mathcal{M}$ with the distance of $d=20~ \mathcal{D}$. The evolution time is $3000~\mathcal{T}$, the initial position of $m_{3}$ is chosen randomly to be within $(x_{0}\pm 0.5x_{0},~y_{0} \pm 0.5y_{0})$ and this particle rotates initially in the same direction of $m_{1} $ and $m_{2}$.   }
  \label{fig:mu04pl05}
\end{figure*}

\begin{figure*}
\centering
\includegraphics[width=17.5cm]{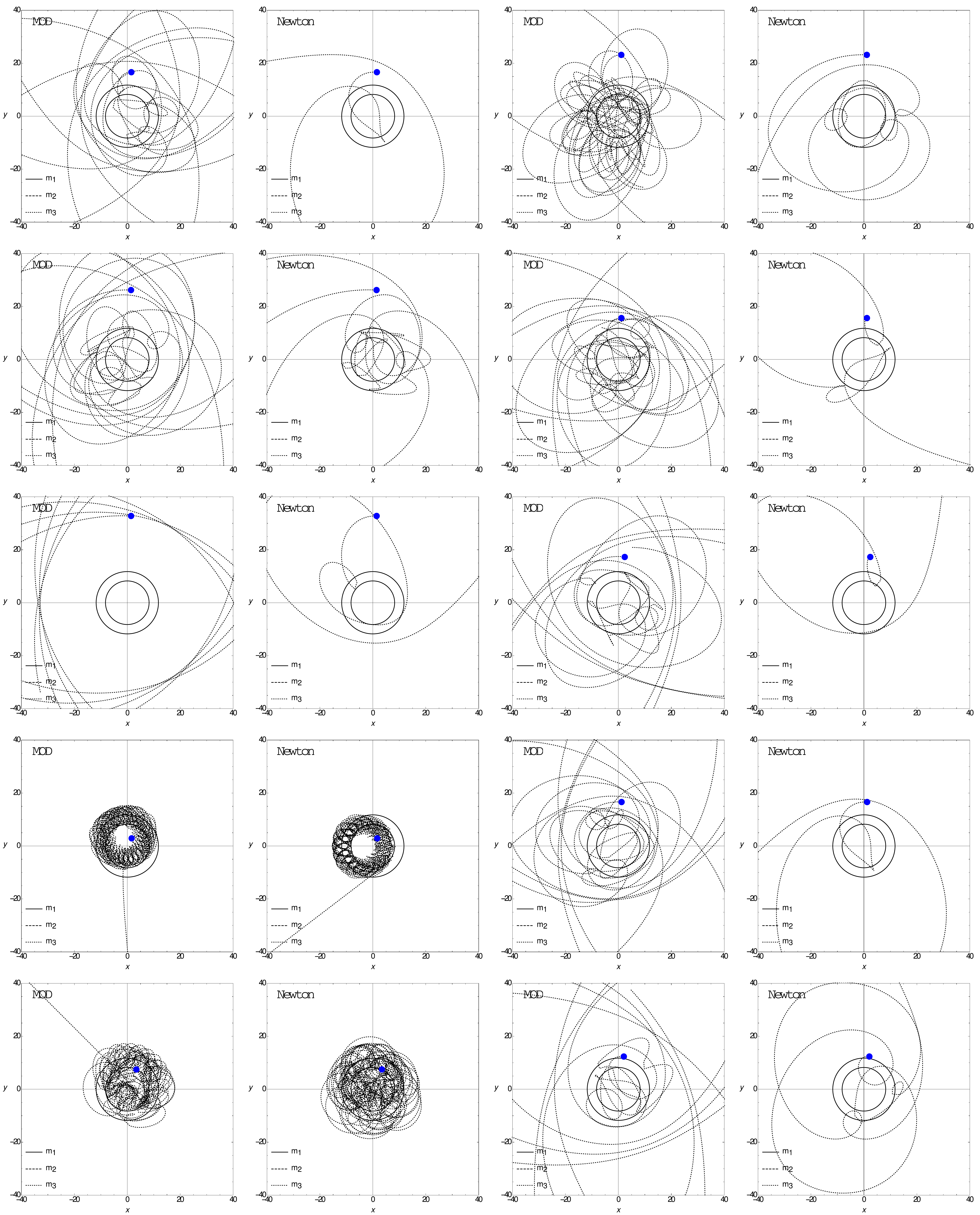}
\\
  \caption{The evolution of a test particle around two massive objects of $m_{1}=10~ \mathcal{M}$ and $m_{2}=7~ \mathcal{M}$ with the distance of $d=20~ \mathcal{D}$. The evolution time is $3000~\mathcal{T}$, the initial position of $m_{3}$ is chosen randomly to be within $(x_{0}\pm x_{0},~y_{0} \pm y_{0})$ and this particle rotates initially in the same direction of $m_{1} $ and $m_{2}$.  }
  \label{fig:mu04pl10}
\end{figure*}

In addition, the initial position of $m_{3}$ is chosen randomly to be within $(x_{0}\pm 0.1x_{0},~y_{0} \pm 0.1y_{0})$.  The results are provided in Fig. \ref{fig:mu04pl01} which includes ten plots of MOD orbits and their Newtonian counterparts. The random initial position in each case is shown by a blue point. The orbits  of $m_{1}$ and $m_{2}$ are shown as two concentric circles while the trajectory of $m_{3}$ is shown by dotted curves ( Using the electronic version of the work, the reader may zoom in to see more details in the plots. ). In this figure, it is seen that most Newtonian orbits diverge after a few revolutions while the MOD orbits are  non-divergent in this case ( Of course, Newtonian orbits could be stable as we see below. ). All initial positions in this figure lie outside of the $m_{2}$ orbit. The trajectory of $m_{2}$  is shown by the circle with the larger radius. Then, in Figs. \ref{fig:mu04pl05} and \ref{fig:mu04pl10} we have expanded the area of random initial position to $(x_{0}\pm 0.5x_{0},~y_{0} \pm 0.5y_{0})$ and $(x_{0}\pm x_{0},~y_{0} \pm y_{0})$ respectively. In these figures too, ejection is more  common in the case of Newtonian model compared to the MOD one. However, the rate of ejection in Newtonian case is less than the Newtonian plots of the previous figure. Also, it is seen that if the initial positions of $m_{3}$ is within the  circle of large radius, the orbits seem to be  more stable ( less ejections in both Newtonian and MOD models ). In one particular case, i.e. the last row in Fig. \ref{fig:mu04pl10}, MOD presents an ejection while the Newtonian orbit, within the time of $3000~\mathcal{T}$, is still bounded. 
 
The test particle  $m_{3}$ and the two massive particles could also rotate in opposite directions. This could be achieved simply by adding a phase factor to the code, as discussed before, to make $m_{3}$ rotate  clockwise initially. The result of such experiments are reported in Figs. \ref{fig:mu04nl01} to  \ref{fig:mu04nl10} in Appendix \ref{supp}. The interesting phenomena here is the appearance of much more regulated orbits in comparison to the cases where all particles rotate counterclockwise. If this is not due to shear chance, it   needs to be investigated in more details. In these figures too, the rate of ejections in Newtonian model is vividly larger than MOD.

Also, we can decrease, or increase the mass ratio and see how the model of RTBP works. In the rest of the figures in the Appendix \ref{supp}, we have used $m_{2}=3~ \mathcal{M}$ ( Figs. \ref{fig:mu02pl01} to \ref{fig:mu02nl10}  ) and $m_{2}=1~ \mathcal{M}$ ( Figs. \ref{fig:mu01pl01} to \ref{fig:mu01nl10}  ). By decreasing $\mu$, it is seen that the orbits become more regulated. This is expected since $m_{1}$ starts to dominate the system now. In all of these figures too, the Newtonian orbits show a larger rate of ejections compared to MOD ones. 
 
 A similar behavior of divergent orbits could also be seen in Figs. \ref{fig:Pythagorean} and \ref{fig:Binary}. From dynamical point of view, this behavior of three-body systems under MOD is rooted in the long-range nature of the fifth force. In fact, in galactic and extragalactic scales this force is quite strong and  it could bound the systems more efficiently. In other words, although the particle $m_{3}$ might be ejected from the system due to close encounters, in MOD it would most probably come back eventually  and make restricted orbits (   due to the long range behavior of the fifth force ).

It is illuminating to consider this behavior from the point of view of statistical mechanics too.  It is now well-known that within Newtonian dynamics, actual astrophysical systems, e.g. star clusters, galaxies etc., are "open" since particles in such systems could in principle fly to infinity \citep{1990PhR...188..285P}. In fact, it could be proved that  the density of states  becomes divergent if the integral defining this function extends to infinity. It should be noted that even in the case of an
ideal gas, one could derive such  divergent quantity unless the gas is contained within  a box. Thus, it is usually assumed that the gravitating system under consideration lives inside a spherical box. As \cite{1990PhR...188..285P} argues, this assumption is justified by presuming a small rate of evaporation for gravitating particles. However, MOD's partition function of an NBody system contains a factor of $\exp (-2 \beta c_{1}a_{0}r )$ which saturates the integrals at large radii $r$ naturally. Here, $\beta $ is proportional to the reciprocal of the thermodynamic temperature. On the other hand, the problem of the statistical mechanics of the three-body problem is somehow different because in this case case, the  number of particles is quite small. Although, using the method of \citet{1976MNRAS.176...63M,1976MNRAS.177..583M,1978MNRAS.184..119N}, it is still possible to build a statistical theory of the three-body problem. We will report such theory based on MOD in future and show that because of the long range nature of the fifth force, the integral defining the density of states would be finite; meaning that the states at far distances would be less favorable than the near ones.

\begin{figure*}
\centering
\includegraphics[width=8.5cm]{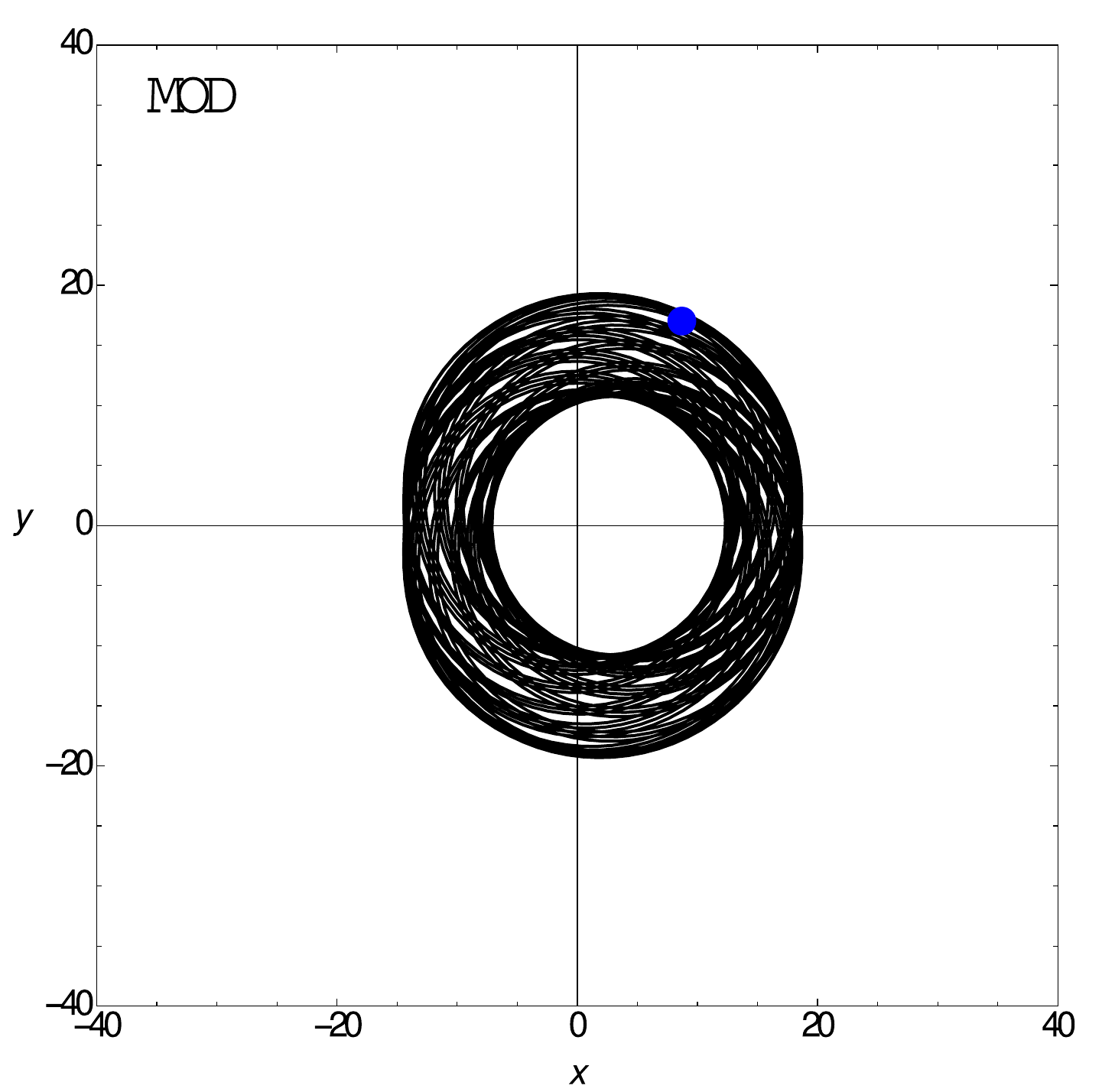}\includegraphics[width=8.5cm]{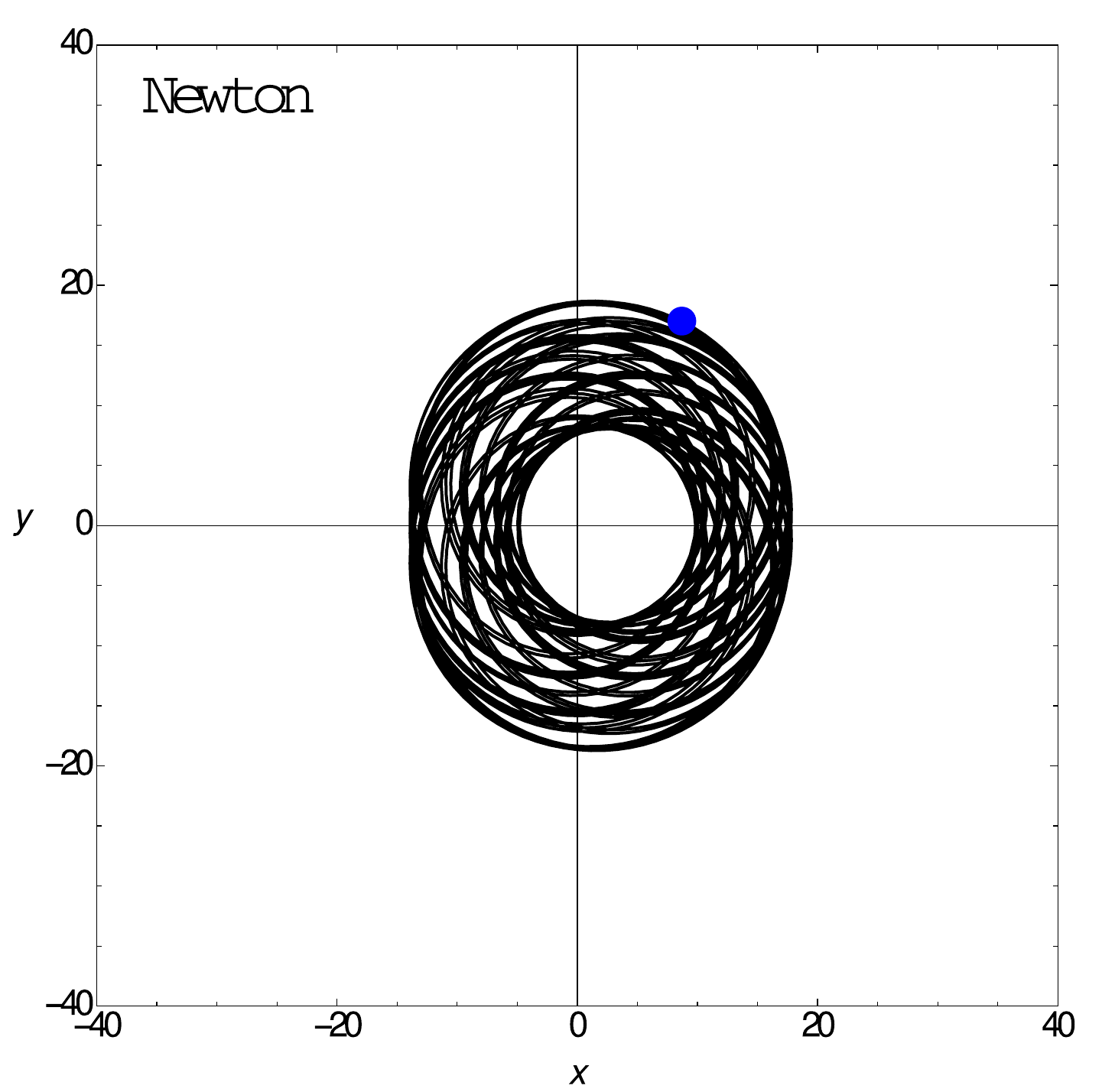}
\\
  \caption{The evolution of a test particle around two massive objects of $m_{1}=10~ \mathcal{M}$ and $m_{2}=7~ \mathcal{M}$ with the distance of $d=20~ \mathcal{D}$ in the rotating frame. The evolution time is $3000~\mathcal{T}$, the initial position of $m_{3}$ is chosen randomly to be within $(x_{0}\pm 0.1x_{0},~y_{0} \pm 0.1y_{0})$ and this particle rotates initially in the same direction of $m_{1} $ and $m_{2}$. The code also provides Jacobi integral and physical energy of each test particle. }
  \label{fig:rotatingframe}
\end{figure*}

 In addition, one may derive the orbits in the rotating frame ( of the two large body ). See Fig. \ref{fig:rotatingframe}. A code which shows orbits in the rotating frame is also available online.     Furthermore, this code presents Jacobi integral and physical energy of each test particle. One could easily see that the Jacobi integral is a constant of the motion ( as it should be ) while the physical energy is not. 

\section{Discussions and Conclusion}
\label{Conclusion}
The main result of this work is that the stability of the three-body systems  depends crucially on the governing dynamics. This has been shown  through analyzing general three-body systems ( Sec. \ref{Review} ), restricted three-body problem ( Sec. \ref{Restricted} ) and some simple PRTBP configurations ( Sec.  \ref{Lab} ), in all of which the difference between Newtonian and MOD models is explicit. The general dependency  might seem obvious at first glance; however, it is important to know exactly  at what scales of length and time  one should expect such effects. The effects of the fifth force are anticipated to be  unambiguous at galactic scales. Although, even at the scales of the solar system one would expect modifications in the long-term stability of the system due to long range nature of the fifth force ( which results in less favorable states at larger distances ).

This study could be   extended in a few directions. For example,  it should be now clear that when a long-range fifth force exists, the rotation curve of any galaxy could be modified due to the presence of   neighboring galaxies.  An analytical approach to this problem, i.e. a test particle in the field of two galaxies, would be challenging; though, it could be also  illuminating. However, at this point, the best strategy to solve this problem seems to be through numerical modeling of the gravitational potential of the neighboring galactic system ( neighboring galaxies could be assumed as point particles ) and then adding the field of the target galaxy to derive the total potential. In this way, one could estimate the effect of the surrounding matter distribution on the rotation curve of a galaxy. 

Another interesting problem arises when we investigate the behavior of an ensemble of particles  under the influence of two major objects, e. g. two galaxies. This is the classic problem of \cite{1972ApJ...178..623T} who first explained the formation of galactic bridges and tails in their classic paper. As it turns out, the method and the code of Toomre $\&$ Toomre could be simply modified to contain the fifth force which is discussed here. The results of the simulations, however, are strikingly different and the main distinctions are as follows: First, the companion could harass the target galaxy even from large distances, i.e. a very close encounter is not necessary for this matter. This is due to the long range nature of the interactions. Second, long-lived filaments could form between the two objects in certain situations, without the  merging of the two objects. We will report these findings in future. As discussed before,   a persistent challenge here would be to model the effect of the neighboring objects on the main ones, especially considering the long range nature of the fifth force.

As Figs. \ref{fig:mu04pl01} onwards show,  test particles  with the same initial position display drastically different orbits under Newtonian and MOD models ( In most cases, the initial velocities are very close too. ). This variety of orbits under different dynamics is good news for investigation of  extended gravities signatures; however, it should be quantified. In fact, the general behavior of the gravitating systems in extended gravities needs to be surveyed through methods of statistical mechanics. This is also true about the statistical behavior of three-body systems; though, here we are dealing with small numbers of particles as \cite{1976MNRAS.176...63M} explains. A successful statistical theory in extended gravities could greatly simplify the analysis of these exotic models and possibly clears the way in finding the governing equation of motion at galactic, and extragalactic, scales. The results of such statistical analysis could then be  compared with catalogues of isolated triplet galaxies in the local universe \citep{2015A&A...578A.110A} to put an upper bound on the constants of the extended gravities or limit their interpolating functions. The anti-gravity nature of $\Lambda$ term in $\Lambda CDM$ model could be tested against the attractive fifth force of MOD in isolated triplets. The main problem here would be  whether  any version of extended gravities could be found to encompass the statistical distribution of  both compact and wide triplet galaxies. See \cite{2016ARep...60..397E} for an investigation of wide triple galaxies in the standard model.

Meanwhile, the theoretical challenges of extended gravities, and the signatures of these theories, are hoped to be gradually understood with further investigations. Especially, in the case of modified dynamics,  one might argue  that \textit{ if there is a connection between local and global physics, the equation of motion might show a dependency to the Hubble parameter and its derivatives, as a result of this connection.} A simple search through literature shows that there are indeed  some reports on this connection; though,  in any occasion there is another explanation too. For example, at the scale of the solar system we might mention an apparent anomalous, weak, long-range acceleration  in Pioneer 10 and Pioneer 11, generally known as Pioneer anomaly which is very close to $\approx cH_{0}$ \citep{1998PhRvL..81.2858A,2002PhRvD..65h2004A,2010LRR....13....4T};  though, there are also reports in favor of thermal origin of the Pioneer anomaly \citep{2011PhRvL.107h1103T,2012PhRvL.108x1101T}.   In addition, at galactic and extragalactic scales, MOND which is based on a fundamental acceleration $ \approx cH_{0}$, has achieved tremendous success in addressing the missing mass problem \citep{2012LRR....15...10F}; though, the community mostly favours CDM paradigm \citep{2022arXiv220104741T}. Even at cosmic scale, there are implications that a term proportional to the Hubble parameter $H$ in Friedmann equations could play the role of dark energy \citep{2011PhRvD..84l3501C,2012PhR...513....1C}. Of course, in this case too there is the cosmological constant $\Lambda$, as well as other forms of dark energy, to explain the accelerated expansion of the universe. These coincidences might disappear with more accurate data in future, or they might be explained through a final successful model of dark matter and dark energy ( The situation is an interesting example of the philosophical question: "How does one distinguish between physical laws and coincidence?", which is also related to the problem of induction and confirmation of physical laws. ). Currently however, these observations open up the possibility of a unification through a new equation of motion; and there have been some interesting theoretical speculations in this regard. For instance,  we might mention an effective theory of gravity by \citet{2010PhRvL.105u1303G,2011PhRvL.106c9901G}, a fourth order conformal gravity by 
\citet{1989ApJ...342..635M,1994GReGr..26..337M,2006PrPNP..56..340M} and a theory of five-dimensional brane world unification of space, time and velocity by \citet{1999astro.ph..7244B,2003NuPhS.124..258C} which all predict a Pioneer-Rindler constant acceleration term. 

Of course, not all extended theories of gravity could be true. In fact, a feasible strategy to determine the viability of extended gravity models is to seek experiments and observations to put upper bounds on the free parameters of the model ( for example, here   $c_{1}$ ) and then pushing this limit to lower values with more accurate experiments and observations ( In the case of MOND, the form of the interpolating function should be tested too, in addition to the free parameter of the model, i.e. $a_{0}$. ). If due to this process the upper bound on the free parameters  converges to a certain value, then the extended gravity under consideration would be considered safe ( at least for now ). However, if  different regimes, experiments and observations report contradictory results on the bound, we may be able to rule out the model. This stage, however, depends crucially on the accuracy of the test which could be best achieved in controlled experiments.  This is why laboratory experiments are the best hope in finding, or rejecting, extended theories of gravity.  However, it is also important to test new models at larger scales, because after all,  these theories are invented to address the discrepancies at galactic and extragalactic scales. Thus, astrophysical data are also an important source in testing the modified models of gravity.  Although, the error bars in these data usually grow at larger distances ( mainly due to uncertainty in the distance itself ) and this problem might restrict our ability to rule out some models of extended gravity based on data at large scales.

Another possibility for an accurate test of MOD model  would be to use binary pulsars. See \cite{2021PhRvX..11d1050K}  who employs these systems and confines   \cite{2004PhRvD..70h3509B} TeVeS theory, which is a relativistic version of MOND, and also \cite{1993PhRvL..70.2220D} theory. However, to use  precise measurements of these objects, one needs to overcome the theoretical challenges ( as discussed in Sec. 2 ) and  rebuild the model for a typical binary system.

\section*{Acknowledgements}
I would like to thank Ramezan Ebrahimi for  checking the equations in Sec. \ref{Restricted}. Also, I appreciate fruitful discussions with members of the astronomy and astrophysics group, Ferdowsi University of Mashhad. I appreciate helpful discussions with Elham Nazari on the nature of the problem of the binary systems. In addition, I benefited from comments and suggestions by the reviewers for which I am very grateful. ( One comment helped me to find an error in Fig. 1 of the earlier version of this manuscript.  )  This research has made use of NASA's Astrophysics Data System.  The MATHEMATICA codes related to orbit integration are provided in my GitHub repository\footnote{https://github.com/Shenavar/ThreeBodyinMOD}.

\bibliographystyle{mnras}
\bibliography{main}

\clearpage

\appendix
\section{Supplementary Figures}
\label{supp}

\begin{figure*}
\centering
\includegraphics[width=17.5cm]{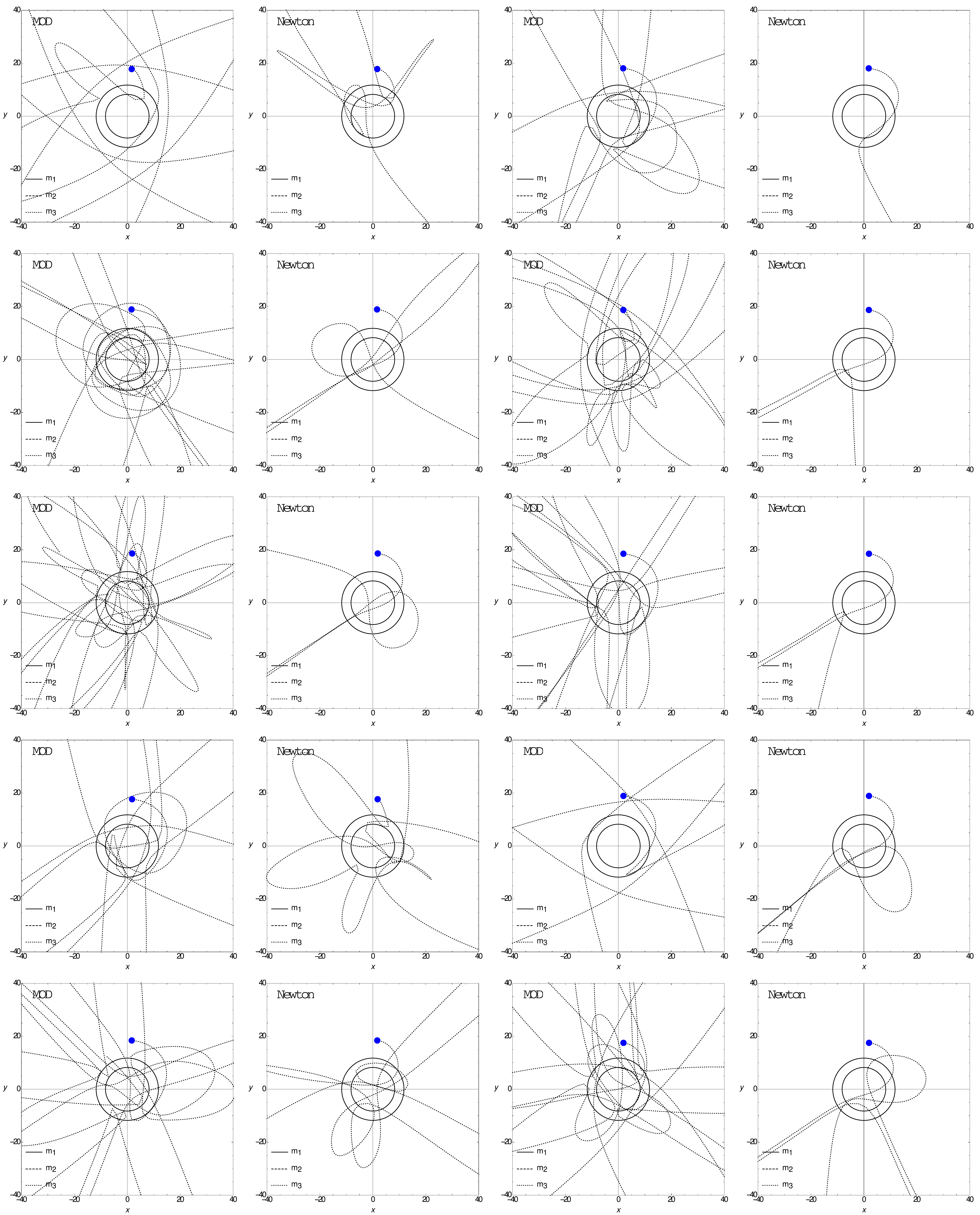}
\\
  \caption{The evolution of a test particle around two massive objects of $m_{1}=10~ \mathcal{M}$ and $m_{2}=7~ \mathcal{M}$ with the distance of $d=20~ \mathcal{D}$. The evolution time is $3000~\mathcal{T}$, the initial position of $m_{3}$ is chosen randomly to be within $(x_{0}\pm 0.1x_{0},~y_{0} \pm 0.1y_{0})$ and this particle rotates initially in the opposite direction of $m_{1} $ and $m_{2}$.  }
  \label{fig:mu04nl01}
\end{figure*}

\begin{figure*}
\centering
\includegraphics[width=17.5cm]{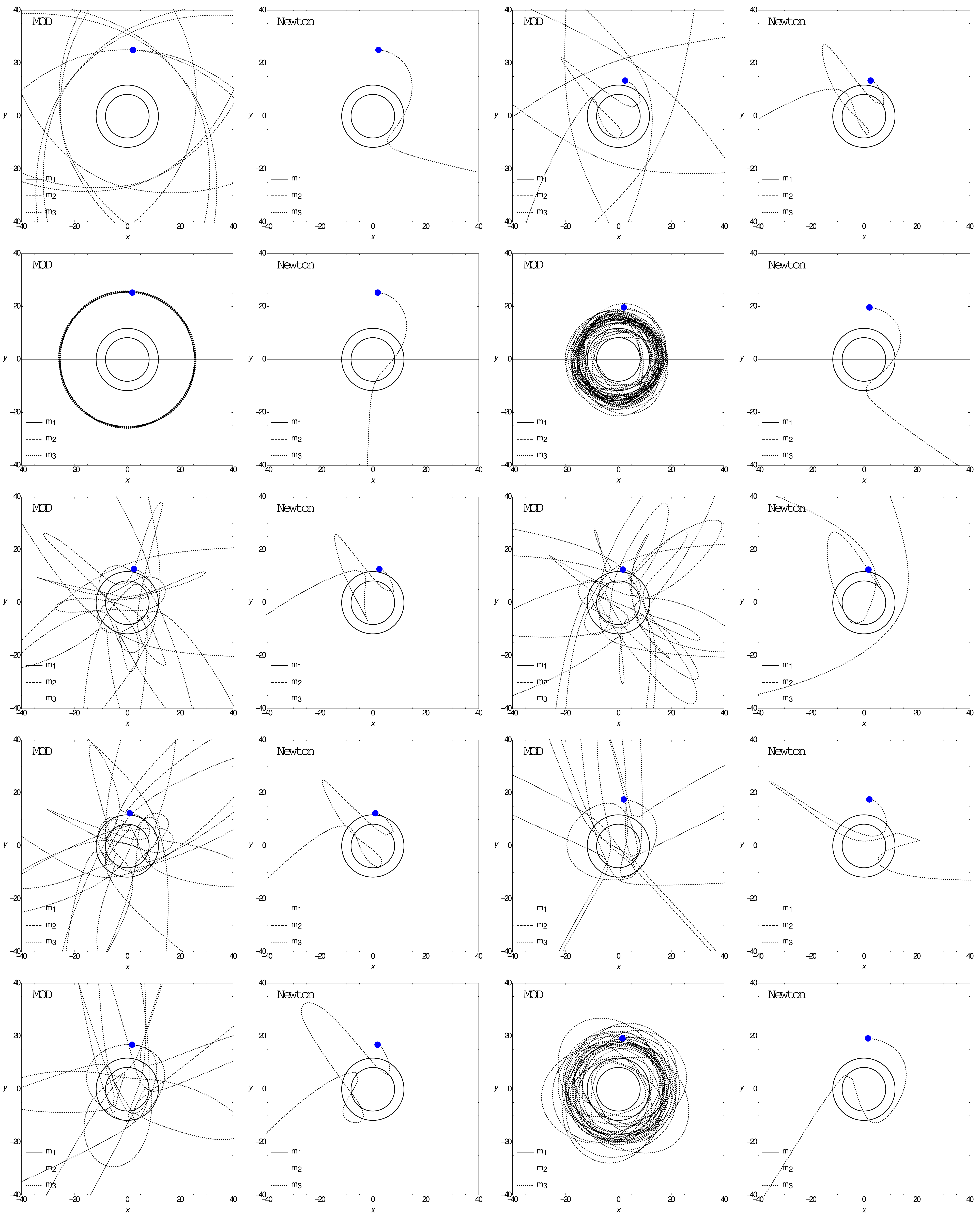}
\\
  \caption{The evolution of a test particle around two massive objects of $m_{1}=10~ \mathcal{M}$ and $m_{2}=7~ \mathcal{M}$ with the distance of $d=20~ \mathcal{D}$. The evolution time is $3000~\mathcal{T}$, the initial position of $m_{3}$ is chosen randomly to be within $(x_{0}\pm 0.5x_{0},~y_{0} \pm 0.5y_{0})$ and this particle rotates initially in the opposite direction of $m_{1} $ and $m_{2}$.  }
  \label{fig:mu04nl05}
\end{figure*}

\begin{figure*}
\centering
\includegraphics[width=17.5cm]{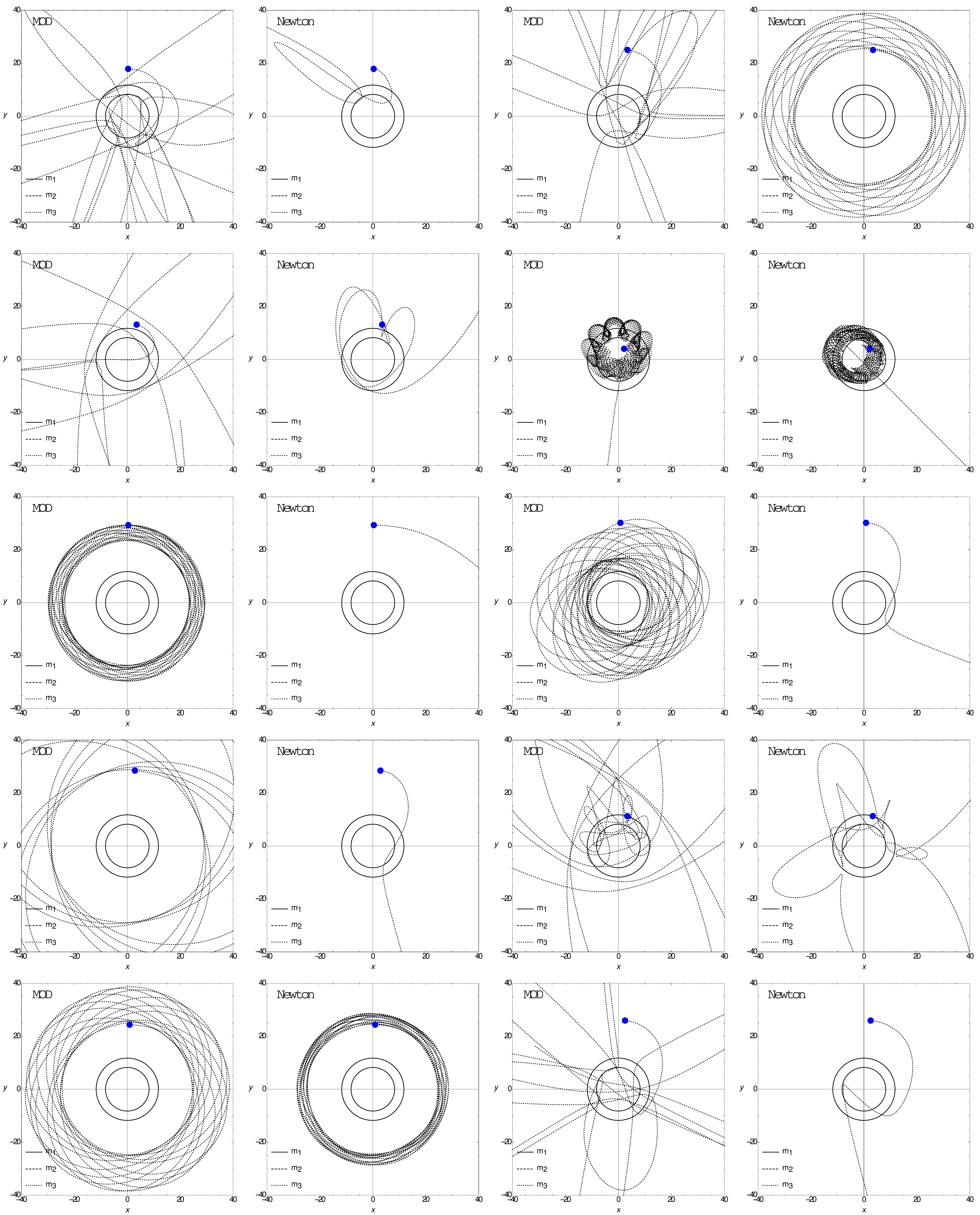}
\\
  \caption{The evolution of a test particle around two massive objects of $m_{1}=10~ \mathcal{M}$ and $m_{2}=7~ \mathcal{M}$ with the distance of $d=20~ \mathcal{D}$. The evolution time is $3000~\mathcal{T}$, the initial position of $m_{3}$ is chosen randomly to be within $(x_{0}\pm x_{0},~y_{0} \pm y_{0})$ and this particle rotates initially in the opposite direction of $m_{1} $ and $m_{2}$.  }
  \label{fig:mu04nl10}
\end{figure*}

\begin{figure*}
\centering
\includegraphics[width=17.5cm]{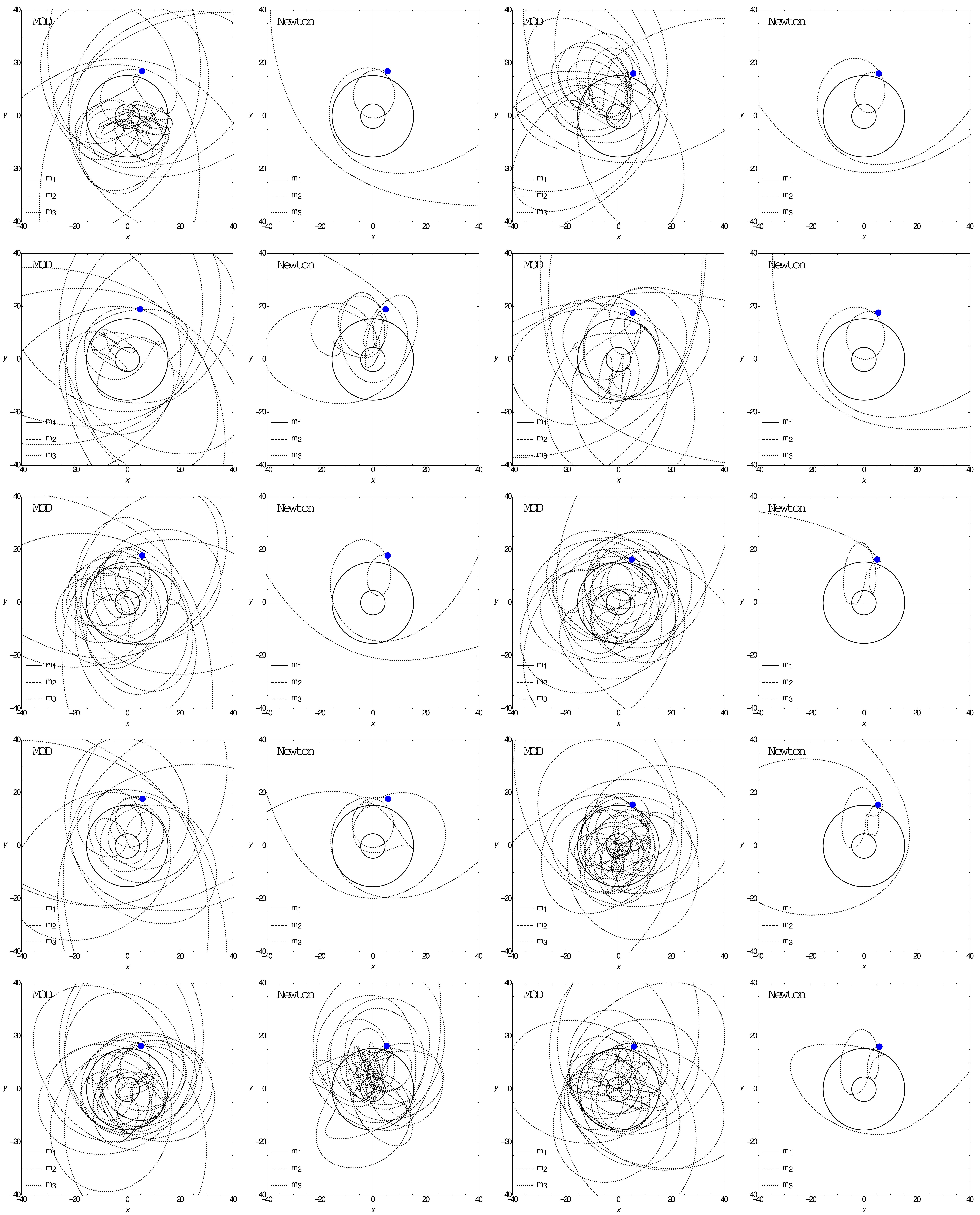}
\\
  \caption{The evolution of a test particle around two massive objects of $m_{1}=10~ \mathcal{M}$ and $m_{2}=3~ \mathcal{M}$ with the distance of $d=20~ \mathcal{D}$. The evolution time is $3000~\mathcal{T}$, the initial position of $m_{3}$ is chosen randomly to be within $(x_{0}\pm 0.1x_{0},~y_{0} \pm 0.1y_{0})$ and this particle rotates initially in the same direction of $m_{1} $ and $m_{2}$.  }
  \label{fig:mu02pl01}
\end{figure*}

\begin{figure*}
\centering
\includegraphics[width=17.5cm]{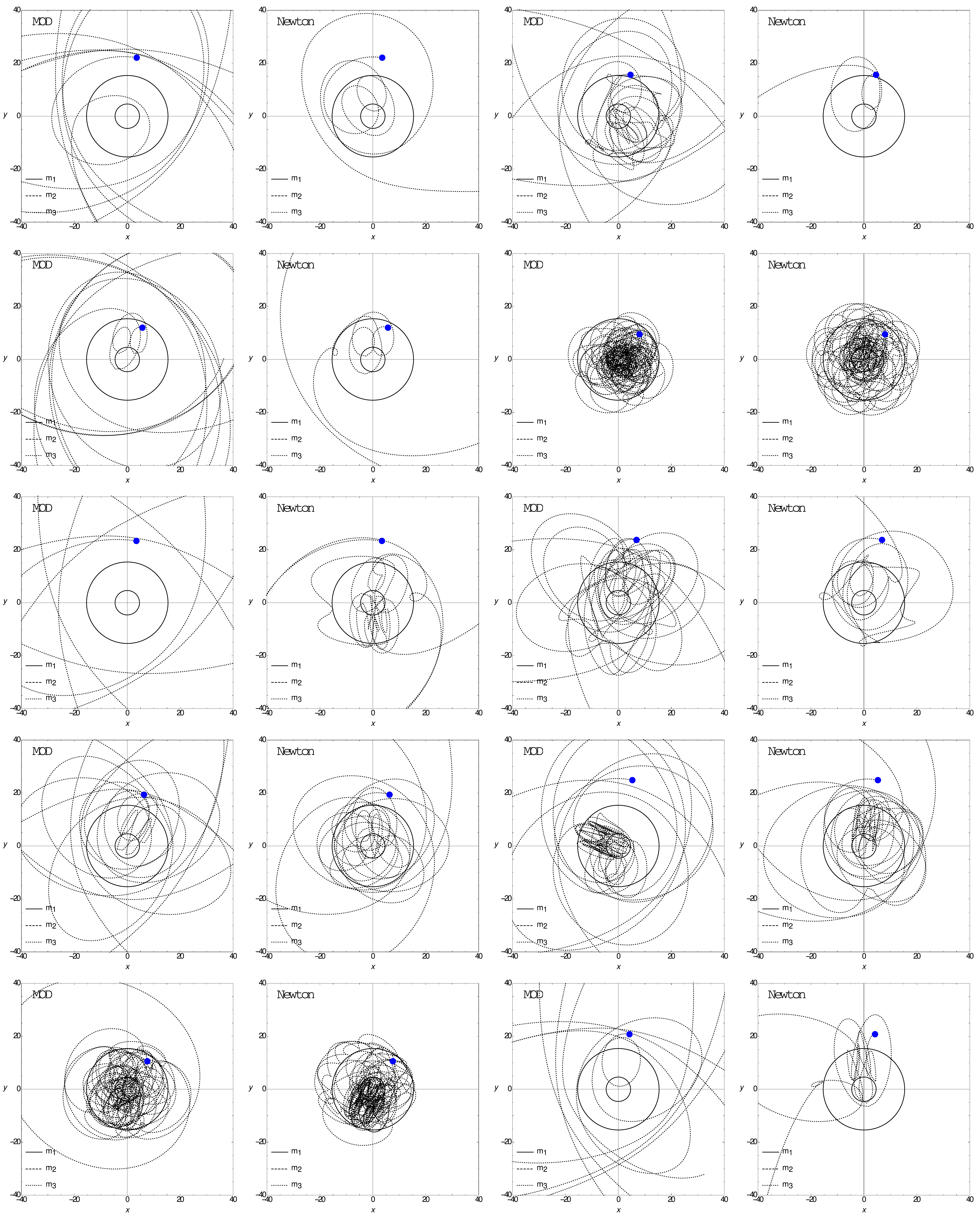}
\\
  \caption{The evolution of a test particle around two massive objects of $m_{1}=10~ \mathcal{M}$ and $m_{2}=3~ \mathcal{M}$ with the distance of $d=20~ \mathcal{D}$. The evolution time is $3000~\mathcal{T}$, the initial position of $m_{3}$ is chosen randomly to be within $(x_{0}\pm 0.5x_{0},~y_{0} \pm 0.5y_{0})$ and this particle rotates initially in the same direction of $m_{1} $ and $m_{2}$.   }
  \label{fig:mu02pl05}
\end{figure*}

\begin{figure*}
\centering
\includegraphics[width=17.5cm]{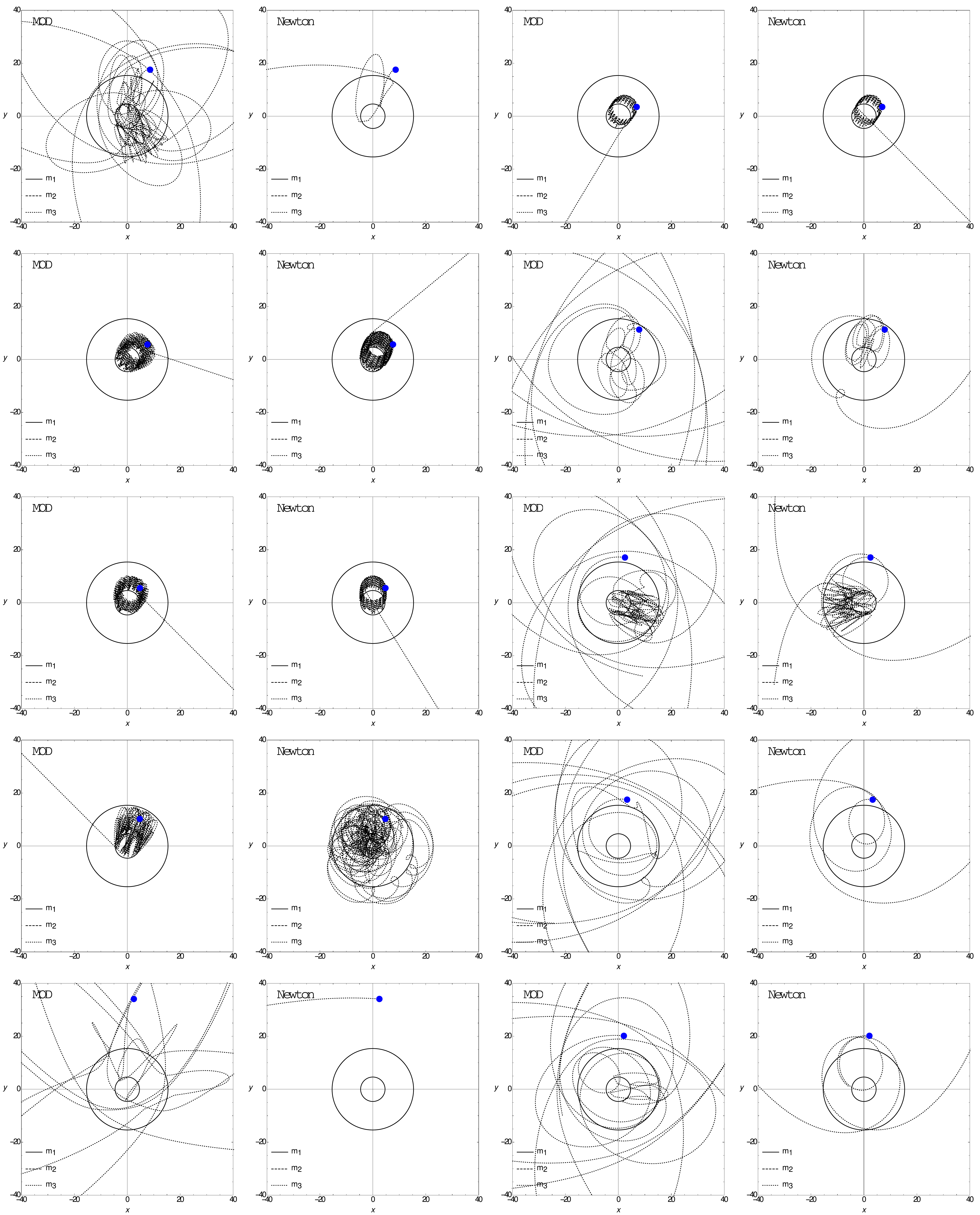}
\\
  \caption{The evolution of a test particle around two massive objects of $m_{1}=10~ \mathcal{M}$ and $m_{2}=3~ \mathcal{M}$ with the distance of $d=20~ \mathcal{D}$. The evolution time is $3000~\mathcal{T}$, the initial position of $m_{3}$ is chosen randomly to be within $(x_{0}\pm x_{0},~y_{0} \pm y_{0})$ and this particle rotates initially in the same direction of $m_{1} $ and $m_{2}$.  }
  \label{fig:mu02pl10}
\end{figure*}

\begin{figure*}
\centering
\includegraphics[width=17.5cm]{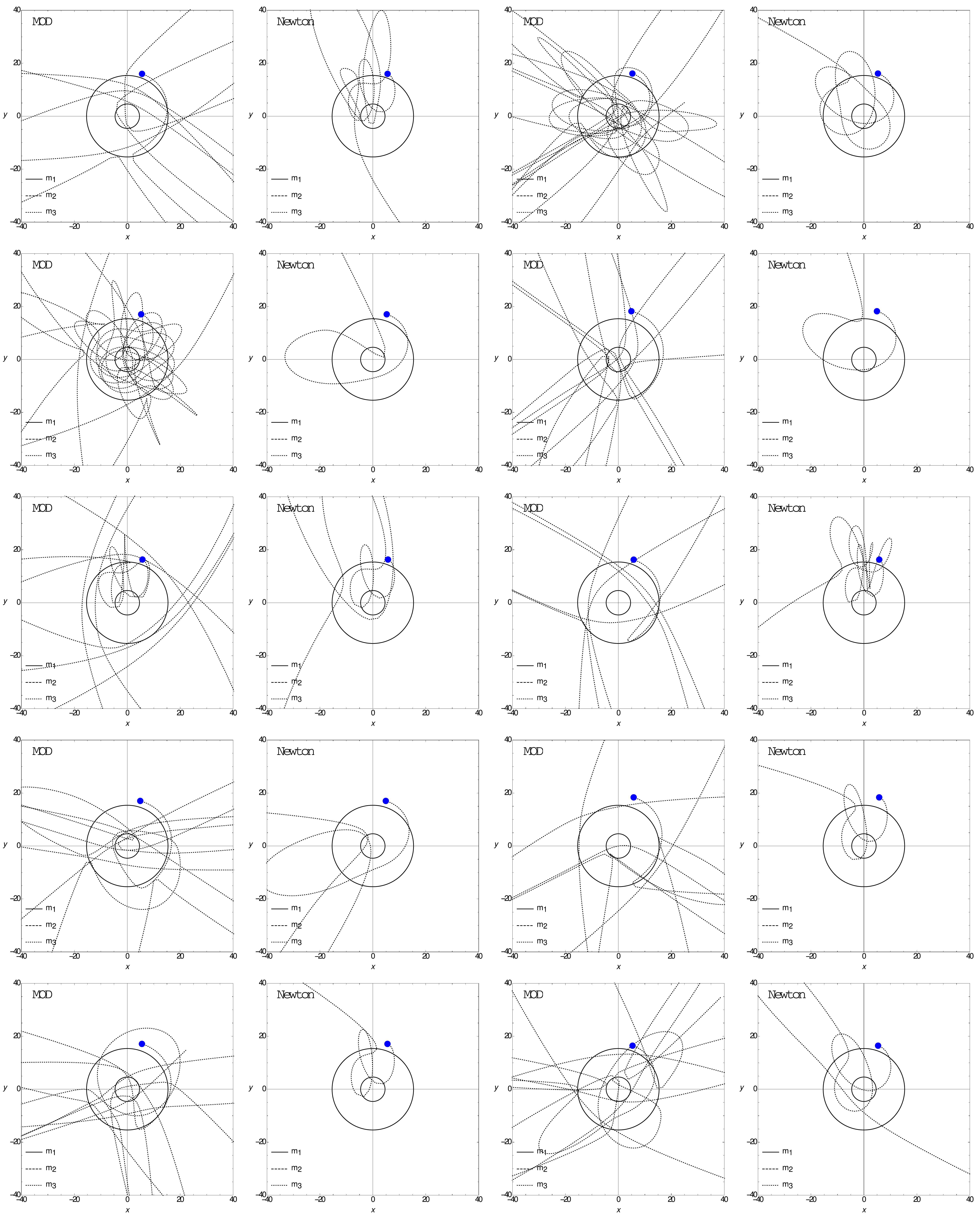}
\\
  \caption{The evolution of a test particle around two massive objects of $m_{1}=10~ \mathcal{M}$ and $m_{2}=3~ \mathcal{M}$ with the distance of $d=20~ \mathcal{D}$. The evolution time is $3000~\mathcal{T}$, the initial position of $m_{3}$ is chosen randomly to be within $(x_{0}\pm 0.1x_{0},~y_{0} \pm 0.1y_{0})$ and this particle rotates initially in the opposite direction of $m_{1} $ and $m_{2}$.  }
  \label{fig:mu02nl01}
\end{figure*}

\begin{figure*}
\centering
\includegraphics[width=17.5cm]{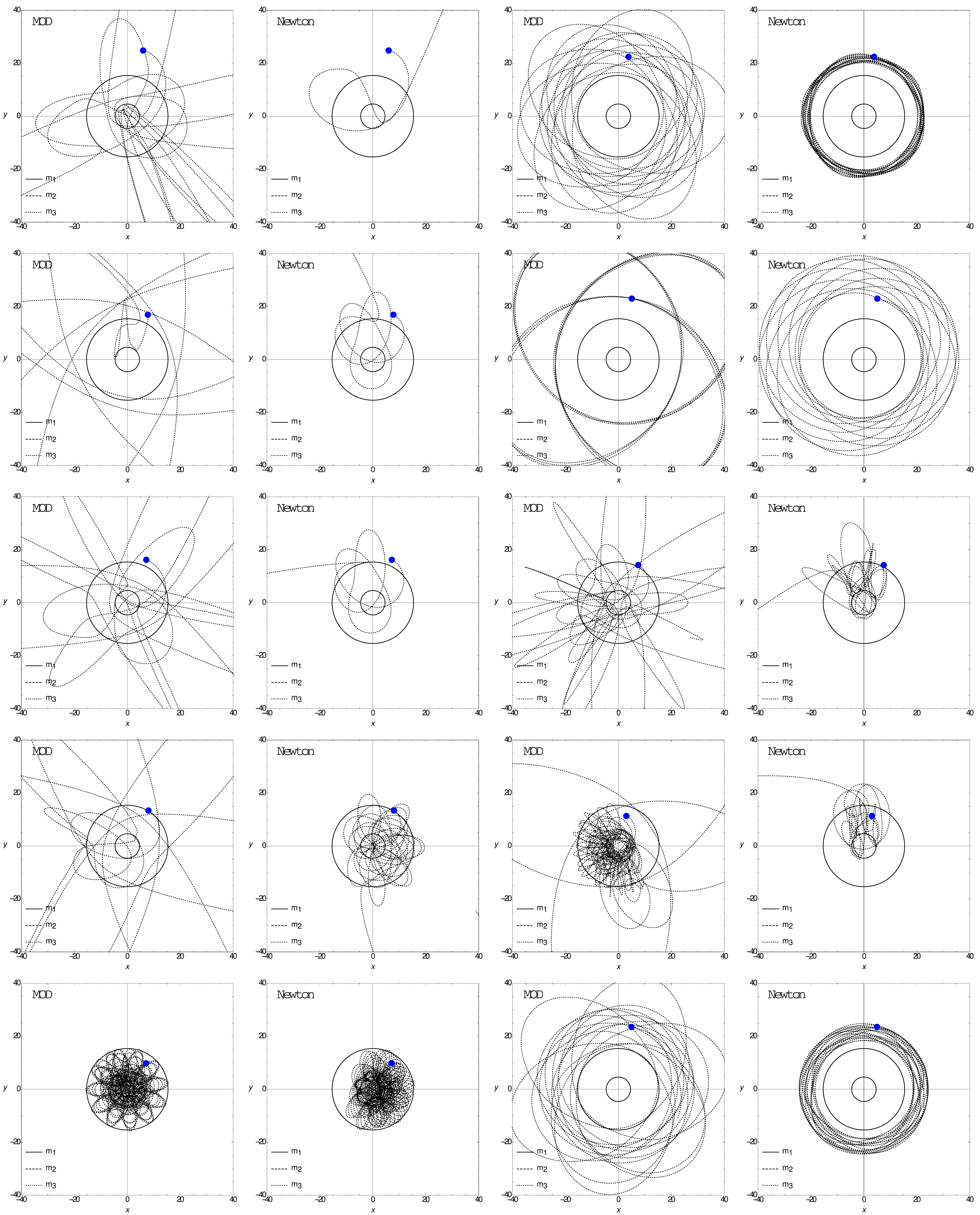}
\\
  \caption{The evolution of a test particle around two massive objects of $m_{1}=10~ \mathcal{M}$ and $m_{2}=3~ \mathcal{M}$ with the distance of $d=20~ \mathcal{D}$. The evolution time is $3000~\mathcal{T}$, the initial position of $m_{3}$ is chosen randomly to be within $(x_{0}\pm 0.5x_{0},~y_{0} \pm 0.5y_{0})$ and this particle rotates initially in the opposite direction of $m_{1} $ and $m_{2}$.  }
  \label{fig:mu02nl05}
\end{figure*}

\begin{figure*}
\centering
\includegraphics[width=17.5cm]{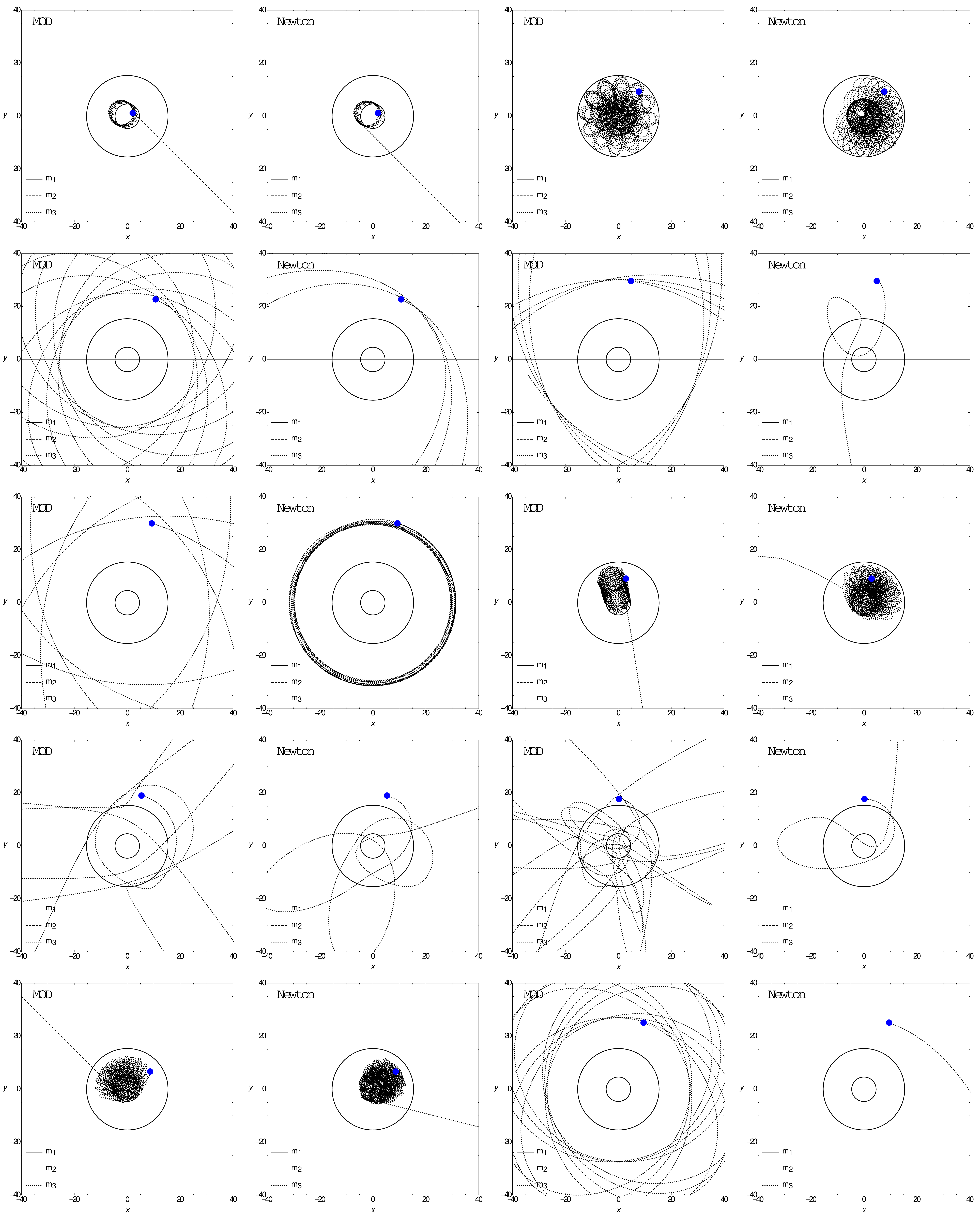}
\\
  \caption{The evolution of a test particle around two massive objects of $m_{1}=10~ \mathcal{M}$ and $m_{2}=3~ \mathcal{M}$ with the distance of $d=20~ \mathcal{D}$. The evolution time is $3000~\mathcal{T}$, the initial position of $m_{3}$ is chosen randomly to be within $(x_{0}\pm x_{0},~y_{0} \pm y_{0})$ and this particle rotates initially in the opposite direction of $m_{1} $ and $m_{2}$.  }
  \label{fig:mu02nl10}
\end{figure*}

\begin{figure*}
\centering
\includegraphics[width=17.5cm]{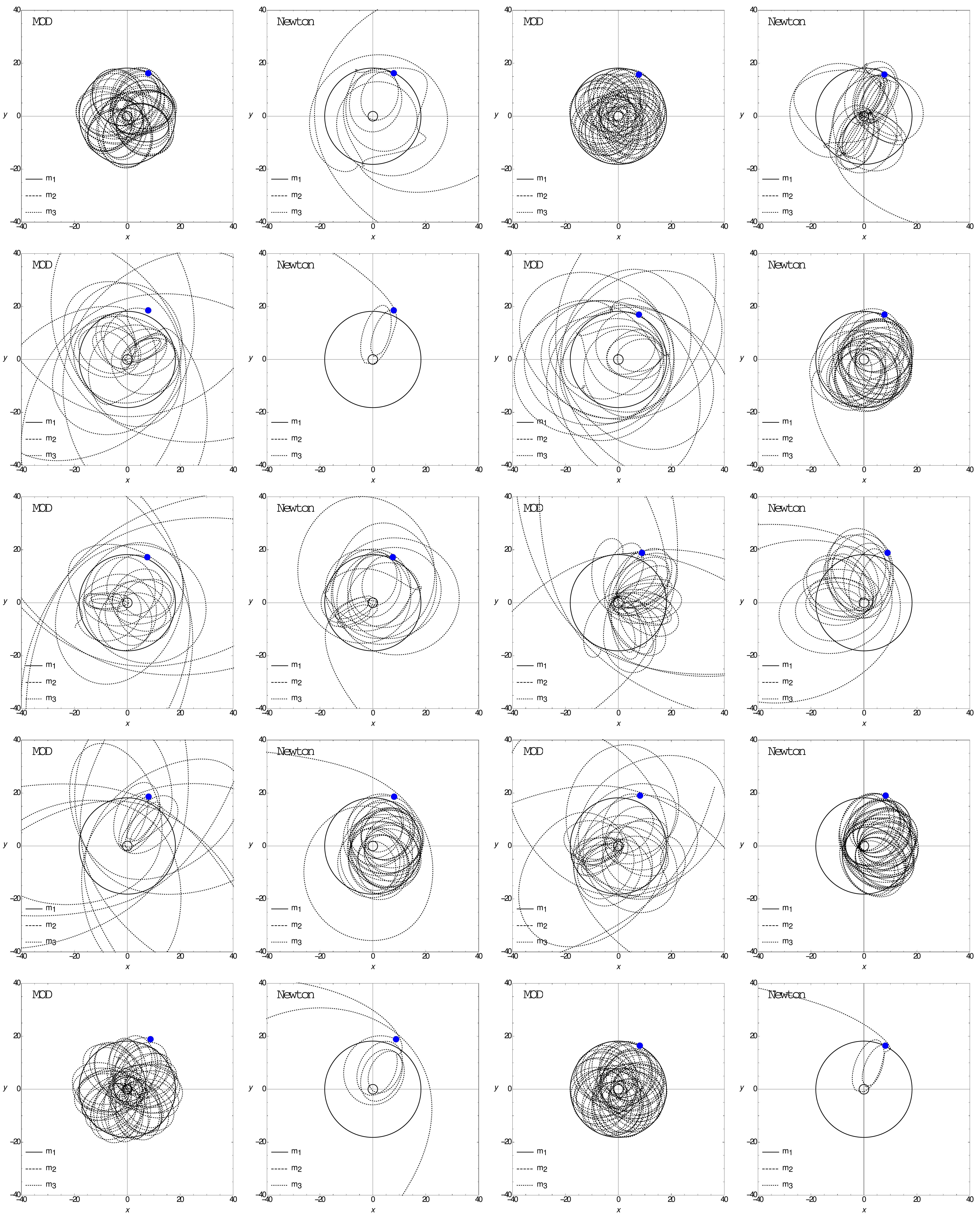}
\\
  \caption{The evolution of a test particle around two massive objects of $m_{1}=10~ \mathcal{M}$ and $m_{2}=1~ \mathcal{M}$ with the distance of $d=20~ \mathcal{D}$. The evolution time is $3000~\mathcal{T}$, the initial position of $m_{3}$ is chosen randomly to be within $(x_{0}\pm 0.1x_{0},~y_{0} \pm 0.1y_{0})$ and this particle rotates initially in the same direction of $m_{1} $ and $m_{2}$.  }
  \label{fig:mu01pl01}
\end{figure*}

\begin{figure*}
\centering
\includegraphics[width=17.5cm]{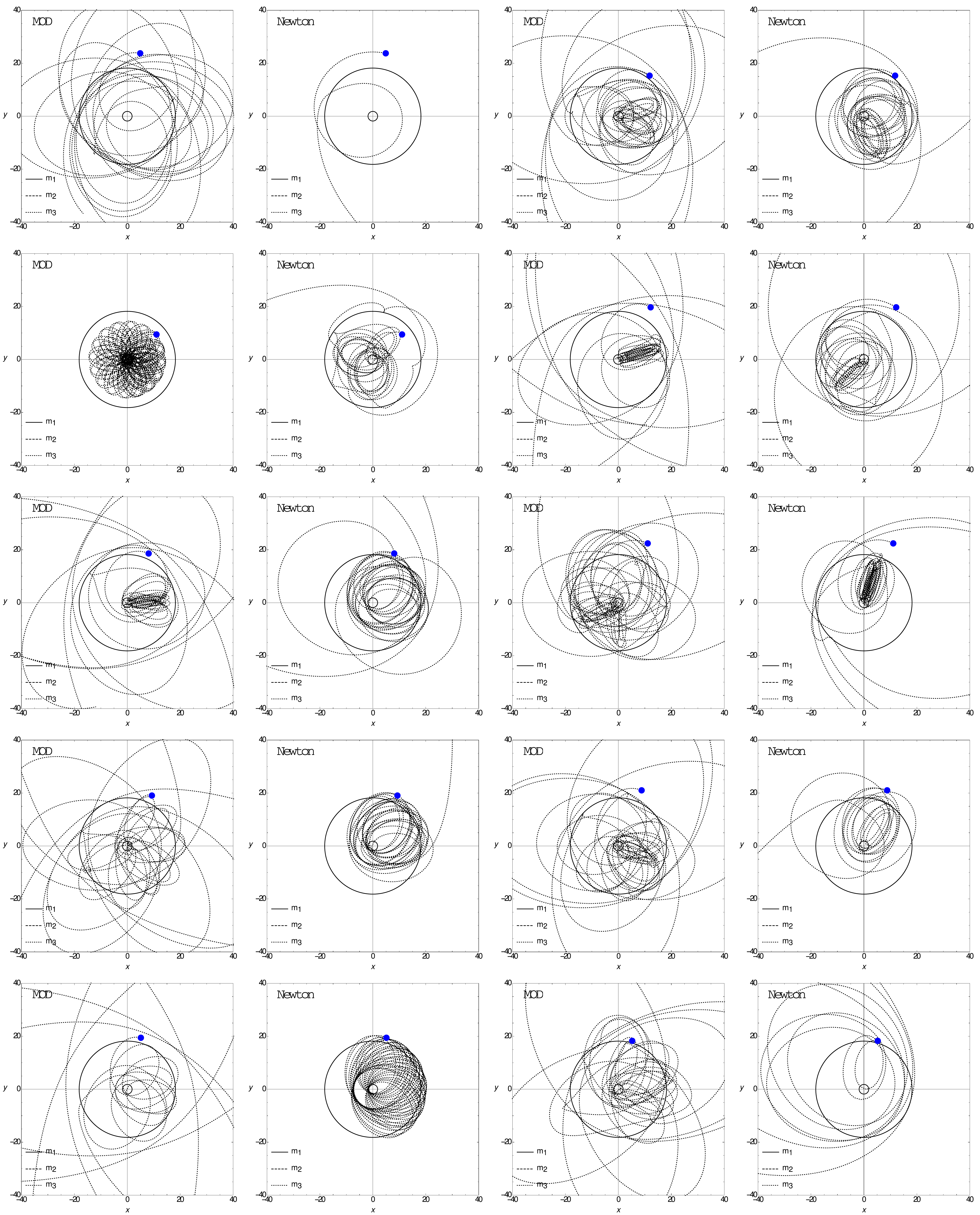}
\\
  \caption{The evolution of a test particle around two massive objects of $m_{1}=10~ \mathcal{M}$ and $m_{2}=1~ \mathcal{M}$ with the distance of $d=20~ \mathcal{D}$. The evolution time is $3000~\mathcal{T}$, the initial position of $m_{3}$ is chosen randomly to be within $(x_{0}\pm 0.5x_{0},~y_{0} \pm 0.5y_{0})$ and this particle rotates initially in the same direction of $m_{1} $ and $m_{2}$.   }
  \label{fig:mu01pl05}
\end{figure*}

\begin{figure*}
\centering
\includegraphics[width=17.5cm]{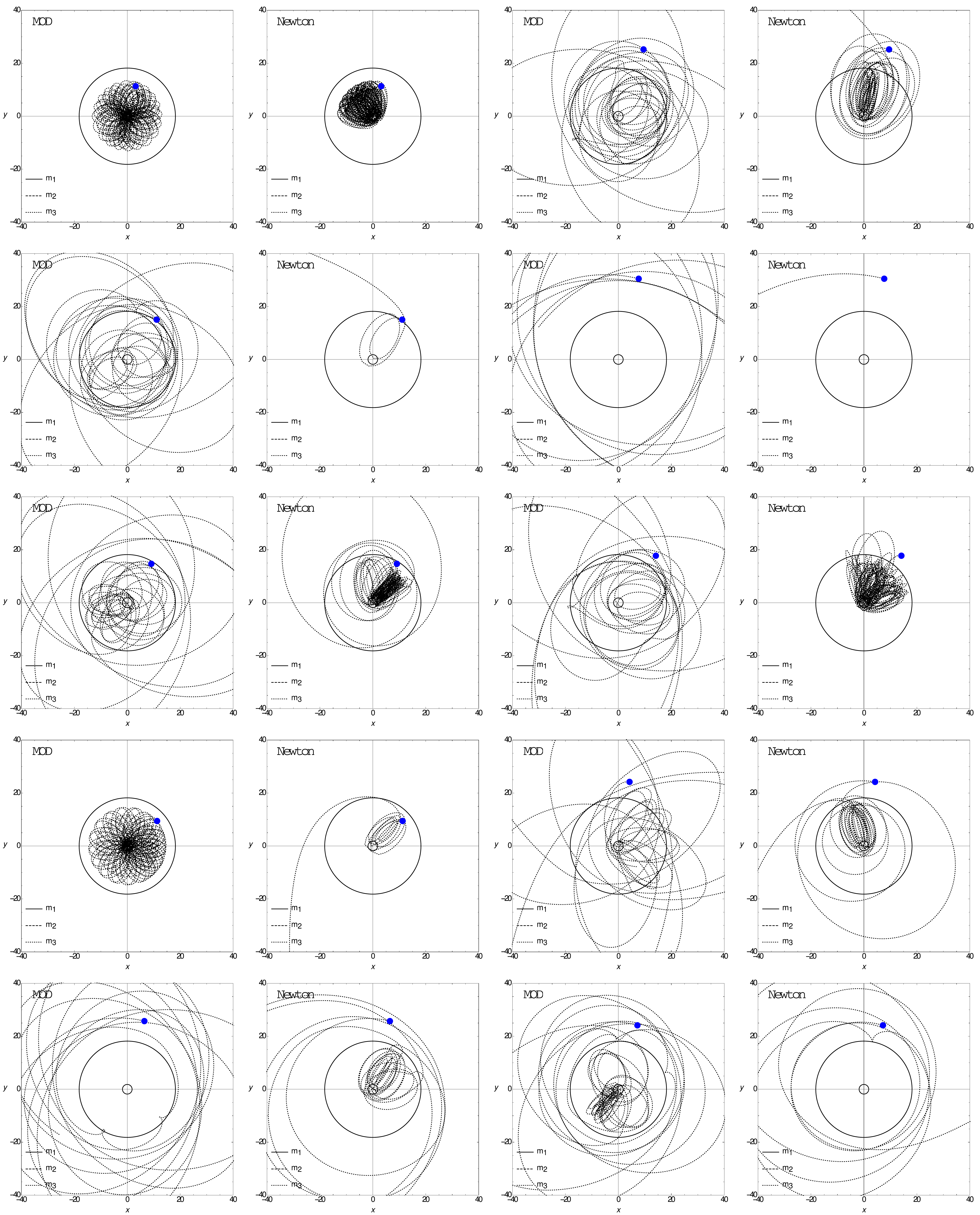}
\\
  \caption{The evolution of a test particle around two massive objects of $m_{1}=10~ \mathcal{M}$ and $m_{2}=1~ \mathcal{M}$ with the distance of $d=20~ \mathcal{D}$. The evolution time is $3000~\mathcal{T}$, the initial position of $m_{3}$ is chosen randomly to be within $(x_{0}\pm x_{0},~y_{0} \pm y_{0})$ and this particle rotates initially in the same direction of $m_{1} $ and $m_{2}$.  }
  \label{fig:mu01pl10}
\end{figure*}

\clearpage

\begin{figure*}
\centering
\includegraphics[width=17.5cm]{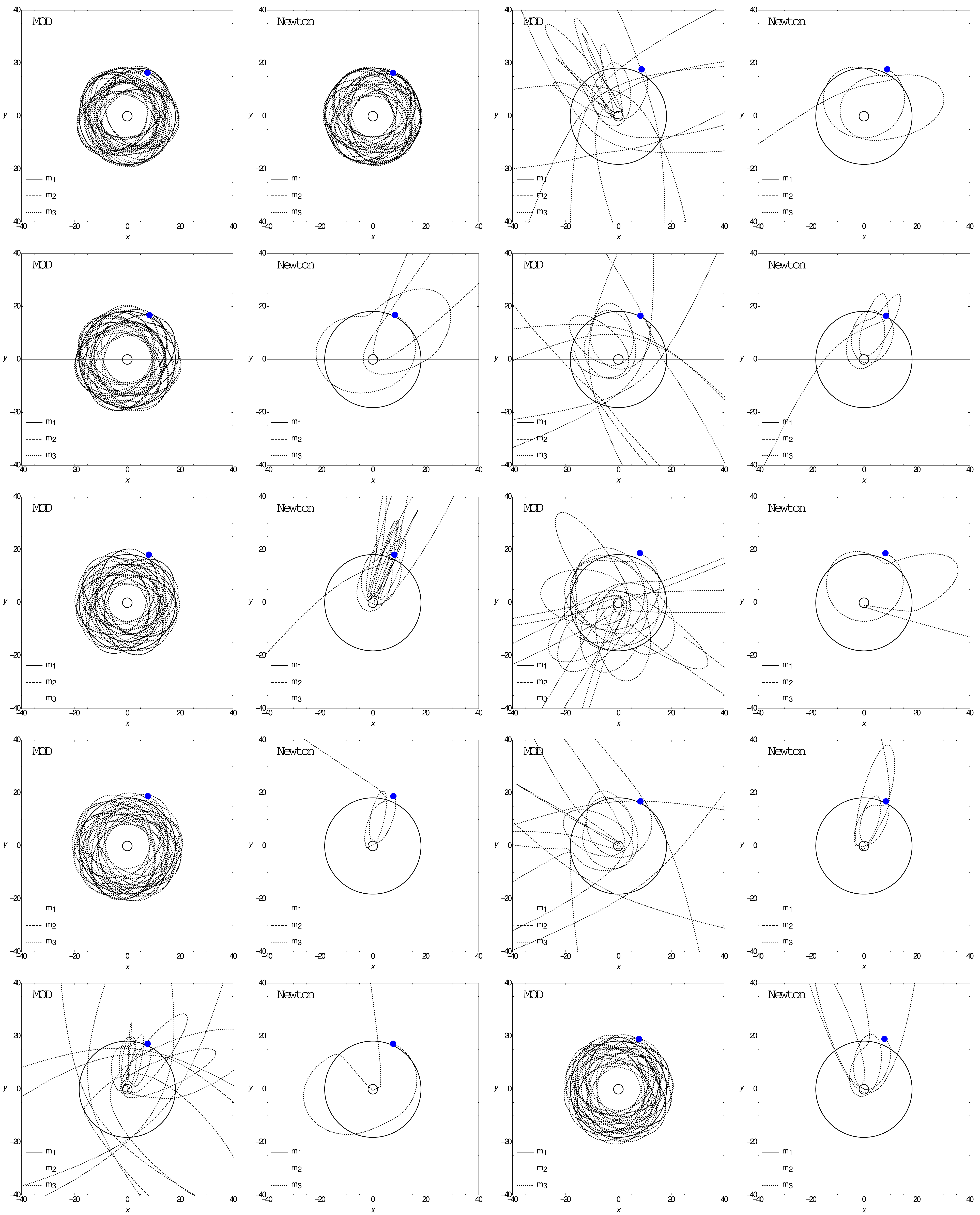}
\\
  \caption{The evolution of a test particle around two massive objects of $m_{1}=10~ \mathcal{M}$ and $m_{2}=1~ \mathcal{M}$ with the distance of $d=20~ \mathcal{D}$. The evolution time is $3000~\mathcal{T}$, the initial position of $m_{3}$ is chosen randomly to be within $(x_{0}\pm 0.1x_{0},~y_{0} \pm 0.1y_{0})$ and this particle rotates initially in the opposite direction of $m_{1} $ and $m_{2}$.  }
  \label{fig:mu01nl01}
\end{figure*}

\begin{figure*}
\centering
\includegraphics[width=17.5cm]{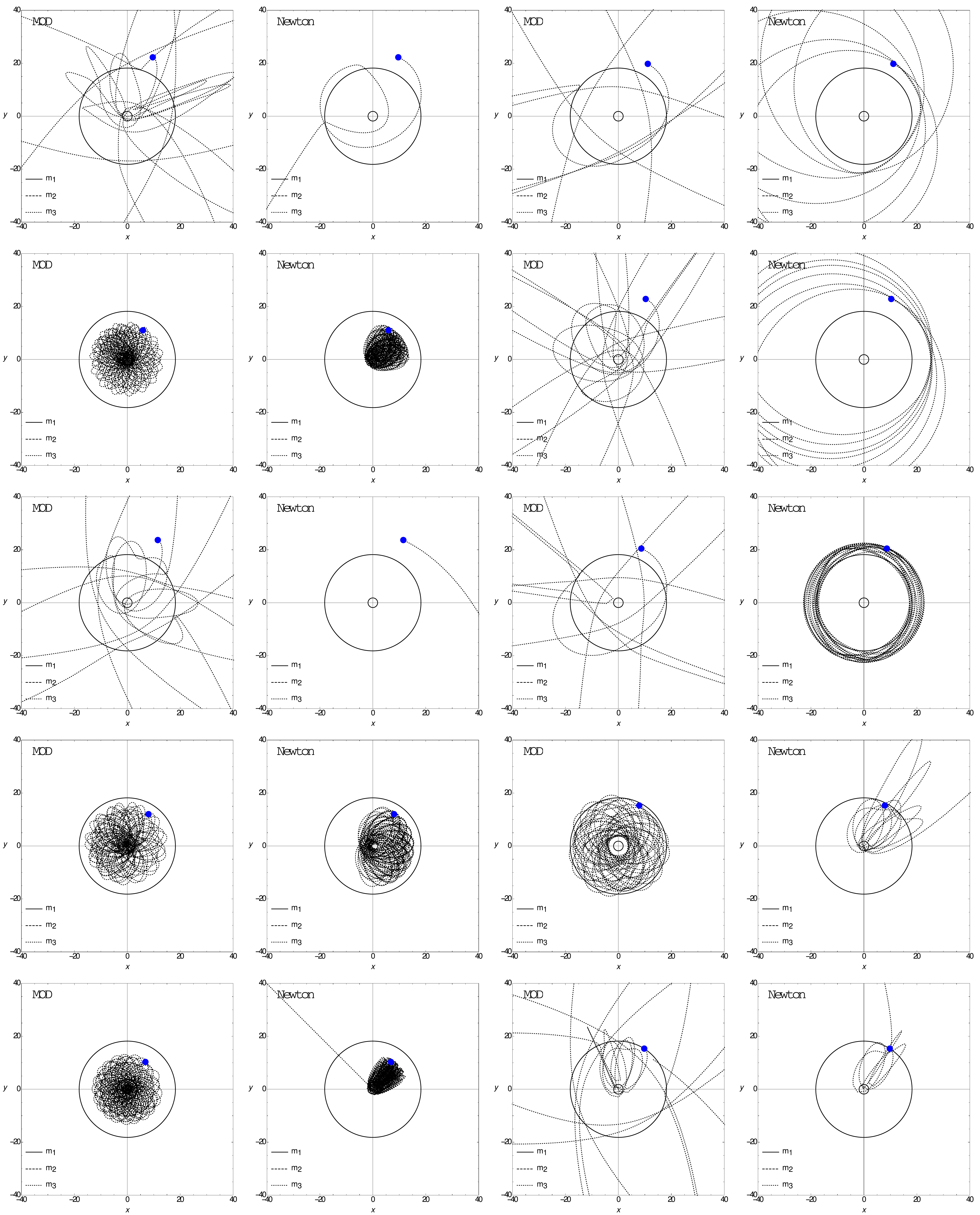}
\\
  \caption{The evolution of a test particle around two massive objects of $m_{1}=10~ \mathcal{M}$ and $m_{2}=1~ \mathcal{M}$ with the distance of $d=20~ \mathcal{D}$. The evolution time is $3000~\mathcal{T}$, the initial position of $m_{3}$ is chosen randomly to be within $(x_{0}\pm 0.5x_{0},~y_{0} \pm 0.5y_{0})$ and this particle rotates initially in the opposite direction of $m_{1} $ and $m_{2}$. }
  \label{fig:mu01nl05}
\end{figure*}

\begin{figure*}
\centering
\includegraphics[width=17.5cm]{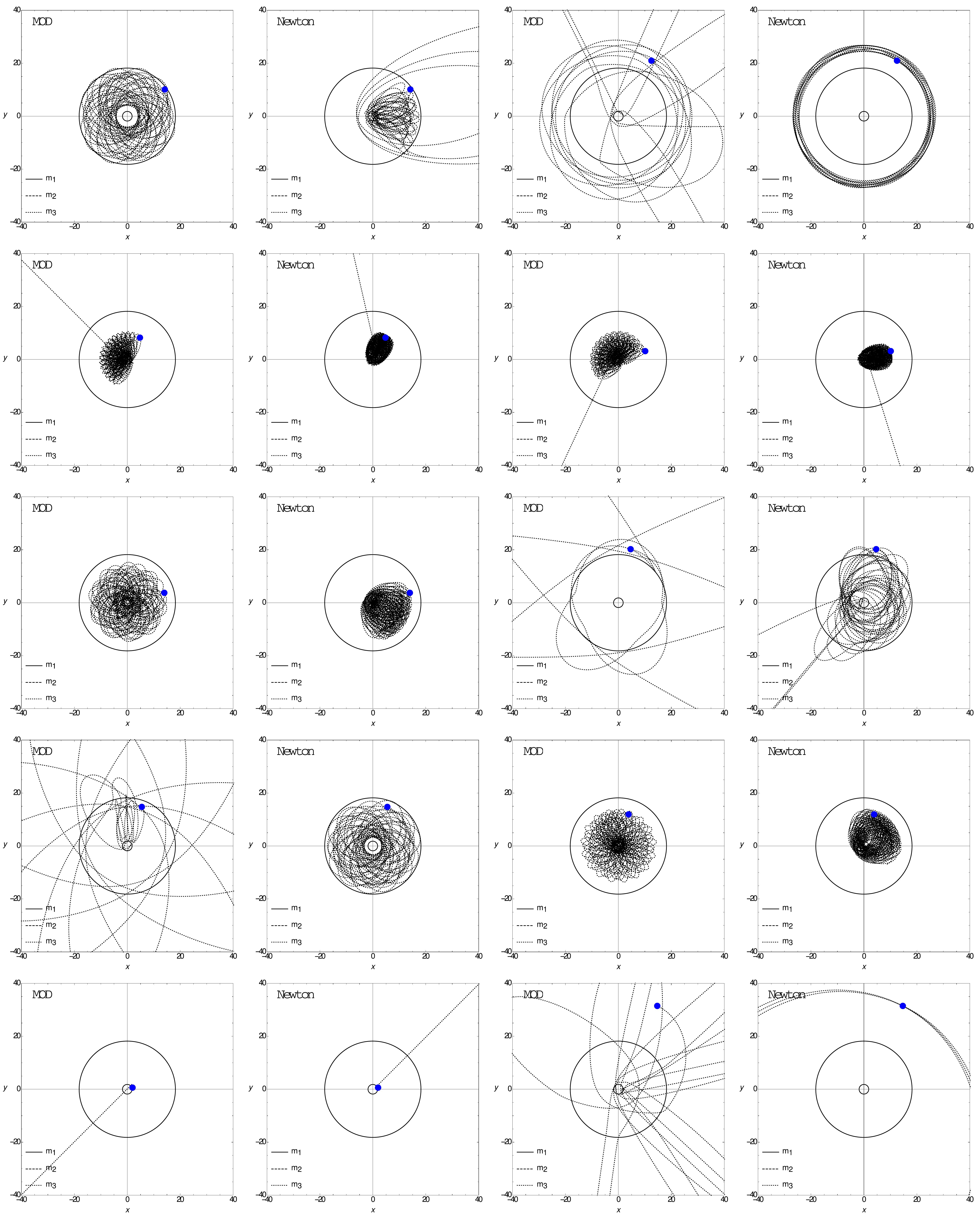}
\\
  \caption{The evolution of a test particle around two massive objects of $m_{1}=10~ \mathcal{M}$ and $m_{2}=1~ \mathcal{M}$ with the distance of $d=20~ \mathcal{D}$. The evolution time is $3000~\mathcal{T}$, the initial position of $m_{3}$ is chosen randomly to be within $(x_{0}\pm x_{0},~y_{0} \pm y_{0})$ and this particle rotates initially in the opposite direction of $m_{1} $ and $m_{2}$.  }
  \label{fig:mu01nl10}
\end{figure*}

\end{document}